\documentclass[11pt]{article}
\pdfoutput=1
\usepackage{amsfonts, amsthm, simplewick, datetime, multirow, array, booktabs,verbatim, jcappub, floatrow, caption, subcaption, mathabx,url}
\usepackage[font=footnotesize, labelfont={sf,bf}, margin=1cm]{caption}
\usepackage{latexsym}
\usepackage{graphicx, graphics, hyperref, amsmath, amssymb, slashed,mathabx}
\usepackage{color}

\newcommand{\nc}{\newcommand}

\nc{\beq}{\begin{equation}}
\nc{\eeq}{\end{equation}}
\nc{\barray}{\begin{eqnarray}}
\nc{\earray}{\end{eqnarray}}
\nc{\barrayn}{\begin{eqnarray*}}
\nc{\earrayn}{\end{eqnarray*}}
\nc{\bcenter}{\begin{center}}
\nc{\ecenter}{\end{center}}
\nc{\ket}[1]{| #1 \rangle}
\nc{\bra}[1]{\langle #1 |}
\nc{\0}{\ket{0}}
\nc{\mc}{\mathcal}
\nc{\er}[1]{(\ref{eq:#1})}
\nc{\onehalf}{\frac{1}{2}}
\nc{\partialbar}{\bar{\partial}}
\nc{\psit}{\widetilde{\psi}}
\nc{\Tr}{\mbox{Tr}}
\nc{\hc}{\mbox{H.c.}}
\nc{\ev}{\;\mathrm{eV}}
\nc{\mev}{\;\mathrm{MeV}}
\nc{\gev}{\;\mathrm{GeV}}
\nc{\tev}{\;\mathrm{TeV}}

\def\chii0{\chi_i^0}
\def\chij0{\chi_j^0}

\newcommand{\be}{\begin{equation}}
\newcommand{\ee}{\end{equation}}

\newcommand{\Tcore}{T_{\text{core}}}

\title{A Solar System Test of  Self-Interacting Dark Matter
}
\author{Cristian Gaidau}
\author{and Jessie Shelton}

\affiliation{Department of Physics, University of Illinois at Urbana-Champaign,\\  1110 W Green St., Urbana, IL 61801, USA}

\emailAdd{gaidau2@illinois.edu}
\emailAdd{sheltonj@illinois.edu}

\date{\today}

\abstract{ Dark matter (DM) self-interactions affect the gravitational
  capture of DM in the Sun and Earth differently as a simple
  consequence of the differing kinematics of collisions within the two potential wells: the dominant effect of
  self-interactions in the Sun is to provide an additional channel for
  capture, while the dominant effect in the Earth is to eject
  previously captured DM. We point out that this simple observation
  can be used to deduce the existence of DM self-interactions by
  comparing the annihilation rates of DM gravitationally bound within
  the Sun and Earth. We compute the Sun and Earth annihilation fluxes
  for DM with spin-independent nuclear cross-sections and thermal
  annihilation cross-sections and demonstrate that, for cross-sections
  allowed by direct detection, self-interactions can easily suppress
  the expected Earth flux by multiple orders of magnitude.  This
  suppression is potentially significant even for self-interaction cross-sections
  orders of magnitude below the Bullet Cluster bounds, making this
  solar system comparison a leading test of dark matter
  self-interactions. Additionally, we consider thermalization of the
  captured DM population with the nuclei of the capturing body in some
  detail, accounting for both nuclear and self-interactions, and point
  out some consequential and broadly applicable considerations.  }

\begin{document}

\maketitle

\section{Introduction}

The nature of dark matter (DM) is one of the biggest outstanding
mysteries in particle physics.  The possibility that DM may have
sizeable self-interactions is especially intriguing.  Sufficiently
strong dark matter self-interactions can affect galaxy formation and
structure \cite{Spergel:1999mh}, and may thereby explain several
outstanding discrepancies between the properties of dwarf galaxies in
observations versus (DM-only) simulations \cite{Tulin:2017ara,
  Bullock:2017xww}.  More broadly, strong self-interactions are a
generic property that DM may have, and establishing the existence of
such self-interactions would be a major step towards understanding the
particle nature of DM.

In this article we point out that the kinematics of our solar system
enable a simple test of dark matter self-interactions through a
comparison of the DM populations gravitationally bound within the Sun
and the Earth.  Dark matter that scatters within a massive body can
lose enough kinetic energy in the scattering to become gravitationally
bound
\cite{Spergel:1984re,griest1987cosmic,Gould:1987ju,Gould:1987ir,gould1988direct}.
Thus massive bodies can build up a population of bound DM particles,
which can be detected either through their impact on the properties of
the massive body, or by observing the products of their annihilation
within the massive body.

The number $N$ of gravitationally captured dark matter particles in the
massive body is given in general by
\begin{align}
\frac{dN}{dt} = C_{\text{c}} + (C_{\text{sc}} - C_{\text{se}}- C_\text{e})N -\left(C_{\text{ann}} + C_{\text{sevap}}\right) N^2.
\label{eq:dNdtgen}
\end{align}
Here the $C_i$ are rates for scattering processes that change
the number of gravitationally bound DM particles.  Most familiarly,
$C_{\text{c}}$ is the rate of capture of DM particles by elastic
scattering on nuclei, $C_{\text{e}}$ is the rate at which captured DM
particles evaporate by scattering against nuclear targets, and
$C_{\text{ann}}$ is the rate at which captured particles annihilate.
In the presence of DM self-interactions, three additional scattering
processes can become relevant: $C_{\text{sc}}$ and $C_{\text{se}}$,
the rate for DM self-capture and self-ejection, respectively
\cite{Zentner:2009is}, and $C_{\text{sevap}}$ \cite{chen2014probing},
the rate for DM self-evaporation.  For the Sun, self-capture dominates
over self-ejection, but the shallowness of the Earth's potential well
ensures that self-ejection dominates.

At fixed DM mass and annihilation cross-section, the annihilation flux
of non-self-interacting DM is entirely determined by the nuclear
cross-section.  In the presence of DM self-interactions, the
annihilation flux also depends on the self-scattering cross-section,
and therefore measurements of the annihilation flux from two different
massive bodies are needed to determine both cross-sections.  As
self-interactions have opposite effects on the populations in the
Earth and the Sun, the imprint of self-interactions can be dramatic
and unambiguous in parameter space of physical interest.

In this paper we will establish the power and utility of this general
observation.  Toward this end we will keep our discussion as
model-independent as possible. In particular, although for
definiteness we will assume that DM is a thermal relic with an
$s$-wave annihilation cross-section, we will remain agnostic about the
final states to which it annihilates.  Of course, directly measuring
the DM annihilation rate in the Sun (Earth) is possible only if some
of the annihilation products are sufficiently weakly interacting to
escape the Sun (Earth). Well-established search strategies rely on the
neutrinos produced by annihilations into Standard Model (SM) final states
\cite{Silk:1985ax, Gaisser:1986ha, Bell:2011sn,
  Rott:2012qb,Bernal:2012qh,Choi:2015ara,Aartsen:2016zhm,Adrian-Martinez:2016gti},
while in multi-state dark sectors long-lived dark states could also
furnish this role \cite{Batell:2009zp, Schuster:2009au,
  Meade:2009mu,Ajello:2011dq,
  Berger:2014sqa,Feng:2015hja,Feng:2016ijc,Adrian-Martinez:2016ujo,
  Smolinsky:2017fvb,Arina:2017sng,Leane:2017vag}.  Such multi-state
dark sectors are often proposed in order to realize strong DM
self-interactions, and are especially well-motivated in this context.

Additionally, we will assume for simplicity that DM interacts
sufficiently strongly with nuclei in the massive body to reach thermal
equilibrium in the core.  This assumption allows the number of
captured particles to be factorized according to $N(t,r) \equiv N(t)
n_\text{c}(r)$, where the DM radial profile $n_\text{c}(r)$ is
time-independent and unit-normalized:
\begin{align}
\label{eq:thermaldist}
n_\text{c}(r) = A\exp{\left(-\frac{M_{\text{DM}}\Phi(r)}{T_{\text{core}}}\right)},
\end{align} 
where $M_{\text{DM}}$ is the DM mass, $\Phi (r)$ is the gravitational
potential, $\Tcore$ is the core temperature, and $A$ is a constant
normalization factor.  When thermalization of captured DM is a good
approximation, all the $C_i$ in Eq.~\ref{eq:dNdtgen} become
constant, vastly simplifying the determination of $N(t)$.  As
DM-nuclear cross-sections $\sigma_\text{p}$ have become more and more
constrained by direct detection experiments, the available parameter
space that allows captured DM to thermalize in the Sun is becoming
notably restricted.  We consider thermalization through
spin-independent nuclear scatterings in detail, both with and without
self-interactions, and establish some general criteria for the
self-consistency of the thermal description.  For nuclear scatterings,
we find more restrictive thermalization criteria than some previous
estimates.  We also find that in the presence of DM
self-interactions, energy exchange between bound and halo DM
populations can be sizeable but, for constant self-interaction cross-sections,
poses a subleading obstacle to achieving
thermalization.

The organization of this paper is as follows. In
Sec.~\ref{sec:gravcap} we discuss the gravitational capture of DM in
the solar system, establishing our conventions.  We quantify the
impact of DM self-interactions on the annihilation flux from the Sun
and the Earth in Secs.~\ref{sec:sunmod} and~\ref{sec:earthmod}
respectively, specializing for illustration to an isotropic
self-interaction cross-section.  Our main results are presented in
Sec.~\ref{sec:selfints}, where we show how comparing annihilation
fluxes from the Sun and the Earth can reveal the presence of DM
self-interactions. In Sec.~\ref{sec:conclusions} we conclude. Two
appendices contain technical aspects of the calculations:
App.~\ref{sec:models} documents our models of the Earth and the Sun,
while App.~\ref{sec:therm} contains a detailed examination of DM
thermalization within a capturing body, both with and without
self-interactions.

\section{Gravitational capture of DM in the solar system}
\label{sec:gravcap}

We begin by reviewing the capture and annihilation of non-self-interacting DM in the Sun
and then discuss how adding self-interactions alters the evolution of
the captured DM population.  We then discuss capture and annihilation
of self-interacting DM in the Earth, highlighting the different impacts of
self-interactions on the populations in the Earth and the Sun.

\subsection{DM capture and annihilation in the Sun}
\label{sec:review}

Here we review the calculation of the nuclear capture,
annihilation, and evaporation rates of DM in the Sun, following the
pioneering treatment of Gould
\cite{Gould:1987ju,Gould:1987ir,gould1988direct}.  

We begin with nuclear capture. We take the local density of DM to be
$\rho_\odot=0.4\,\mathrm{GeV}/\mathrm{cm}^3$, and assume that far from
the Sun the DM halo has a Maxwell-Boltzmann distribution $f(u)$ of
speeds $u$,
\begin{align}
f(u)d^3u = 4\pi\left(\frac{3}{2\pi\bar{v}^2}\right)^{3/2}\exp{\left(-\frac{3u^2}{2\bar{v}^2}\right)}u^2du = \frac{4}{\sqrt{\pi}}x^2e^{-x^2}dx,
\end{align}
where $\bar{v}^2$ is the DM rms speed, and we have introduced the
dimensionless speed $x^2 = 3/(2\bar{v}^2)\, u^2$. We assume that the
DM halo is virialized so that $\bar{v}^2 = 3/2 \,v^2_{\text{LSR}}$,
where $v_{\text{LSR}}$ is the Local Standard of Rest at the position
of the Sun. Recent estimates place $v_{\text{LSR}} = 235 \text{
  km}/\text{s}$ \cite{bovy2009galactic, reid2009trigonometric,
  mcmillan2010uncertainty}, so that $\bar{v} = 288 \text{
  km}/\text{s}$.  Meanwhile, given the Sun's peculiar velocity
${\bf{v}}_{\text{pec}}=(11,12,7) \text{ km}/\text{s}$
\cite{schonrich2010local}, it moves at $\tilde{v} = 247 \text{
  km}/\text{s}$ with respect to the DM halo. These velocities are
collected for easy reference in Table~\ref{table:Parameters} in the
Appendix. We define $\vec{\eta}= \sqrt{
  3/(2\bar{v}^2)}\,\vec{\tilde{v}}$ for the dimensionless velocity of
the Sun, and perform a Galilean transformation $\vec{x} \rightarrow
\vec{x}+\vec{\eta}$ to express the DM velocity distribution in the
Sun's rest frame,
\begin{align}
\label{eq:halosunframe}
f_{\eta}(x)d^3x = \frac{4}{\sqrt{\pi}}x^2e^{-x^2}e^{-\eta^2}\frac{\sinh{(2x\eta)}}{2x\eta}dx,
\end{align}
where averaging over the solid angle has been performed.\footnote{We
  neglect the galactic escape velocity, as capture rates are
  insensitive to the high-velocity tail of the DM distribution.}

As the DM particle falls into the Sun's potential well, its speed
increases. Letting $r$ denote the distance between the DM particle and
the center of the Sun, the instantaneous speed $w$ of the DM particle
is given in terms of the local escape velocity $v_{\text{esc}}(r)$ by $
w^2(r) = u^2 + v^2_{\text{esc}}(r)$.  Once the particle reaches the surface
of the Sun, it can scatter off of nuclei in the Sun.  In this work we will assume for simplicity a constant spin-independent DM-nucleon
cross-section $\sigma_\text{p}$. Spin-independent cross-sections follow from generic scenarios such as  
e.g. Higgs portal or kinetic mixing portal interactions between DM and the SM,
and are important for obtaining sizeable nuclear capture rates in the Earth, as we will discuss in Sec.~\ref{sec:earth} below.

For a spherical shell of
radius $r$ within the Sun, the nuclear capture rate due to species $i$
per unit shell volume can then be written as \cite{Gould:1987ir}
\beq
\frac{dC_{\text{c},i}}{dV} = n_{\text{DM}} \int_0^\infty du \,f_\eta(u) \frac{w}{u}  \Omega_{i} (w),
\eeq
where $n_{\text{DM}} = \rho_\odot/M_{\text{DM}}$ is the local number density of DM, and $
\Omega_{i} (w)$ is the capture rate for a DM particle with velocity $w(r)$,
\beq
\Omega_{i} (w) = n_{i}(r) \sigma_{\text{cap}} w (r).
\eeq
Here $n_i(r)$ is the local density of target nuclei of species
$i$, and $\sigma_ {\text{cap}}$ is the cross-section for DM to undergo
a nuclear scattering that results in capture.  This cross-section can
be found straightforwardly by integrating the DM-nucleus cross-section
over the portion of phase space where the DM particle loses enough
energy in a nuclear collision to be gravitationally bound.  To ensure
that a nuclear scattering leads to gravitational capture, we must have
the final velocity of the DM particle be less than the local escape
velocity, or, in other words, the DM particle's fractional energy loss
$\Delta E/E$ must satisfy
\begin{align}
\label{eq:ncapmin}
\frac{\Delta E}{E} \geq \frac{u^2}{w^2(r)}.
\end{align}
Meanwhile the maximum possible energy loss is fixed by kinematics,
\begin{align}
\label{eq:ncapmax}
\frac{\Delta E}{E} \leq   \frac{\mu_{i}}{\mu^2_{i, +}},
\end{align}
where, following Gould, we have defined
\beq
\mu_{i} \equiv \frac{M_{\text{DM}}}{m_{i}},\phantom{space} \mu_{i,\pm} \equiv \frac{\mu_{i} \pm 1}{2},
\eeq
where $m_i$ is the mass of element $i$.  To describe nuclear
scattering, we employ an exponential (Helm) form factor following
Ref.~\cite{Jungman:1995df},
\begin{align}
\label{eq:dsigmadE}
\left(\frac{d\sigma}{d(\Delta E)}\right)_{i} = \sigma_{0,i}\frac{\mu^2_{i,+}}{\mu_{i}}\frac{2}{M_{\text{DM}}w^2}F_{i}^2(\Delta E),
\end{align}
where $\Delta E$ is again the energy transfer in the collision and the form
factor is
\begin{align}
F_{i}^2(\Delta E) = \exp\left(-\frac{\Delta E}{E_{0,i}}\right). 
\end{align}
Here $E_{0,i}$ is given by
\beq
\quad E_{0,i} \equiv \frac{3}{2m_{i}R^2_{0,i}},
\eeq
where the nuclear radius is modeled as \cite{Jungman:1995df}
\begin{align}
 R_{0,i} = 10^{-13}\,\mathrm{cm}\times\left( 0.3 + 0.91\left(\frac{m_{i}}{\text{GeV}}\right)^{1/3}\right).
\end{align}
Finally, we specify $\sigma_{\text{0}}$ as follows:
\begin{align}
\sigma_{\text{0},i} = \int_0^{4m_{\text{r}}^2w^2}\left(\frac{d\sigma(q=0)}{d|\vec{q}|^2}\right)d|\vec{q}|^2 = \frac{4m_{\text{r}}^2}{\pi}[Z_{i}f_{\text{p}} +(A_{i}-Z_{i})f_{\text{n}}]^2,
\end{align}
where $m_{\text{r}} =
m_{i}M_{\text{DM}}/(m_{i}+M_{\text{DM}})$ is the reduced
mass.  We restrict ourselves to the isospin-conserving case of
$f_{\text{n}} =f_{\text{p}}$ for simplicity, and write $\sigma_{\text{0},i}$
in terms of the effective per-nucleon cross section
$\sigma_{\text{p}}$,
\begin{align}
 \sigma_{\text{0},i} = \sigma_{\text{p}}A_{i}^2\frac{m_{\text{r},i}^2}{m_{\text{r},\text{p}}^2} =  \sigma_{\text{p}}A_{i}^2\frac{M_{\text{DM}}^2m_{i}^2}{(M_{\text{DM}}+m_{i})^2}\frac{(m_{\text{p}}+M_{\text{DM}})^2}{m_{\text{p}}^2M_{\text{DM}}^2},
\end{align}
where $m_{\text{p}}$ is the proton mass. 

For an isotropic scattering cross-section $\sigma_{\text{p}}$, $\Delta
E$ is uniformly distributed between $0$ and the kinematic maximum,
meaning that, collision by collision, elements closest in mass to DM
will most efficiently contribute to the capture rate.  (The slight departure from isotropy in {\em nuclear} scatterings resulting from the form factor suppression of the most energetic scatterings does not alter this conclusion.) The sheer
abundance of helium in the Sun ensures that it dominates the capture
rate for light ($M_{\text{DM}} < 36.6$ GeV) DM, while oxygen dominates
for heavier DM.  For further detail, see Fig.~\ref{plot:Sun_CCapture}
in Appendix~\ref{sec:appsun}.  The full capture rate is obtained by
integrating the capture rate for each radial shell over the volume of
the Sun and summing over all relevant nuclear species,
\begin{align}
C_{\text{c}} = \sum_{i}\left(\int_0^{R_{\odot}}d^3r\frac{dC_{\text{c},i}}{dV}\right).
\end{align}
The nuclear capture coefficient $C_{\text{c}}$ is directly
proportional to the DM-nucleon cross-section $\sigma_{\text{p}}$ and
to the local DM number density $n_{\text{DM}}$.  For heavy DM
($M_{\text{DM}}\gtrsim$ 100 GeV), $C_{\text{c}}$ falls off as
$1/M_{\text{DM}}^2$, which can be simply understood: one factor of
$1/M_{\text{DM}}$ comes from the kinematic suppression in the capture
probability $\Omega_{i}(w)$, while the other factor comes from
the decreasing number density of DM with increasing mass,
$n_{\text{DM}} = \rho_{\odot}/M_{\text{DM}}$.

The evaporation coefficient $C_{\text{e}}$ is also directly
proportional to the DM-nucleon cross-section, and additionally depends
on the spatial distribution of DM within the Sun.  Assuming this
distribution is (near-)thermal, $C_{\text{e}}$ becomes important for
DM masses below $M_{\text{DM}}\lesssim 4$ GeV
\cite{Gould:1987ju,Busoni:2013kaa}, and falls off exponentially with
increasing mass.

Meanwhile, the annihilation rate $\Gamma_{\text{ann}}$ is directly
proportional to the thermally-averaged DM annihilation cross-section
$\langle\sigma v\rangle_{\text{ann}}$.  We define
\begin{align}
\Gamma_{\text{ann}}(t) = C_{\text{ann}}N^2(t),
\end{align}
and, in the case of constant $s$-wave annihilations, 
\begin{align}
C_{\text{ann}} =\frac{1}{2}\langle  \sigma v \rangle_{\text{ann}} \int_0^{R_{\odot}}
d^3r A^2\exp\left(\frac{-M_{\text{DM}} \Phi(r)}{T_{\text{core}}}\right).
\end{align}
Unless otherwise specified, we will adopt a fixed reference value of
$\langle\sigma v\rangle_{\text{ann}} = 3\times
10^{-26}\,\text{cm}^3/\text{s}$, corresponding to the assumption that
DM is a thermal relic constituting the main component of DM in the
universe.  In this case, $C_{\text{ann}} \propto M_{\text{DM}}^{3/2}$
for large $M_{\text{DM}}$, with all the mass dependence coming from
the spatial distribution of DM within the Sun.

For a traditional WIMP, elastic self-scattering is negligible, and for
masses $M_{\text{DM}} \gtrsim 4$ GeV evaporation can also be
neglected.  The DM number abundance in the Sun is then simply given by
the balance of capture and annihilation,
\begin{align}
\frac{dN}{dt} = C_{\text{c}}  -C_{\text{ann}}  N^2.
\label{eqn:Nsimple}
\end{align}
The solution to this equation is 
\begin{align}
N_{\text{0}}(t) = \sqrt{\frac{C_{\text{c}}}{C_{\text{ann}}} }\tanh \left( \frac{t}{\tau_{\odot}} \right),
\label{eqn:N0}
\end{align}
where $\tau_{\odot}^{-1} \equiv\sqrt{C_{\text{c}} C_{\text{ann}}}$ is
the timescale for the system to reach equilibrium.  When the solar
population has achieved steady-state, the present-day annihilation
rate is directly proportional to the nuclear cross-section,
\begin{align}
\Gamma_{\text{ann}} (t_0) = C_{\text{c}} \tanh^2 \left( \frac{t_0}{\tau_{\odot}} \right) = C_{\text{c}}.
\end{align}
However, in much of the parameter space allowed by direct detection
constraints, the equilibration timescale is longer than the age of the
Sun, and the present-day annihilation rate therefore depends on the
annihilation cross-section as well as the nuclear cross-section.  For
reference, Fig.~\ref{fig:SunAndEarthThermalization} in the Appendix
shows the minimum value of $\sigma_{\text{p}}$ yielding equilibration as a
function of $M_{\text{DM}}$.

\subsection{DM capture and annihilation in the Sun with self-interactions}
\label{sec:sun}

\begin{figure}[t]
\begin{subfigure}{0.45\textwidth}
  \hspace{-1.2cm}
 \includegraphics[width=1.12\linewidth]{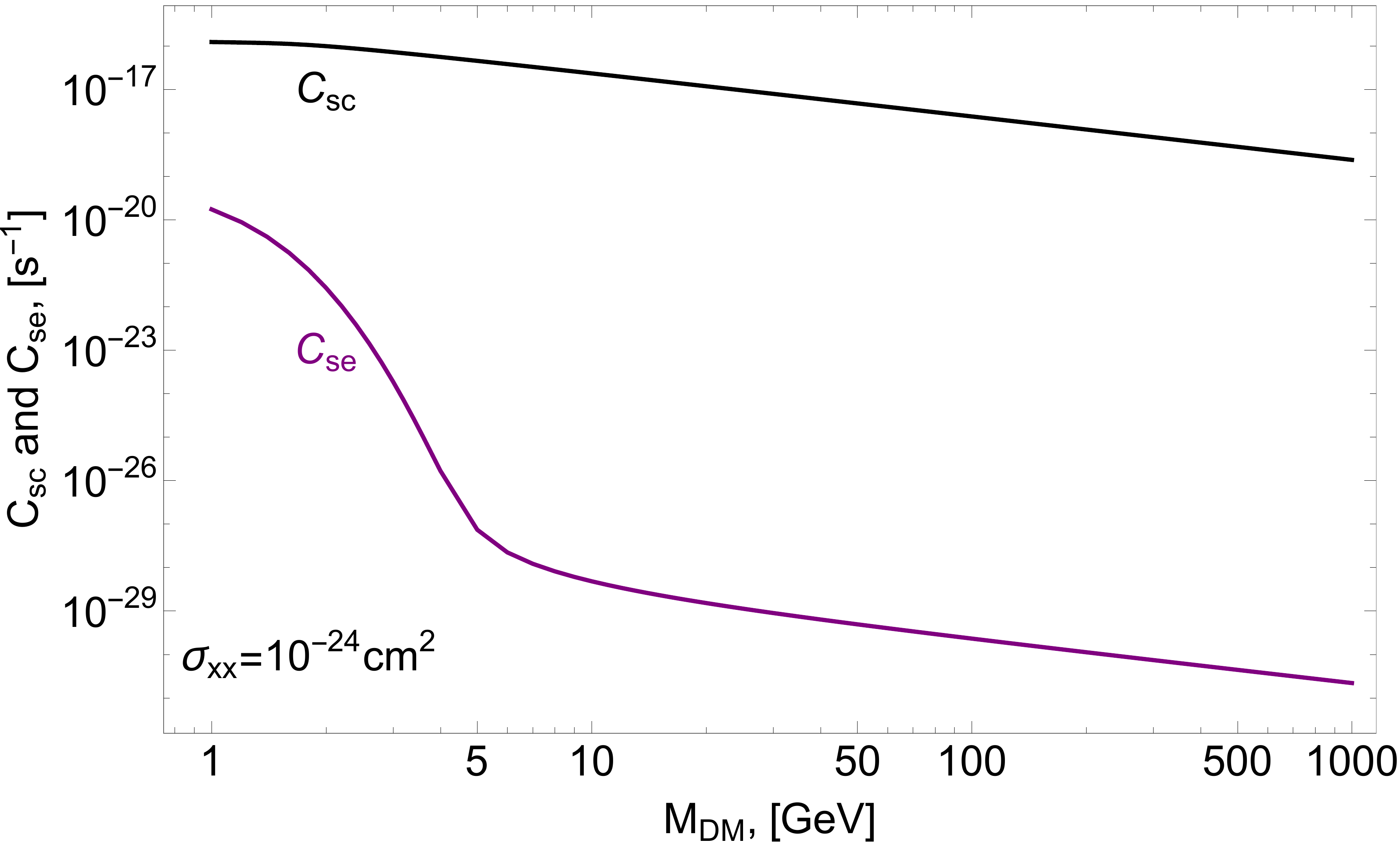}
\end{subfigure}
\begin{subfigure}{0.45\textwidth}
 \includegraphics[width=1.12\linewidth]{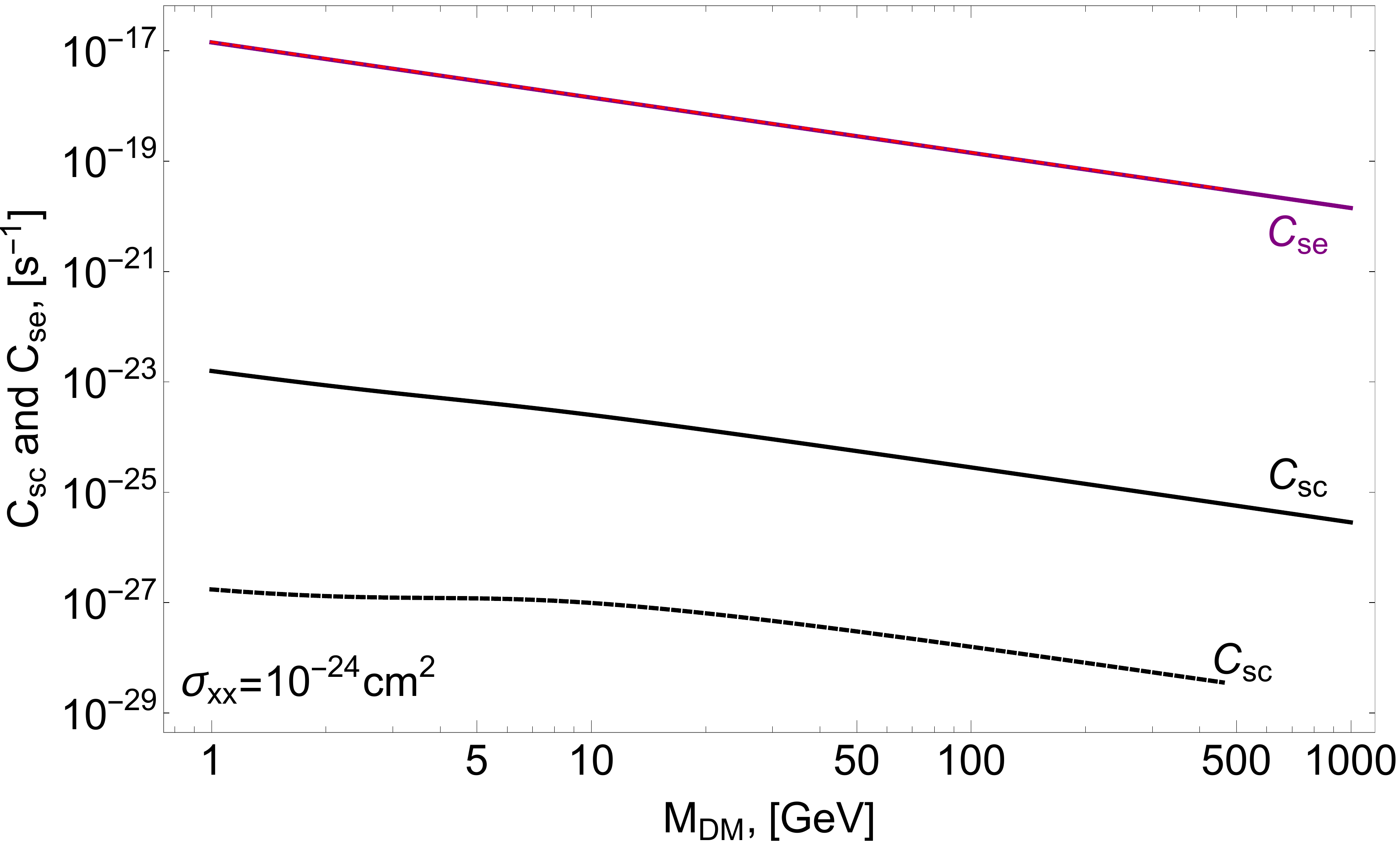}
\end{subfigure}
\caption{{\bf Left:} Self-capture (black) and self-ejection (purple) rates
  in the Sun as a function of DM mass, for constant self-interaction 
  cross-section $\sigma_{\text{xx}}= 10^{-24} \text{cm}^2$. {\bf Right:}  
  Self-capture and self-ejection coefficients in the Earth as a
  function of DM mass again with $\sigma_{\text{xx}}= 10^{-24} \text{cm}^2$. Self-capture is shown in black
  for direct capture (dashed) and free space  (solid).
  Self-ejection is shown in dashed red for direct capture and solid 
  purple for free space. }
\label{fig:selfcap}
\end{figure}

Next we discuss the processes that depend on DM's elastic
self-scattering.  For illustration, we consider an isotropic and
velocity-independent self-scattering cross section
$\sigma_{\text{xx}}$.  The construction of the self-capture and
self-ejection rates $C_{\text{sc}}$, $C_{\text{se}}$ is very similar
to the calculation of the nuclear capture rate $C_{\text{c}}$
\cite{Zentner:2009is}, with two important differences.  The first is
that the target nuclei of the Sun, $n_{i} (r)$, are replaced
with the numerically much smaller population of captured DM
$n_{\text{c}}(r)$ (again assumed to be thermalized within the Sun).
Collision by collision, momentum transfer is much more efficient for
self-capture than for nuclear capture, and therefore the relative
importance of self-capture depends on the DM mass. The other important
difference is that in evaluating the probability of self-capture we
must require not only that the incoming DM particle become
gravitationally bound, but also that the target particle remain bound
to the Sun after the collision:
\begin{align}
\label{eq:scap}
\frac{\Delta E}{E} \geq \frac{u^2}{w^2(r)} \quad \text{and} \quad \frac{\Delta E}{E} \leq \frac{v^2_{\text{esc}}(r)}{w^2(r)}.
\end{align}
Conversely, for self-ejection, we require that the incoming particle
remain unbound to the Sun, while the target becomes unbound after the
collision:
\begin{align}
\label{eq:sej}
\frac{\Delta E}{E} \leq \frac{u^2}{w^2(r)} \quad \text{and} \quad \frac{\Delta E}{E} \geq \frac{v^2_{\text{esc}}(r)}{w^2(r)}.
\end{align}
Both rates are directly proportional to the self-interaction
cross-section $\sigma_{\text{xx}}$.  The Sun's potential well is deep---the escape velocity
at the Sun's surface, $v_{\text{esc}}(R_\odot) = 618 \,\mathrm{ km}/\mathrm{s}$, is far larger than the typical DM speed in the halo, $\bar v = 288$ km$/$s---and
consequently self-ejection is everywhere negligible compared to
self-capture \cite{Zentner:2009is}.  This result can be seen from
Fig.~\ref{fig:selfcap} for constant self-interactions, and also holds for long-range
self-interactions $\sigma_{\text{xx}} \propto v^{-4}$
\cite{Fan:2013bea}.
%
%

For DM masses above 10 GeV, both $C_{\text{sc}}$ and $C_{\text{se}}$
follow very closely a $1/M$ power law. This dependence is simply due
to the local DM number density, $n_{\text{DM}}=\rho_{\odot}/M_{\text{DM}}$. The complicated
velocity and volume integrals have a very weak mass dependence.

This calculation assumes that the Sun is not optically thick to DM.
As self-interaction cross-sections of astrophysical interest can be
very large, it is instructive to check the validity of this
assumption.  To do so, we require that the self-capture cross-section
multiplied by the number of targets, i.e., the number of captured DM
particles, be smaller than the cross-sectional area of the captured DM
ball. Defining $r_X$ as the radius containing $95\%$ of the captured
DM, we thus require
\begin{align}
\label{eq:opticallythin}
\langle\sigma_{\text{sc}}\rangle N(\tau_{\odot}) < \pi r_X^2,
\end{align}
where the average self-capture cross-section
$\langle\sigma_{\text{sc}}\rangle $ is defined by integrating over the
incident DM velocity distribution,
\begin{align}
\langle\sigma_{\text{sc}}\rangle\equiv \frac{\int_0^{r_X}dV\int d^3uf_{\eta}(u)\frac{w}{u}\sigma_{\text{sc}}}{\int_0^{r_X}dV\int d^3uf_{\eta}(u)\frac{w}{u}}.
\end{align}
Here $\sigma_{\text{sc}}$ is defined to incorporate the kinematic
restrictions on the recoil energy that lead to self-capture.  Through
$N(\tau_\odot)$, Eq.~\ref{eq:opticallythin} depends on the nuclear
cross-section $\sigma_{\text{p}}$ and the DM annihilation
cross-section as well as the DM self-interaction cross-section.  We
find for DM with a thermal annihilation cross-section that we are
always in the optically thin regime for nuclear cross-sections allowed
by direct detection and self-interaction cross-sections allowed by the
Bullet Cluster.  A similar calculation for self-ejection in the Earth
shows that the Earth is always optically thin to DM when the Sun is.

\begin{figure}[t]
\begin{subfigure}{0.45\textwidth}
  \hspace{-1.2cm}
  \includegraphics[width=1.12\linewidth]{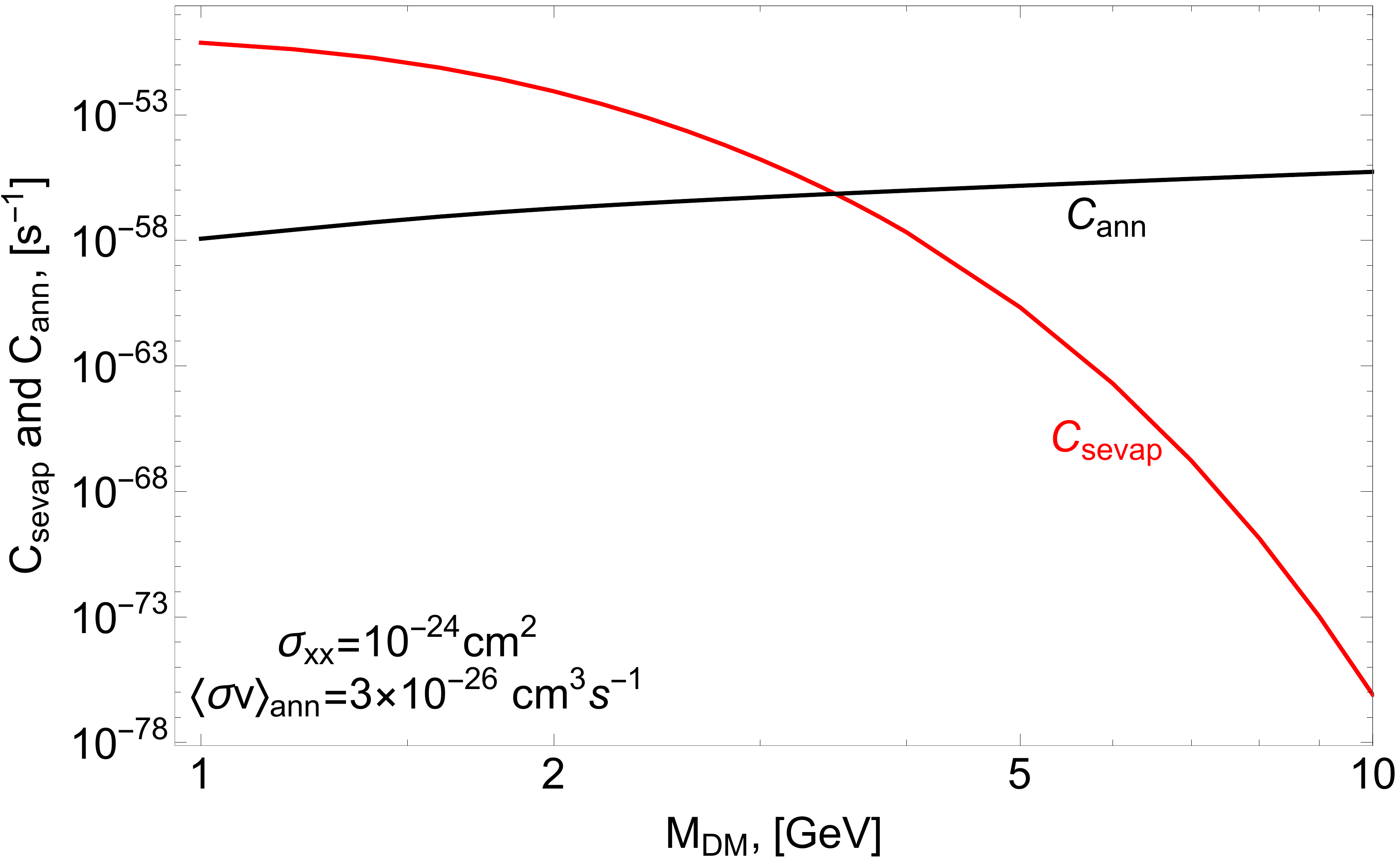}
\end{subfigure}
\begin{subfigure}{0.45\textwidth}
 \includegraphics[width=1.12\linewidth]{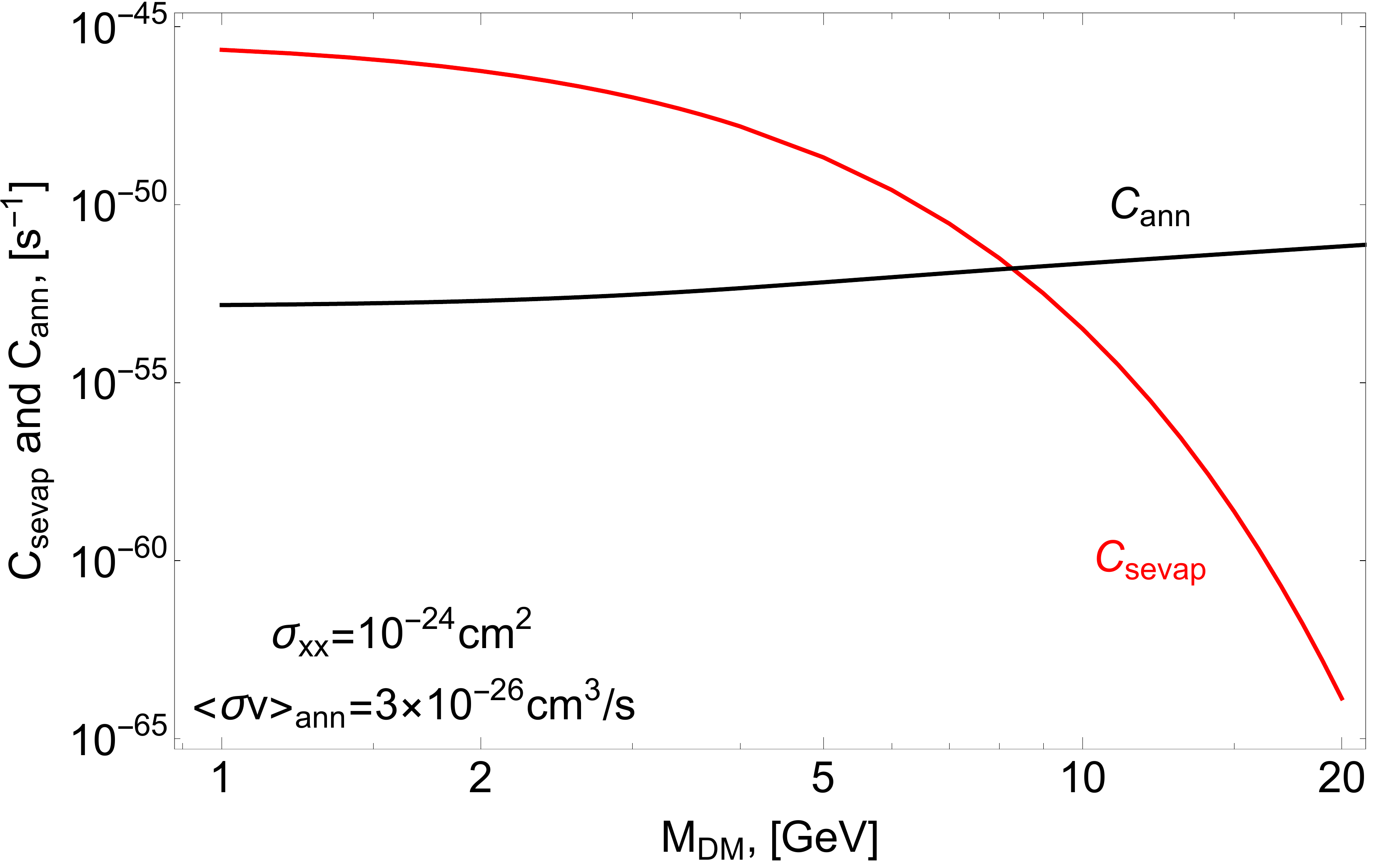}
\end{subfigure}
\caption{Self-evaporation and annihilation rates in the Sun (left) and Earth
  (right). The self-evaporation coefficient $C_{\text{sevap}}$ is
  shown in black and the annihilation coefficient $C_{\text{ann}}$ in
  red for fixed $\sigma_{\text{xx}} = 10^{-24}\, \text{cm}^2$ and $\langle\sigma v\rangle_{\text{ann}} = 3\times 10^{-26}\text{cm}^3/s$.}
\label{plot:CSelfEvaporationAndCAnnihilation}
\end{figure}

Meanwhile, the self-evaporation of DM through isotropic,
velocity-independent scattering can become important for
$M_{\text{DM}}\lesssim 4$ GeV \cite{chen2014probing}; see the left
panel of Fig.~\ref{plot:CSelfEvaporationAndCAnnihilation}.  Our main
regime of interest will be in the range $M_{\text{DM}}>10$ GeV, where
self-evaporation can be neglected in comparison to annihilation.

\subsection{DM capture and annihilation in the Earth}
\label{sec:earth}

Computing dark matter capture rates in the Earth differs from the
corresponding calculations in the Sun in two important ways.  First,
the Earth is situated deep within the Sun's potential well,
complicating the determination of the DM velocity distribution at the
location of the Earth.  Second, the Earth's potential well is much
shallower than the potential well of the Sun, which means that in the
presence of strong self-interactions the dominant process is
self-ejection, rather than self-capture
\cite{Zentner:2009is,Fan:2013bea}.

When DM from the Galactic halo arrives at the Earth, it has already
been accelerated by the Sun.  This solar acceleration combines with
the Earth's orbital velocity to determine a finite minimum velocity
$u$ for DM falling into Earth's potential well.  Since the fractional
energy loss of a DM particle is limited by kinematics, a minimum
incident velocity in turn determines a maximum (minimum) possible DM
mass above (below) which gravitational capture is not possible
\cite{gould1988direct,Lundberg:2004dn}.  However this `direct' capture
of DM from the halo population is supplemented by `indirect' capture
from the population of DM already gravitationally bound to the solar
system \cite{gould1988direct,gould1991gravitational}.  Understanding
the phase space distribution of DM bound to the solar system is a
complicated dynamical question.  In
Ref.~\cite{gould1991gravitational}, Gould argued that a detailed
balance should hold between capture and ejection processes in gravitational DM
interactions with Jupiter, Earth, and Venus, bringing the total
distribution of DM seen at Earth's orbit to an effective free-space
distribution for the Earth as considered in isolation.  Subsequent
numerical work by Peter \cite{peter2009dark} in a simplified model of the solar system showed that this
`detailed balance' description holds to good approximation at low DM
velocities ($v\lesssim 30\,\mathrm{ km/s}$), but overestimates the DM
distribution at higher velocities.  In our calculations we show
results for both direct capture and capture from the detailed balance
free-space distribution, which bracket expectations for Earth capture
of DM.  As low DM velocities are most important for self-capture and
high DM velocities more relevant for self-ejection, we expect that the
free space distribution gives a reasonable description of self-capture
but may underestimate the self-ejection rate.  Thus our choice here is
conservative.

Details of our Earth model are given in Appendix~\ref{sec:appearth},
where we also show in Fig.~\ref{plot:Earth_CCapture_FD_vs_Dir_Capt}
the resulting nuclear capture rates in the Earth, for both direct
capture and free space velocity distributions, assuming constant spin-independent nuclear interactions.  Away from nuclear
resonances, the dominant nuclear capture rate is due to oxygen at low
masses, while at higher DM masses iron dominates.  The choice of DM
velocity distribution has a major impact on the nuclear capture rate
at both high ($M\gtrsim 100$ GeV) and low ($M\lesssim 10$) GeV DM
masses, with less sensitivity for DM masses in the intermediate regime
where nuclear resonances dominate. Finally, we caution that our
simplifying approximation that the captured DM population thermalizes
at a uniform temperature corresponding to the temperature of the
Earth's core is less warranted than for the Sun, as the captured DM
occupies a larger fraction of the Earth's volume; this could be
straightforwardly refined in a more detailed treatment.  

For DM masses below $M_{\text{evap}}\lesssim 10$ GeV, evaporation of
the Earth's captured population through nuclear scattering becomes
important \cite{Gould:1987ir}.  The evaporation mass is larger in the
Earth than in the Sun, thanks to the Earth's more massive constituent
nuclei and its shallower potential well. We show our calculation of
the evaporation rate in the Earth in
Fig.~\ref{plot:Earth_CEvaporation}.

\subsection{DM capture and annihilation in the Earth with self-interactions}
\label{sec:earthself}

The rate of self-ejection and self-capture in the Earth can be
computed analogously to the discussion for the Sun above.
The resulting self-ejection and self-capture coefficients are shown in
Fig.~\ref{fig:selfcap} for specific choices of self-interaction
strength.  Self-ejection is everywhere dominant over self-capture by
many orders of magnitude.  While we consider constant self-interaction
cross-sections here, the dominance of self-ejection over self-capture in the Earth also holds for long-range Rutherford
cross-sections, despite the relatively larger self-capture rates in
that case \cite{Fan:2013bea}. Thus the essential result of this
paper---i.e., that the differering kinematics of gravitational capture within the Sun and Earth's potential wells imply differing and potentially sizeable impacts of self-scattering on the bound DM populations---apply to long-range self-interaction cross-sections as well.\footnote{It is worth observing in this context that substantially enhanced and/or environmentally dependent annihilation cross-sections can also affect the bound Sun and Earth DM populations differently \cite{Delaunay:2008pc}.}
Self-capture can only occur in the low-velocity tail of the DM
distribution, and thus the self-capture coefficient is highly
sensitive to the choice of DM velocity distribution.  The
self-ejection coefficient, on the other hand, is insensitive to the
choice of DM velocity distribution.

Finally, we consider self-evaporation in the Earth.  The shallowness
of the Earth's potential well ($v_{\text{esc}\oplus}(R_\oplus) =11
\,\mathrm{km}/\mathrm{s} $) makes self-evaporation more important in
the Earth than in the Sun.  We find that self-evaporation dominates
over annihilation for $M_{\text{DM}}\lesssim 8$ GeV for constant
self-interactions; see the right panel of
Fig.~\ref{plot:CSelfEvaporationAndCAnnihilation}.

\section{Self-interactions and DM annihilations in the Sun}
\label{sec:sunmod}

In the presence of self-interactions, the equation governing the
evolution of the captured DM population in the Sun is
\begin{equation}
\label{eq:dNdtsun}
\frac{dN}{dt} = C_{\text{c}} + (C_{\text{sc}} - C_{\text{e}})N -\left(C_{\text{ann}}+C_{\text{sevap}}\right)N^2.
\end{equation} 
Here we have neglected the highly subdominant contribution of
self-ejection.  Although our main interest will be in DM masses
$M_{\text{DM}}>10$ GeV, we have retained the contributions of
evaporation and self-evaporation, in order to get a more complete
picture of the impact of self-interactions in the Sun.  With the
initial condition $N(0)=0$, the general solution to
Eq.~\ref{eq:dNdtsun} is \cite{Zentner:2009is}
\begin{align}
N(t) = \frac{C_{\text{c}}\tanh{\left(\frac{t}{\xi}\right)}}{\frac{1}{\xi} -\frac{(C_{\text{sc}}-C_{\text{e}})}{2}\tanh{\left(\frac{t}{\xi}\right)}},
\end{align} 
where 
\beq
\xi^{-1} = \sqrt{C_{\text{c}}(C_{\text{ann}}+C_{\text{sevap}})
  +\frac{(C_{\text{sc}}-C_{\text{e}})^2}{4}}
\eeq
is the inverse of the equilibration time for the captured population.

In the absence of self-interactions, the equilibration time is given by 
\beq
\xi_0^{-1} =
\sqrt{C_{\text{c}}C_{\text{ann}}+ \frac{C_{\text{e}}^2}{4}}
\eeq
and the DM population is
\begin{align}
\label{eq:N0evap}
N_{\text{0}}(t) = \frac{C_{\text{c}}\tanh{\left(\frac{t}{\xi_\text{0}}\right)}}{\frac{1}{\xi_0} +\frac{C_{\text{e}}}{2}\tanh{\left(\frac{t}{\xi_0}\right)}}.
\end{align}
We can quantify the impact of self-interactions on the solar
population by defining an ``enhancement'' factor $\beta$, given by the
ratio of the present-day DM annihilation rate in the presence of
self-interactions to the annihilation rate without self-interactions
for fixed $M_{\text{DM}}$ and $\sigma_{\text{p}}$:
\begin{align}
\label{eq:betadef}
\beta \equiv \frac{\Gamma_{\text{ann}}}{\Gamma_{\text{ann},0}}=\frac{C_{\text{ann}}N(\tau_\odot)^2}{C_{\text{ann},0} N_0(\tau_\odot)^2}.
\end{align}
Here $N_0(t)$ is given by Eq.~\ref{eq:N0evap}, i.e., is the solar
population in the absence of self-interactions, but including the
effect of evaporation.  When the annihilation cross-section is taken
to be independent of the elastic self-scattering cross-section, as
here, $C_{\text{ann}}=C_{\text{ann},0}$, and the definition of $\beta$
then reduces to a comparison of the total captured population in the
cases with and without self-interactions \cite{Zentner:2009is},
\begin{align}
\beta \equiv \left(\frac{N(\tau_\odot)}{N_0(\tau_\odot)}\right)^2.
\end{align}
Since self-capture dominates over self-ejection in the Sun, the main
effect of self-scattering (at fixed annihilation cross-section) is to
increase the total solar population and thus enhance DM annihilations
in the Sun \cite{Zentner:2009is, Albuquerque:2013xna}.

\begin{figure}[t]
\begin{subfigure}{0.5\textwidth}
  \hspace{-0.5cm}
  \includegraphics[width=1.0\linewidth]{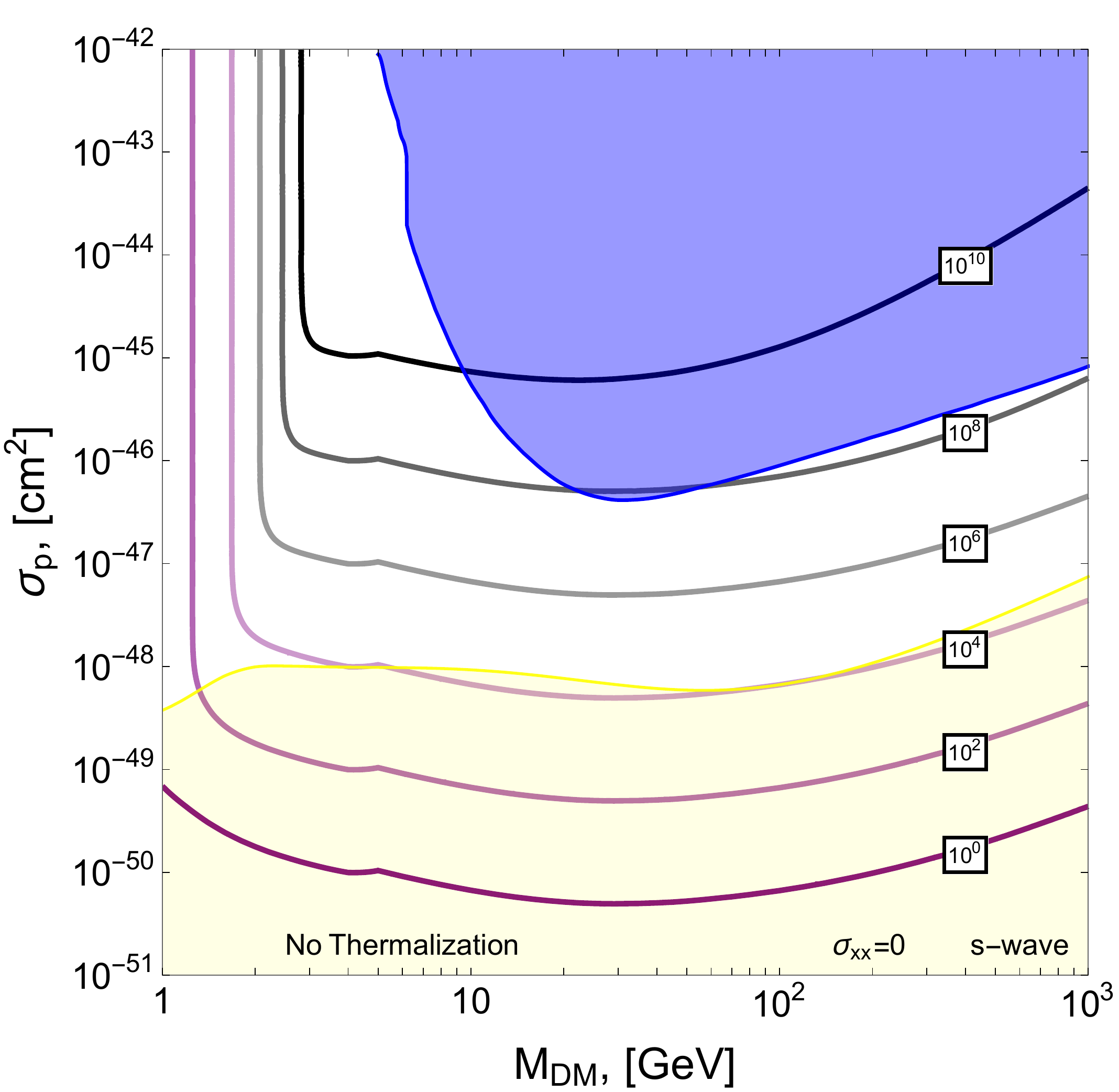}
\end{subfigure}%
\begin{subfigure}{0.5\textwidth}
  \hspace{-0.5cm}
  \includegraphics[width=1.0\linewidth]{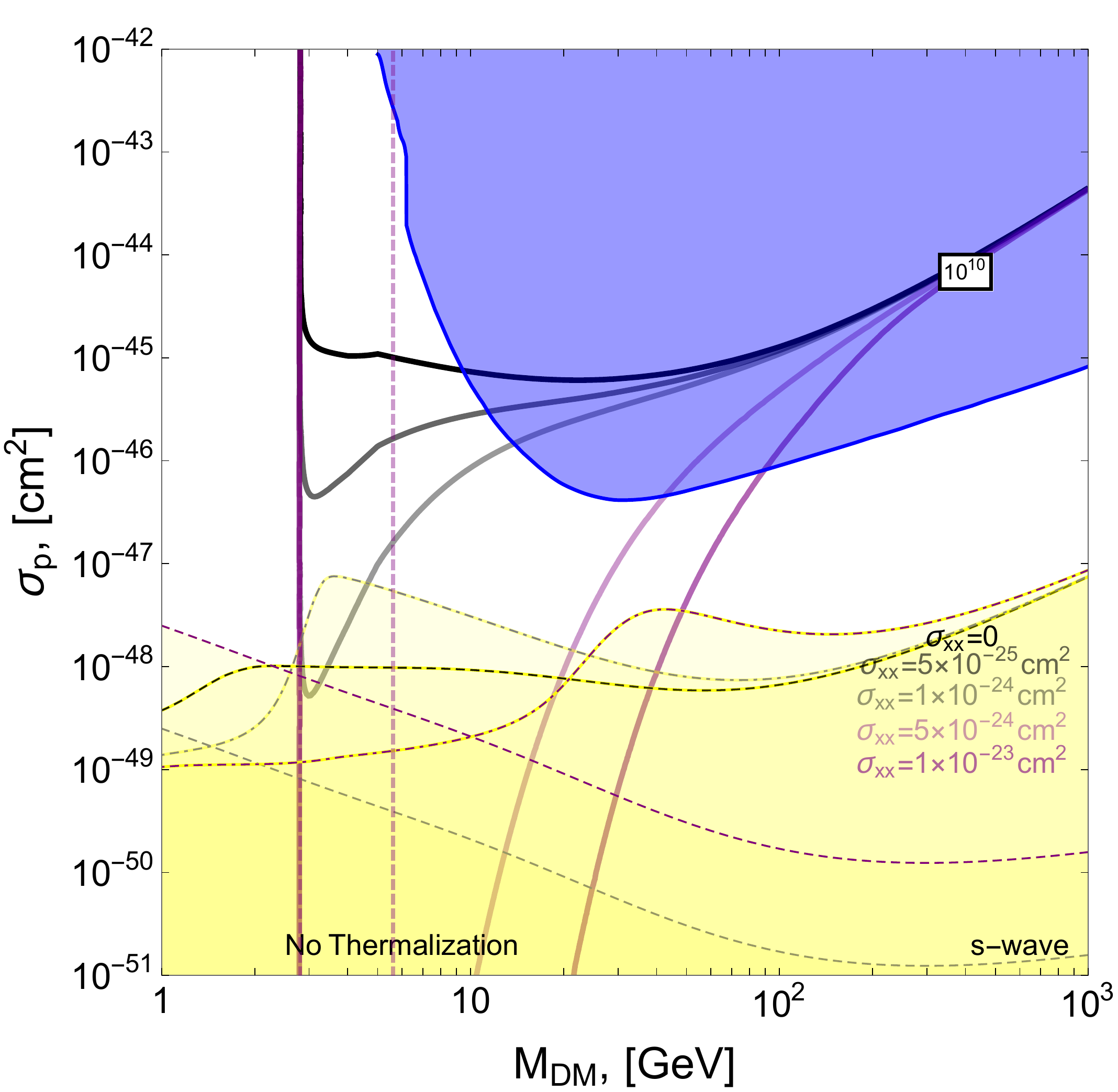}
\end{subfigure}%
\caption{Solar annihilation flux in
  $\text{km}^{-2}\,\text{yr}^{-1}$. Colored lines represent contours
  of constant flux.  In the left plot we show the annihilation fluxes
  in the absence of self-interactions. In the right plot we show how a
  given annihilation flux contour is modified by the introduction of
  increasing constant self-interaction cross-section. The blue shaded
  region is excluded by the PandaX, LUX and XENON1T direct detection
  experiments.  The parameter space to the left of the dashed purple
  lines is in tension with the Bullet Cluster. In the yellow regions
  $\sigma_{\text{p}}$ is too small for the captured DM population to
  be thermalized in the Sun: (left) less than 90\% of the captured DM
  population is thermalized; (right) below the dot-dashed contours,
  less than 90\% of the captured DM population is thermalized for the
  indicated value of $\sigma_{\text{xx}}$, while below the dashed
  contours energy injection from halo DM prevents the captured DM
  population from thermalizing at the Sun's temperature.}
\label{plot:Sun_Annihilation_Flux_1}
\end{figure}

As a prelude to our numerical results, we first establish some scales.
To remain as model-independent as possible, we will show results for
the {\em annihilation flux}, defined as the annihilation rate divided
by the geometric dilution factor:
\begin{align}
\Phi_\Sun = \frac{\Gamma_{\text{ann}}}{4\pi D^2},
\end{align}
where $\text{D} = 1 \,\text{A.U.}$ for the solar annihilation flux.
Contours of constant annihilation flux in the Sun are shown in the
$(M,\sigma_{\text{p}})$ plane in Fig.~\ref{plot:Sun_Annihilation_Flux_1},
without (left) and with (right) constant self-interactions.

Whether or not a given DM annihilation flux is observable depends
strongly on the specific annihilation mode(s) in a given model.  For
DM that annihilates promptly to SM species, current neutrino
telescopes can observe annihilation fluxes in the range
$10^9-10^{13}\,\mathrm{km}^{-2}\mathrm{yr}^{-1}$, depending on the DM
mass and annihilation channel ($bb$, $\tau\tau$, or $WW$)
\cite{Choi:2015ara,Adrian-Martinez:2016gti, Aartsen:2016zhm}.  Greater
sensitivities can be achieved in models where DM annihilates to
long-lived dark particles that escape the Sun; solar $\gamma$ rays can
probe annihilation fluxes on the order of
$10^7-10^8\,\mathrm{km}^{-2}\mathrm{yr}^{-1}$ \cite{Leane:2017vag},
while for a striking enough signal annihilation fluxes as low as $\sim
10^2\,\mathrm{km}^{-2}\mathrm{yr}^{-1}$ can be tested
\cite{Adrian-Martinez:2016ujo}.  Depending on the DM mass and
detection channel, cosmic rays scattering within the solar atmosphere
may provide important backgrounds
\cite{Zhou:2016ljf,Arguelles:2017eao, Ng:2017aur,Edsjo:2017kjk}.

%
\begin{figure}[h!]
\begin{subfigure}{0.45\textwidth}
  \hspace{-1.0cm}
  \includegraphics[width=1.0\linewidth]{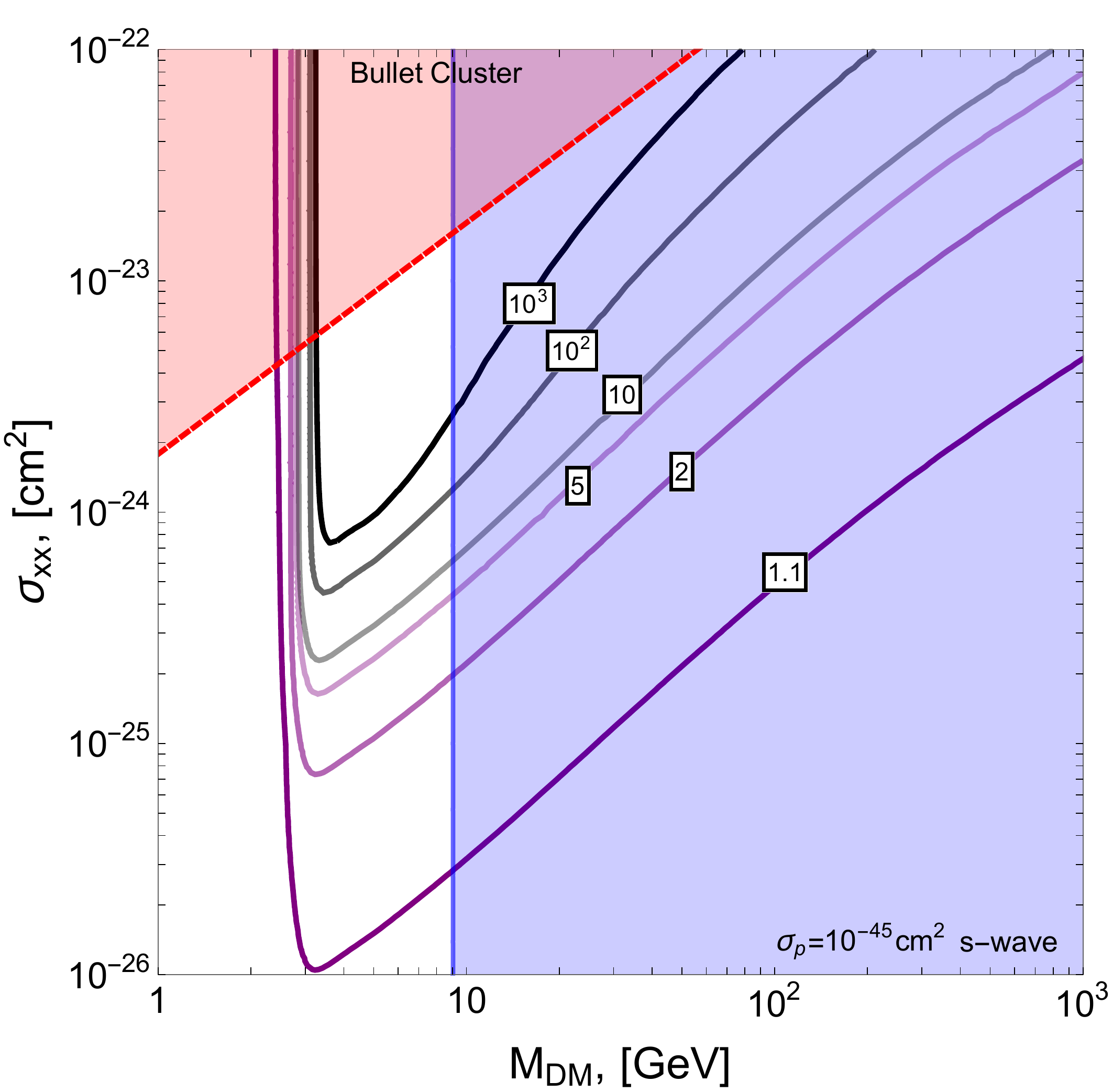}
\end{subfigure}
\begin{subfigure}{0.45\textwidth}
  \hspace{-1.0cm}
  \includegraphics[width=1.0\linewidth]{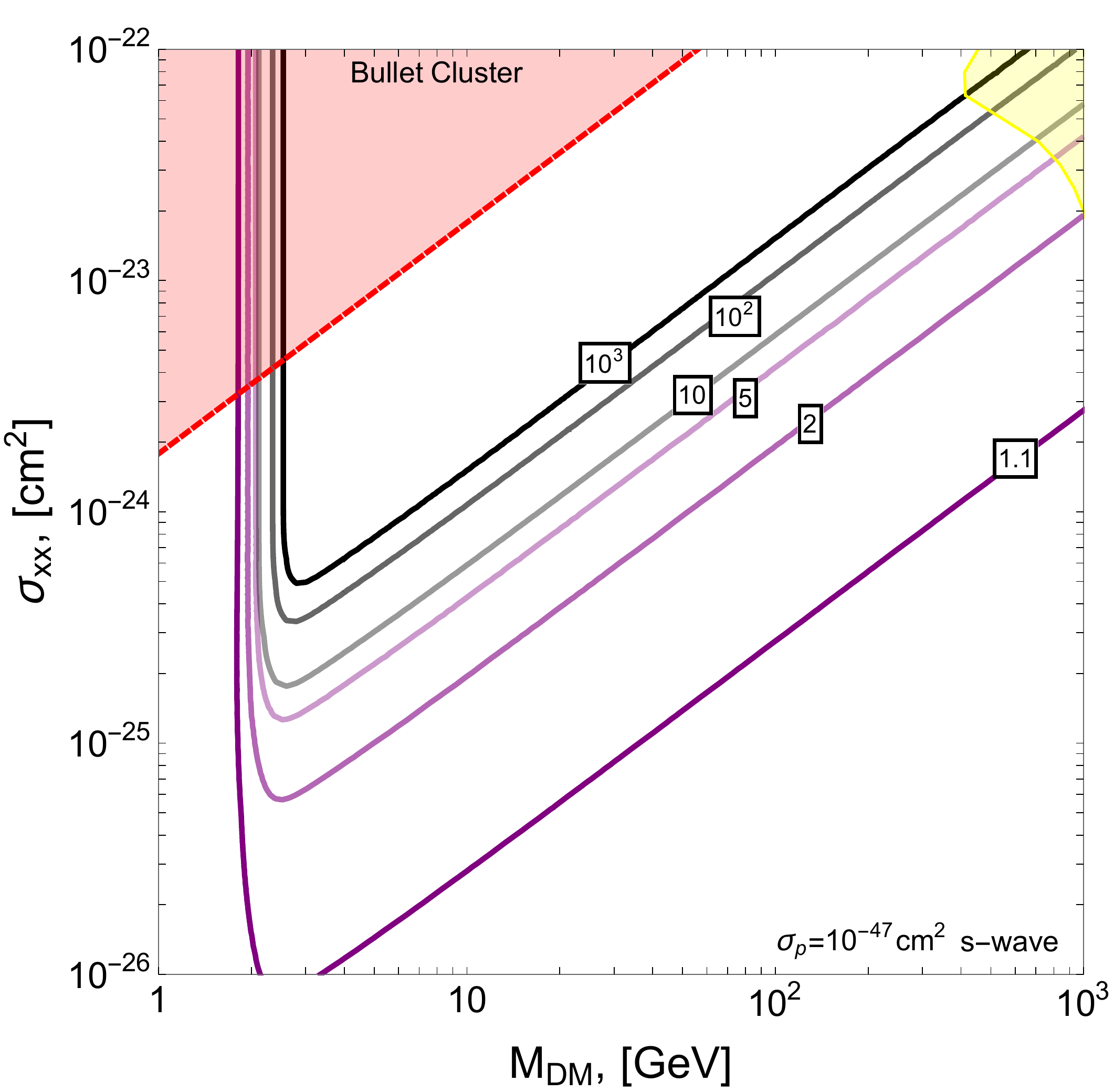}
\end{subfigure}
\caption{Solar annihilation flux ratio $\beta$ in the Sun for constant
  self-interaction cross-section. Colored lines represent contours of
  constant $\beta$.  We show results for fixed nuclear cross-sections
  $\sigma_{\text{p}} = 10^{-45}\text{cm}^2$ (left) and
  $\sigma_{\text{p}} = 10^{-47}\text{cm}^2$ (right). The red shaded
  region is excluded by the Bullet Cluster constraint and the blue
  shaded region is excluded by the PandaX, LUX and XENON1T direct
  detection experiments.  In the yellow region, more than 10\% of the
  captured DM population is not thermalized today.}
\label{plot:Sun_Enhancement_1}
\end{figure}
%
We fix the annihilation cross-section to the reference (near-)thermal
value, $\langle\sigma v\rangle_{\text{ann}} =3\times
10^{-26}\,\mathrm{cm}^3/\mathrm{s}$, and plot contours of $\beta$ in
the $(M_{\text{DM}}, \sigma_{\text{xx}})$ plane for fixed values of
$\sigma_{\text{p}}$ in Fig.~\ref{plot:Sun_Enhancement_1}. Here the red
shaded region indicates self-interaction cross-sections in excess of
the Bullet Cluster bound, $\sigma_{\text{xx}}/M < 1
\,\text{cm}^2/\mathrm{g}$ \cite{Randall:2007ph}.  The blue shaded
regions are excluded by the results of LUX \cite{akerib2016results},
PandaX-II \cite{tan2016dark} and XENON1T \cite{Aprile:2017iyp, Aprile:2018dbl}.

Note that large enhancements, $\beta\sim10^{2} - 10^{3}$, are possible
in the parameter space allowed by all current constraints. As
$\sigma_{\text{p}}$ decreases, a smaller value of $\sigma_{\text{xx}}$
is needed for $C_{\text{sc}}$ to compete with $C_{\text{c}},
C_{\text{e}}$ to obtain the same fixed enhancement $\beta$, as is
evident by comparing the left and right panels of
Fig.~\ref{plot:Sun_Annihilation_Flux_1}.  The decrease in $\beta$ with
increasing mass at fixed $\sigma_{\text{xx}}$ arises from the
parametric dependence of the capture and annihilation coefficients on
DM mass $M_{\text{DM}}$. Recall from Sec.~\ref{sec:gravcap} that at
large $M_{\text{DM}}$, $C_{\text{c}}\sim 1/M_{\text{DM}}^2$, while for
constant self-interactions $C_{\text{ann}}\sim M_{\text{DM}}^{3/2}$
and $C_{\text{sc}}\sim 1/M_{\text{DM}}$. Hence in the fundamental
timescale of the system, $\xi$, the $C^2_{\text{sc}}$ term decreases
faster than $C_{\text{c}}C_{\text{ann}}$ with increasing DM mass.
Thus at a given $\beta$, even for large values of
$\sigma_{\text{xx}}$, nuclear capture will always come to dominate at
sufficiently large DM masses.

From the right panel of Fig.~\ref{plot:Sun_Enhancement_1}, we also
note that for DM with $M_{\text{DM}}\gtrsim$ 10 GeV and fixed $\beta$,
the requisite value of $\sigma_{\text{xx}}$ scales approximately
linearly with $M_{\text{DM}}$.  This feature appears for sufficiently
small $\sigma_{\text{p}}$ (we show results for a fixed value
$\sigma_{\text{p}} =10^{-47}\,\mathrm{cm}^2$) and is explained as
follows. In this regime evaporation and self-evaporation are
negligible and self-interactions dominate over nuclear capture. In the
absence of self-interactions, the small $\sigma_{\text{p}}$ means that
the equilibration time is larger than the age of the Sun. Hence, we
can approximate $\tanh{(\tau_\odot/\xi_0)}\approx
\tau_\odot/\xi_0$. Meanwhile, for large $\beta$, the equilibration
time $1/\xi$ is dominated by self-capture, $C_{\text{sc}}^2 \gg
C_{\text{c}}C_{\text{ann}}$. Hence, we can approximate
\begin{align}
\frac{1}{\xi} =\sqrt{C_{\text{c}}C_{\text{ann}} +\frac{C_{\text{sc}}^2}{4}} \approx  \frac{C_{\text{sc}}}{2} + \frac{C_{\text{c}}C_{\text{ann}}}{C_{\text{sc}}}.
\end{align}
With these approximations, $\beta$ takes on the simpler form
\begin{align}
\beta^{1/2} \approx \frac{1}{\tau_\odot C_{\text{sc}}}\frac{2\tanh{\frac{\tau_\odot}{\xi}}}{(1-\tanh{\frac{\tau_\odot}{\xi}}+ 2\frac{C_{\text{c}}C_{\text{ann}}}{C_{\text{sc}}^2})}
\phantom{spacer} (\beta\gg 1).
\end{align}
The $1/C_{\text{sc}}$ factor gives the dominant mass scaling of
$\beta$ via the mass dependence of $C_{\text{sc}} \sim
1/M_{\text{DM}}$, giving $\beta^{1/2}\sim M_{\text{DM}}$. Conversely,
for a constant $\beta$, we must scale $\sigma_{\text{xx}}\sim
M_{\text{DM}}$, so that $C_{\text{sc}}$ has no strong net mass
dependence. On the other hand, for weak enhancements, the
equilibration time is larger than the age of the Sun both in the
presence and absence of self-interactions. Therefore, taking
$\tanh{\left(\tau_\odot/\xi_0\right)}\approx \tau_\odot/\xi_0$ and
$\tanh{\left(\tau_\odot/\xi\right)}\approx \tau_\odot/\xi$ is
justified. With these approximations, the enhancement becomes
\begin{align}
\beta^{1/2}\approx\frac{\frac{\tau_\odot}{\xi}}{\frac{\tau_\odot}{\xi_0}}\frac{\frac{1}{\xi_0}}{\frac{1}{\xi}-\frac{C_{\text{sc}}}{2}\frac{\tau_\odot}{\xi}} = \frac{1}{1-\frac{1}{2}C_{\text{sc}}\tau_\odot}
\phantom{spacer} (\beta\sim \mathcal{O}( 1) ) .
\end{align}
Thus once again the contours appear nearly linear in the
$(M_{\text{DM}}, \sigma_{\text{xx}})$ plane.
As we move to lower masses, both evaporation and self-evaporation 
start to become important, and the parametric behavior of the
contours changes sharply as both processes turn on exponentially
quickly.

\begin{figure}[t]
\begin{subfigure}{0.5\textwidth}
  \hspace{-0.5cm}
  \includegraphics[width=1.0\linewidth]{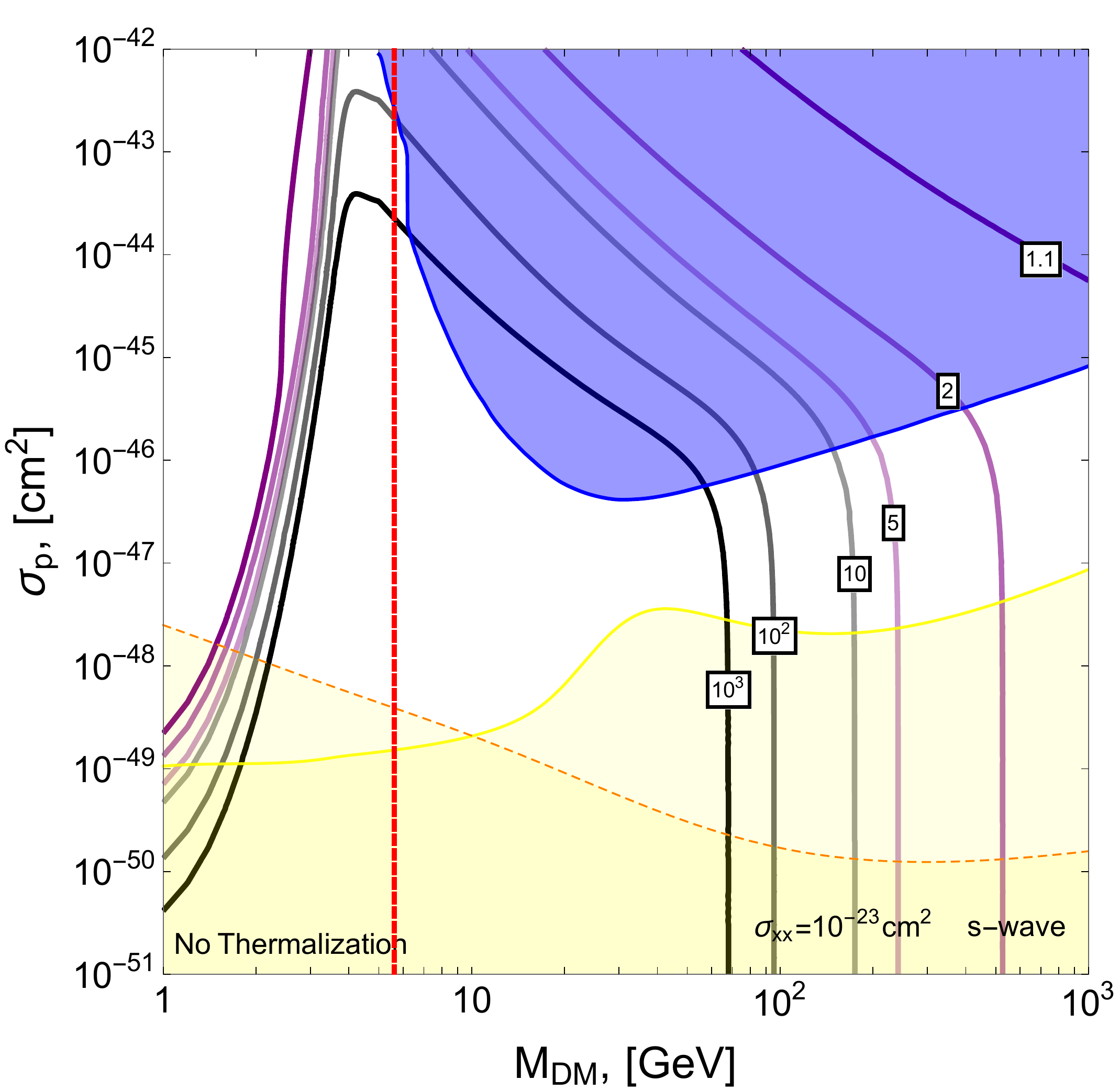}
\end{subfigure}%
\begin{subfigure}{0.5\textwidth}
  \hspace{-0.5cm}
  \includegraphics[width=1.0\linewidth]{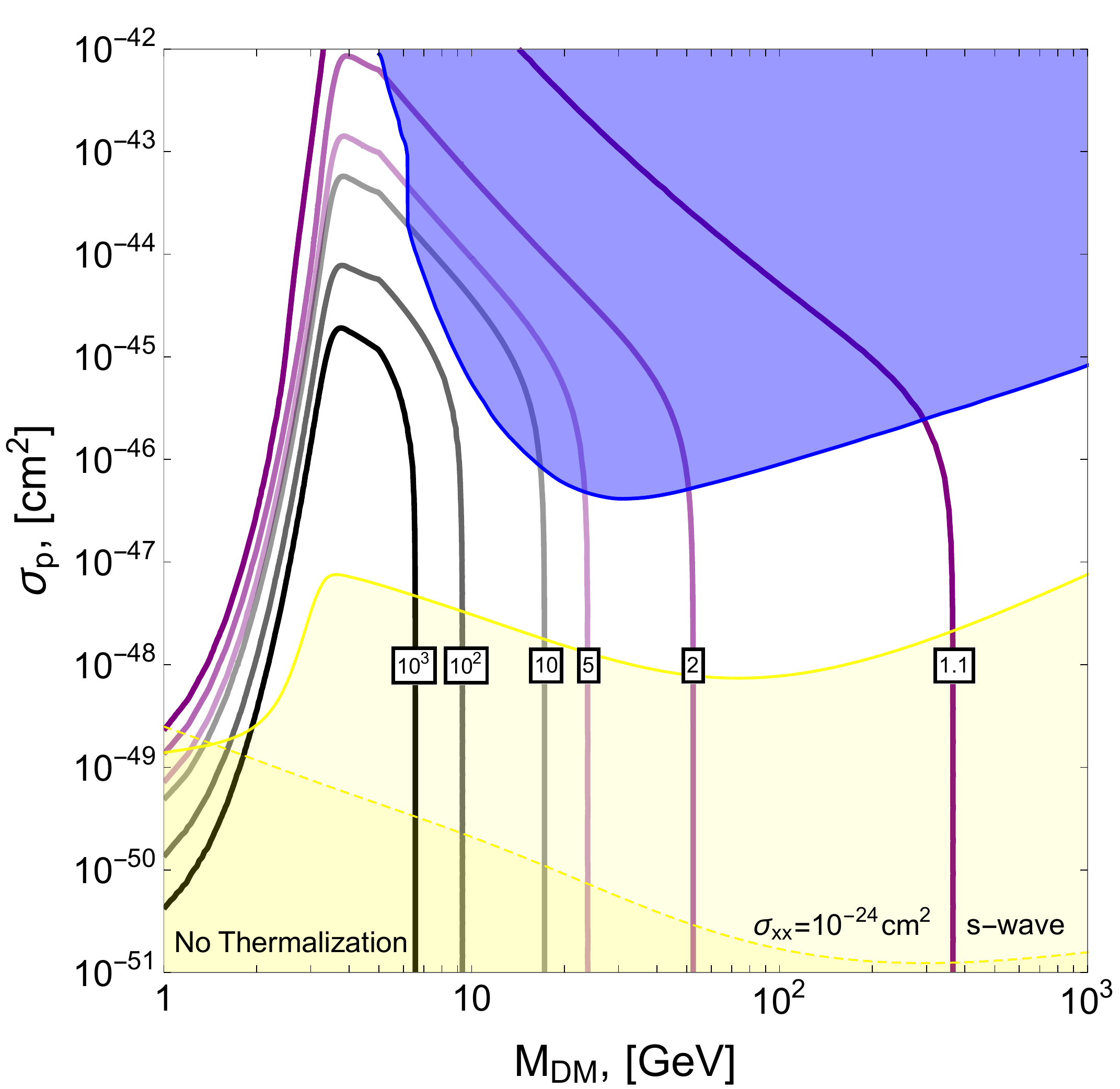}
\end{subfigure}
\caption{Solar annihilation flux ratio $\beta$ in the Sun for constant
  self-interaction cross-section. Colored solid lines represent
  contours of constant $\beta$.  We show results for
  $\sigma_{\text{xx}} = 10^{-23}\text{cm}^2$ (left) and
  $\sigma_{\text{xx}} = 10^{-24}\text{cm}^2$ (right). The blue shaded
  region is excluded by the PandaX , LUX and XENON1T direct detection
  experiments. In the yellow region $\sigma_{\text{p}}$ is too small
  for the captured DM population to be thermalized in the Sun. The
  parameter space to the left of the dashed red line is in tension
  with the Bullet cluster constraint.}
\label{plot:Sun_Enhancement_2}
\end{figure}

Next we consider contours of fixed $\beta$ in the $(M_{\text{DM}},
\sigma_{\text{p}})$ plane in Fig.~\ref{plot:Sun_Enhancement_2}, for
fixed values of $\sigma_{\text{xx}}$.  Here we see again that larger
enhancements are obtained for smaller $\sigma_\text{p}$, and observe
the sharp turnover in the contours at low DM masses as evaporation
becomes important.  From the figure we can further note the important
point that $\beta$ is largely insensitive to $\sigma_{\text{p}}$, for
$\sigma_{\text{p}} \lesssim 10^{-46} \,\text{cm}^2$ and sufficiently
large DM masses. In this regime the $C_{\text{c}}$ term becomes
completely negligible once $\sigma_{\text{p}} \lesssim 10^{-46}\,
\text{cm}^2$ and the population dynamics is controlled entirely by
self-capture. Hence, further decreases in $\sigma_{\text{p}}$ will not
affect the enhancement.

\section{Self-interactions and DM annihilation in the Earth}
\label{sec:earthmod}

Next we turn to quantifying the effect of DM self-interactions on the
annihilation signal from the Earth.  Here self-ejections dominate over
self-capture, giving
\begin{equation}
\frac{dN}{dt} = C_{\text{c}} -(C_{\text{se}} + C_{\text{e}})N - \left(C_{\text{ann}}+C_{\text{sevap}}\right)N^2,
\end{equation}
which has the general solution 
\begin{align}
N(t) = \frac{C_{\text{c}}\tanh{\left(\frac{t}{\xi_{\text{E}}}\right)}}{\frac{1}{\xi_{\text{E}}} +\frac{C_{\text{se}}+C_{\text{e}}}{2}\tanh{\left(\frac{t}{\xi_{\text{E}}}\right)}},
\end{align}
where $\xi^{-1}_{\text{E}} =\sqrt{C_{\text{c}}(C_{\text{ann}}+C_{\text{sevap}}) + (C_{\text{se}}+C_{\text{e}})^2/4}$. 
The general solution in the absence of self-interactions is
\begin{align}
\label{eq:N0evap2}
N_0(t) = \frac{C_{\text{c}}\tanh{\left(\frac{t}{\xi_0}\right)}}{\frac{1}{\xi_0} +\frac{C_{\text{e}}}{2}\tanh{\left(\frac{t}{\xi_0}\right)}},
\end{align}
where $\xi_0^{-1} = \sqrt{C_{\text{c}}C_{\text{ann}} +
  C_{\text{e}}^2/4}$.  Analogously to our discussion of the
Sun above, we quantify the importance of self-interactions by
introducing the ``depletion'' factor, the ratio of the annihilation
flux in the presence of self-interactions to the annihilation flux
expected for non-interacting DM:
\begin{align}
\label{eq:gammadef}
\gamma \equiv 
\frac{\Gamma_{\text{ann}}}{\Gamma_{\text{ann},0}}=\frac{C_{\text{ann}}N(\tau_\Earth)^2}{C_{\text{ann},0} N_0(\tau_\Earth)^2},
\end{align}
where $\tau_{\Earth} =4.5\times 10^9 \text{ years }$ is the age of the Earth and $N_0(t)$ is given by Eq.~\ref{eq:N0evap2}, i.e., the
expected Earth population in the absence of self-interactions, but
including the effect of evaporation.  Again, for fixed $C_{\text{ann}}$, 
$\gamma$ reduces to a comparison
of the total captured population in the cases with and without
self-interactions.

%
\begin{figure}[t]
\begin{subfigure}{0.5\textwidth}
  \hspace{-0.5cm}
  \includegraphics[width=1.0\linewidth]{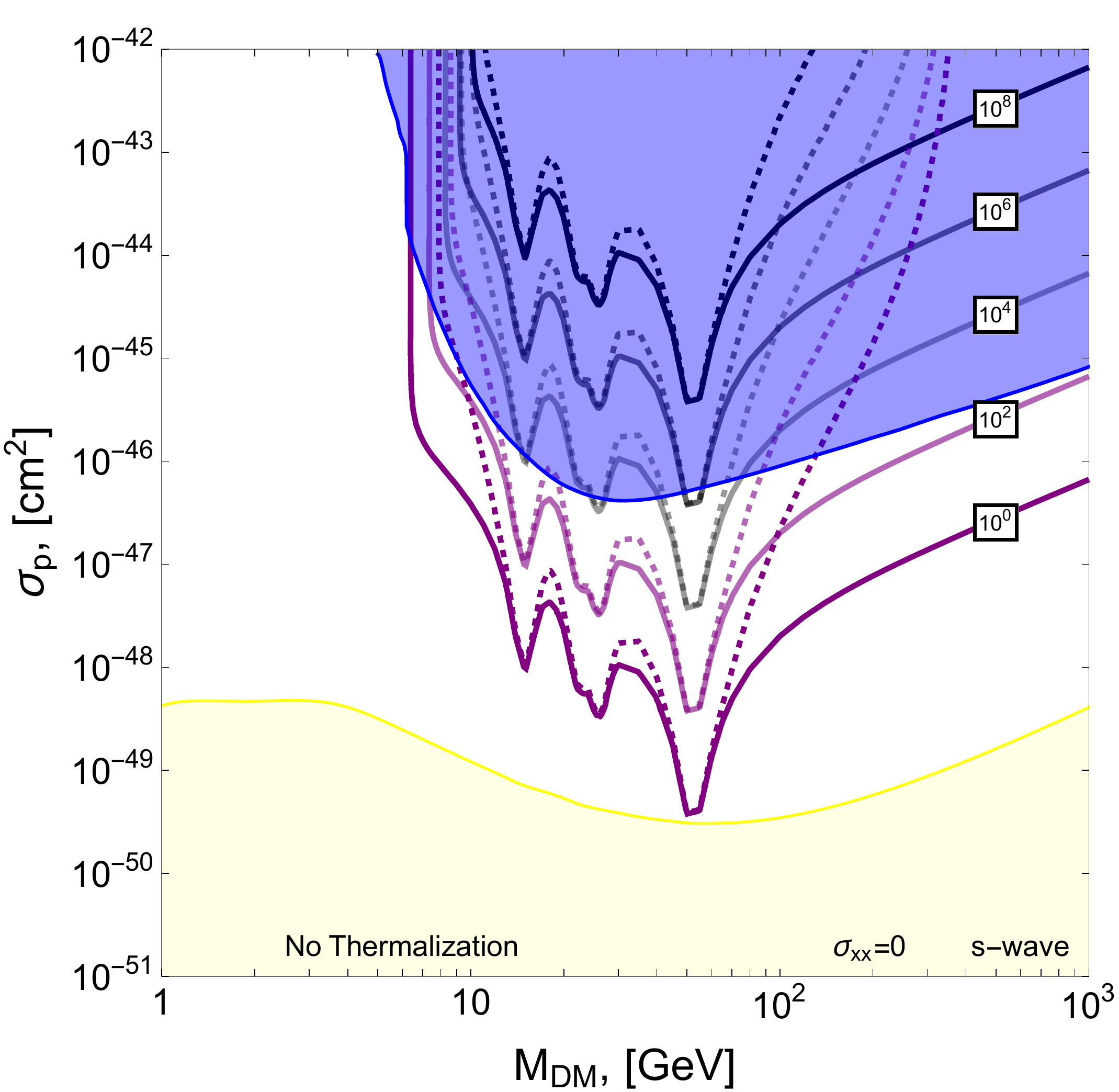}
\end{subfigure}%
\begin{subfigure}{0.5\textwidth}
  \hspace{-0.5cm}
  \includegraphics[width=1.0\linewidth]{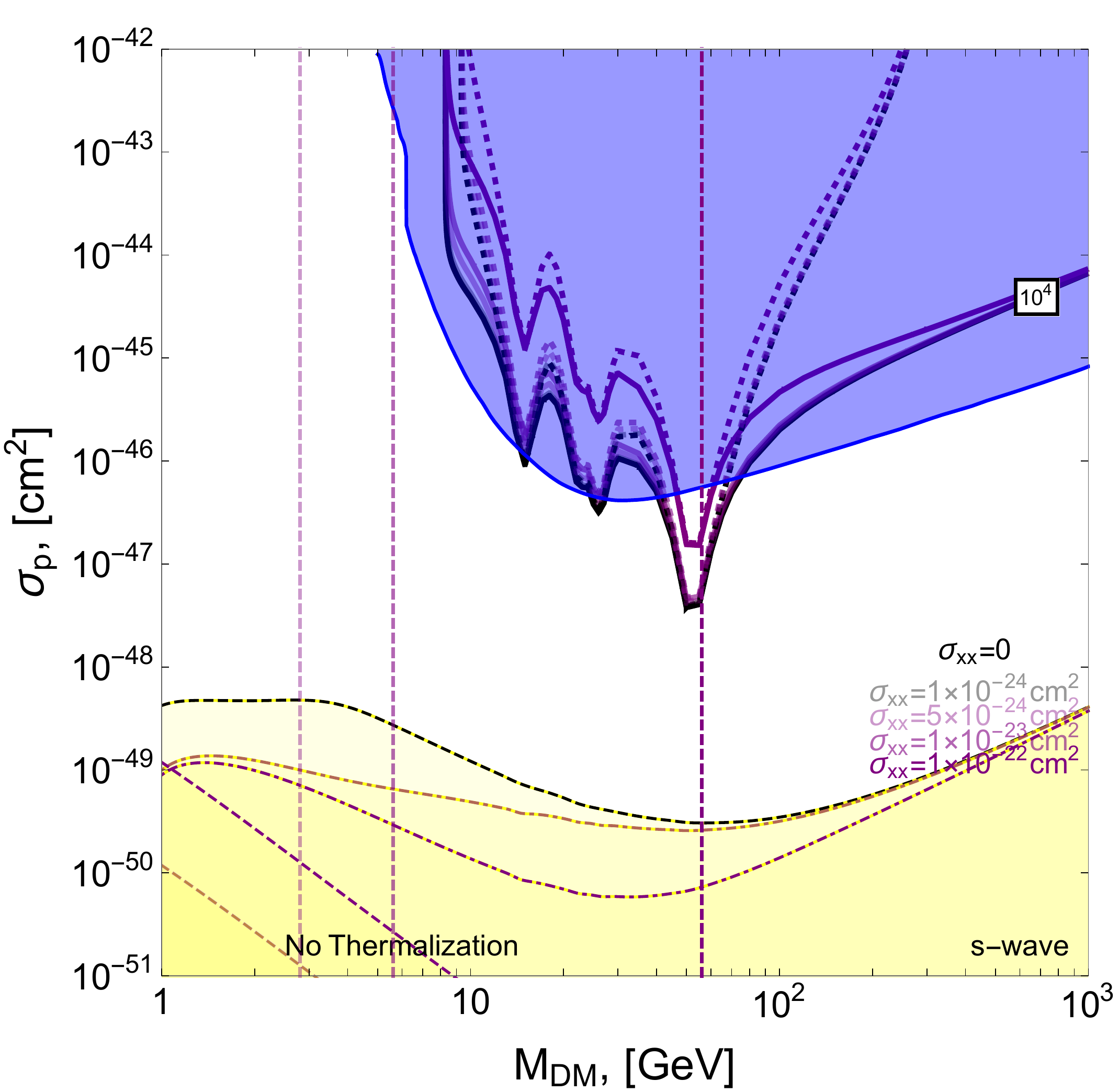}
\end{subfigure}%
\caption{Annihilation flux in the Earth for constant self-interaction
  cross-section, in $\text{km}^{-2}\text{yr}^{-1}$. Colored lines
  represent contours of constant flux. Solid and dashed contours
  correspond to Free Space and Direct Capture, respectively. In the
  left plot we show the annihilation fluxes in the absence of
  self-interactions. In the right plot we show how a given
  annihilation flux contour is modified by the introduction of
  increasing (constant) self-interaction cross-section. The blue
  shaded region is excluded by the PandaX, LUX and XENON1T direct
  detection experiments. The parameter space to the left of the dashed
  purple lines are in tension with the Bullet Cluster. In the yellow
  regions $\sigma_{\text{p}}$ is too small for the captured DM
  population to be thermalized in the Earth: (left) less that 90 \% of
  the captured DM population is thermalized; (right) below the
  dot-dashed contours, less than 90 of the captured DM population is
  thermalized for the given choice of $\sigma_{\text{xx}}$, while
  below the dashed contours the captured DM population does not
  thermalize at the Earth's temperature.}
\label{plot:Earth_Annihilation_Flux_1}
\end{figure}
%
Since self-ejections will dominate over self-capture in the Earth, the
dominant effect of self-scattering is to deplete the Earth population
and thus (at fixed $C_{\text{ann}}$) suppress the annihilation flux.
The annihilation flux from the Earth,
\begin{align}
\Phi_\Earth = \frac{\Gamma_{\text{ann}}}{4\pi D^2},
\end{align}
where $\text{D} = R_{\Earth}$, is plotted in the $(M_{\text{DM}},\sigma_{\text{p}})$
plane in Fig.~\ref{plot:Earth_Annihilation_Flux_1}, without (left) and
with (right) constant self-interactions.  Comparing
Figs.~\ref{plot:Sun_Annihilation_Flux_1}
and~\ref{plot:Earth_Annihilation_Flux_1}, it is clear that the Earth
annihilation flux is numerically subdominant to the flux from the Sun.

Neutrino telescopes are sensitive to annihilation fluxes from the
Earth on the order of
$10^9-10^{13}\,\mathrm{km}^{-2}\mathrm{yr}^{-1}$, typically a factor
of a few worse sensitivity than for the Sun
\cite{Aartsen:2016fep,Albert:2016dsy}, while sensitivity can be much
greater for models where DM annihilates to dark states,
e.g. $10^3\,\mathrm{km}^{-2}\mathrm{yr}^{-1}$ in a model that
annihilates to dark photons \cite{Feng:2015hja}.

\begin{figure}[t]
\begin{subfigure}{0.5\textwidth}
  \hspace{-1.0cm}
  \includegraphics[width=1.0\linewidth]{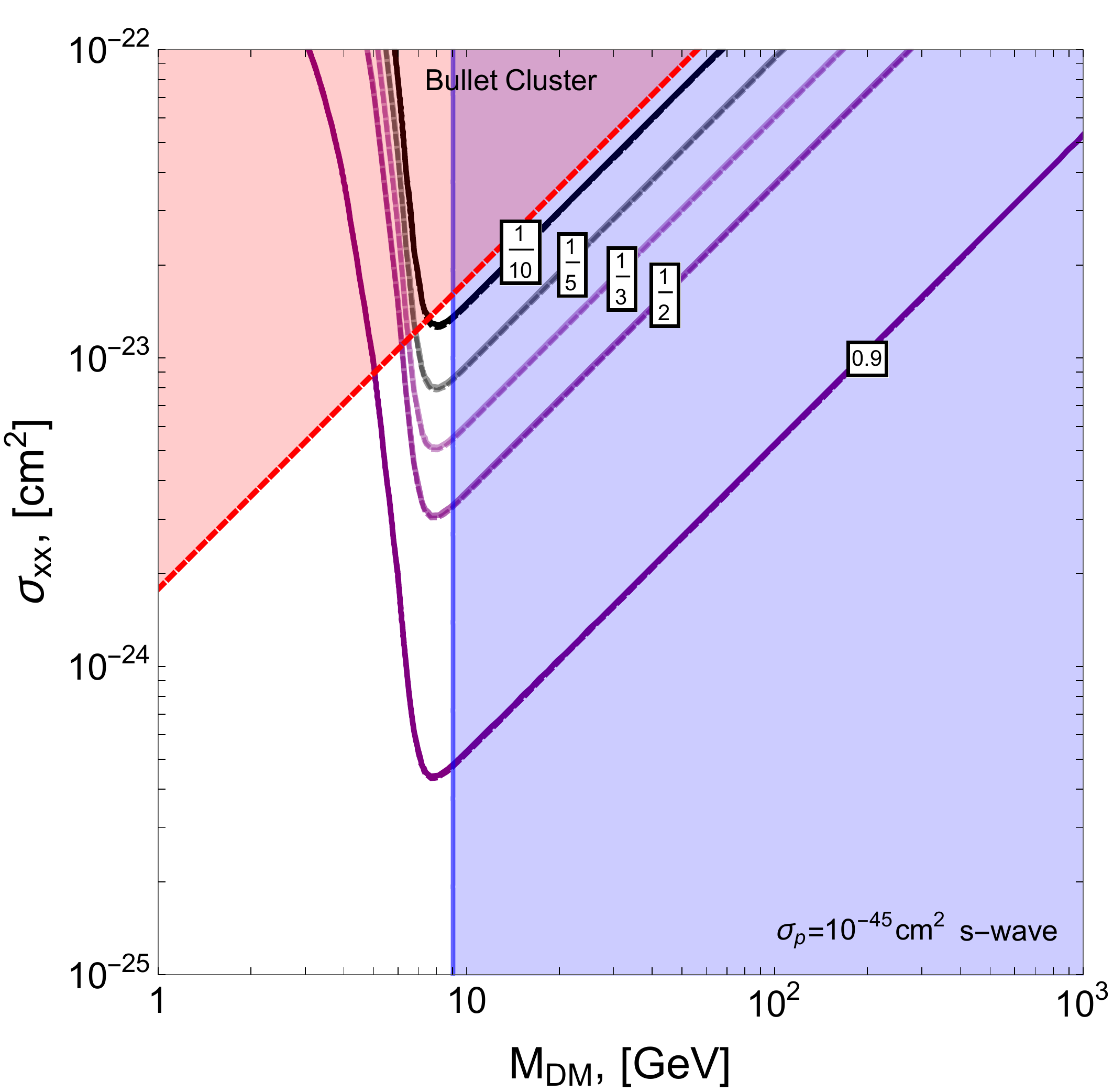}
\end{subfigure}%
\begin{subfigure}{0.5\textwidth}
  \hspace{-1.0cm}
  \includegraphics[width=1.0\linewidth]{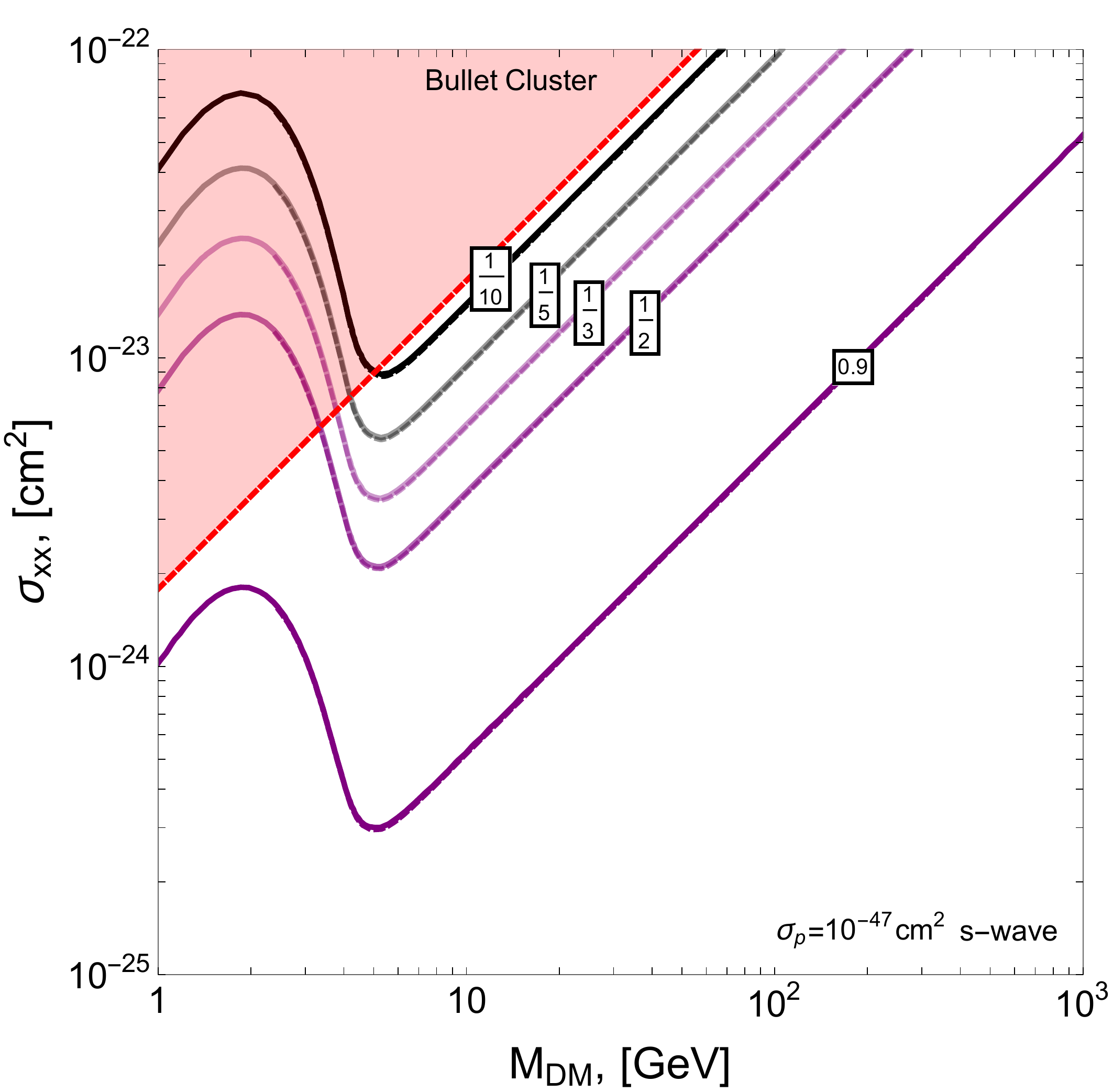}
\end{subfigure}
\caption{Earth annihilation flux ratio $\gamma$ for constant
  self-interaction cross-section. Colored lines represent contours of
  constant $\gamma$: solid and dashed lines correspond to Free Space
  and Direct Capture velocity distributions, respectively.  The solid
  and dashed lines are nearly coincident on the plot.  We show results
  for fixed nuclear cross-sections $\sigma_{\text{p}} =
  10^{-45}\text{cm}^2$ (left) and $\sigma_{\text{p}} =
  10^{-47}\text{cm}^2$ (right). The red shaded region is excluded by
  the Bullet Cluster constraint and the blue shaded region is excluded
  by the PandaX, LUX, and XENON1T direct detection experiments.}
\label{plot:Earth_Depletion_1}
\end{figure}

\begin{figure}[h]
\begin{subfigure}{0.5\textwidth}
  \hspace{-0.5cm}
  \includegraphics[width=1.0\linewidth]{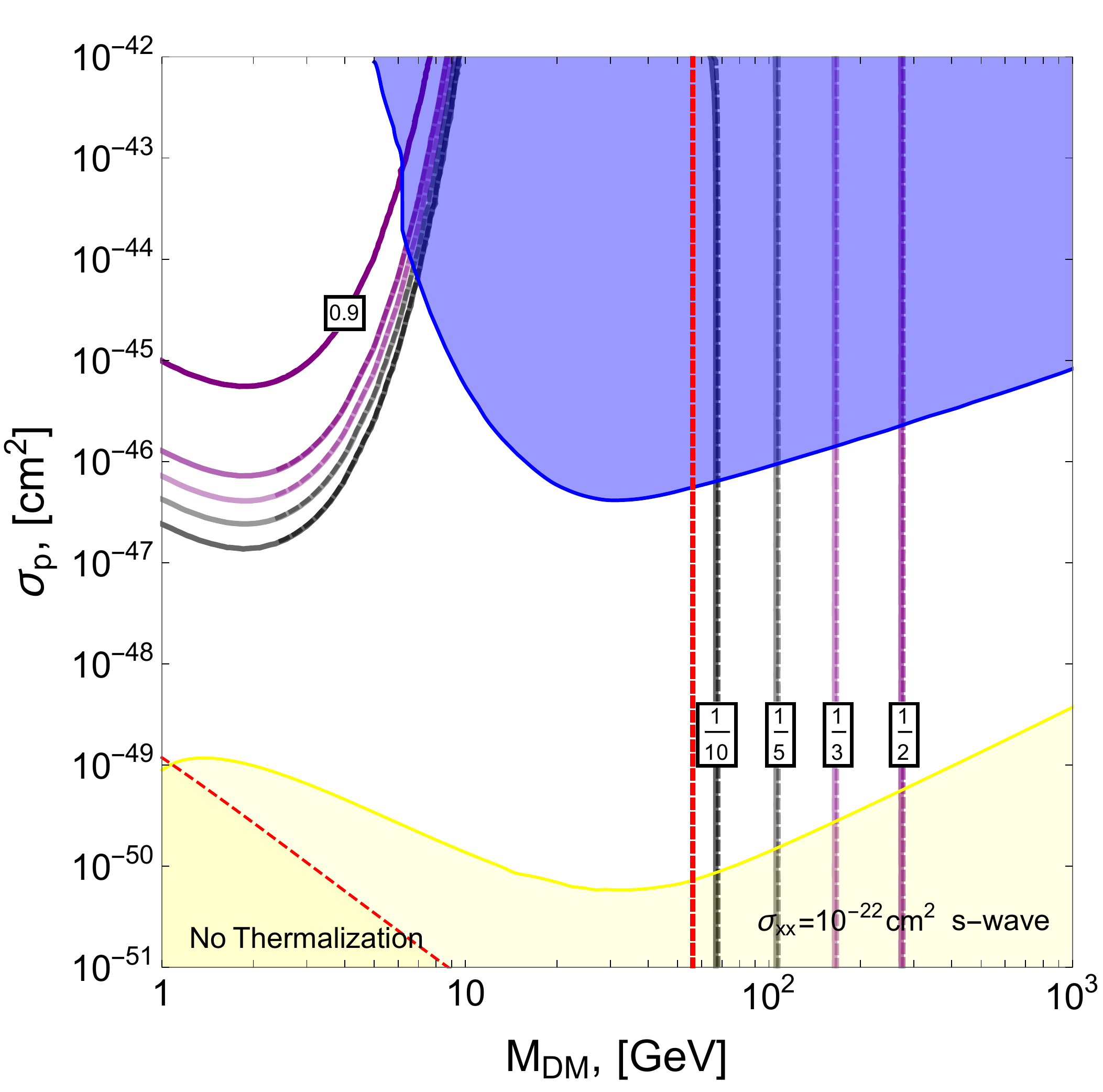}
\end{subfigure}%
\begin{subfigure}{0.5\textwidth}
  \hspace{-0.5cm}
  \includegraphics[width=1.0\linewidth]{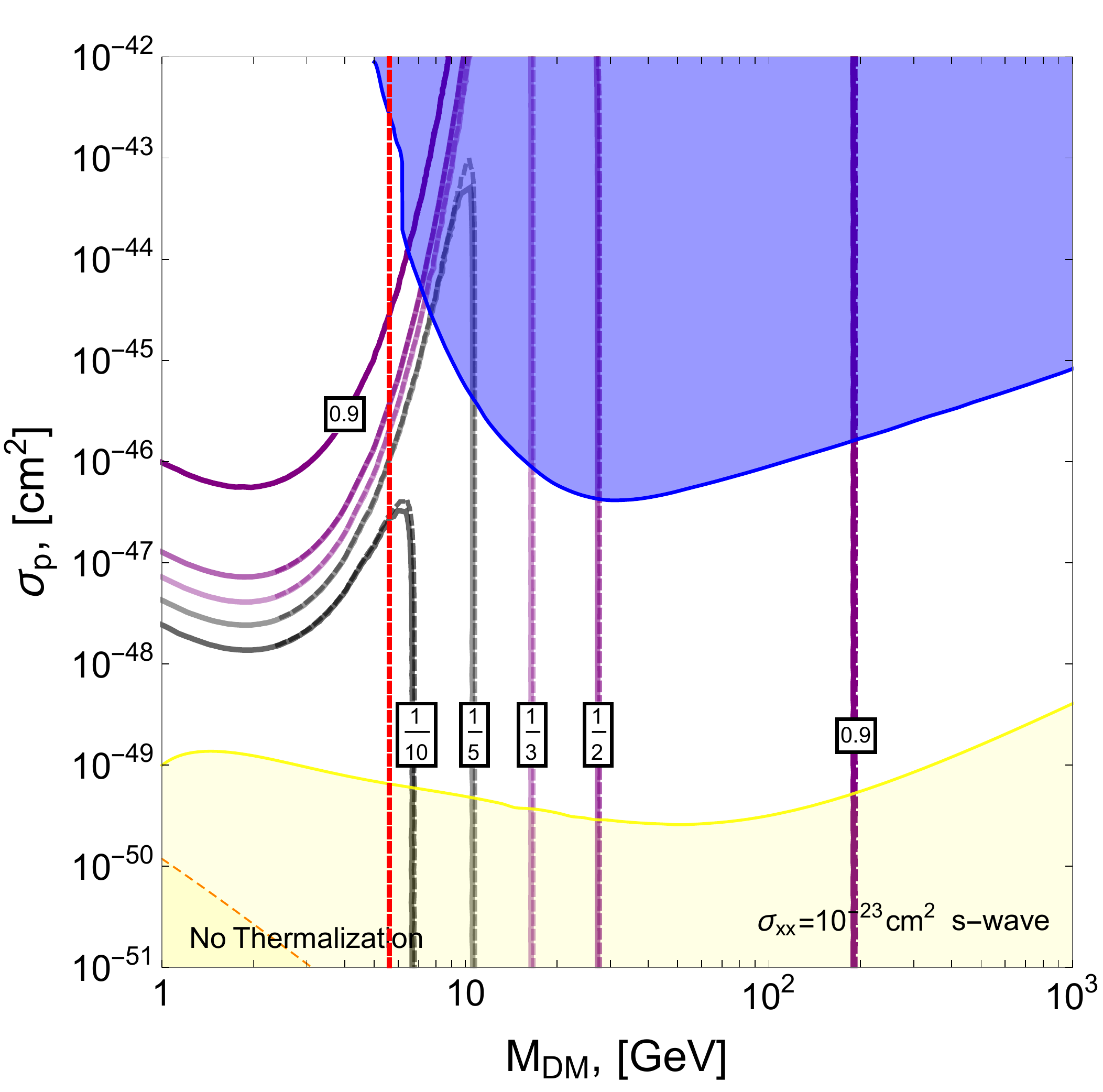}
\end{subfigure}
\caption{Earth annihilation flux ratio $\gamma$ for constant
  self-interaction cross-section. Colored lines represent contours of
  constant $\gamma$: solid and dashed lines correspond to Free Space
  and Direct Capture, respectively.  We show results for fixed
  self-interaction cross-sections $\sigma_{\text{xx}} =
  10^{-22}\text{cm}^2$ (left) and $\sigma_{\text{xx}} =
  10^{-23}\text{cm}^2$ (right). The blue shaded region is excluded by
  the PandaX, LUX and XENON1T direct detection experiments. In the
  yellow region $\sigma_{\text{p}}$ is too small for the captured DM population
  to be thermalized in the Earth. The parameter space to the left of
  the dashed red line is in tension with the Bullet cluster
  constraint.}
\label{plot:Earth_Depletion_2}
\end{figure}

We again fix the annihilation cross-section to the reference value
$\langle\sigma v\rangle_{\text{ann}} =3\times
10^{-26}\,\mathrm{cm}^3/\mathrm{s}$, and show contours of the Earth
flux ratio $\gamma$ in the $(M_{\text{DM}}, \sigma_{\text{xx}})$ and
$(M_{\text{DM}}, \sigma_{\text{p}})$ planes in
Figs.~\ref{plot:Earth_Depletion_1} and~\ref{plot:Earth_Depletion_2}
respectively.

We begin by noting that depletions in the Earth are more modest in
magnitude than the enhancements in the Sun. The strongest depletions
consistent with Bullet Cluster bounds are $\sim 1/10$.  Otherwise,
many features of the contours in Fig.~\ref{plot:Earth_Depletion_1},
such as the linearity at high DM mass, are the same as observed for
the Sun in Fig.~\ref{plot:Sun_Enhancement_1}, and can be similarly
understood.

One new feature that we can observe from
Figs.~\ref{plot:Earth_Depletion_1} and~\ref{plot:Earth_Depletion_2} is
that the depletion factor $\gamma$ does not appreciably depend on the
choice of the DM velocity distribution at the Earth's orbit.  As
discussed in Sec.~\ref{sec:earth}, the greatest impact of the choice
of velocity distribution is on the nuclear capture coefficient
$C_{\text{c}}$ outside the region of resonant capture
(i.e. $M_{\text{DM}} < 10 \,\text{GeV}$ and $M_{\text{DM}} >100\,
\text{GeV}$). But in the low mass region $M_{\text{DM}} < 10\,
\text{GeV}$, evaporation dominates, $\frac{1}{4}C_{\text{e}}^2 \gg
C_{\text{c}}(C_{\text{ann}}+C_{\text{sevap}})$, while in the high-mass
region $M_{\text{DM}} >100\, \text{GeV}$, self-ejection dominates,
$\frac{1}{4}C_{\text{se}}^2 \gg
C_{\text{c}}(C_{\text{ann}}+C_{\text{sevap}})$.  Therefore away from
the regions dominated by resonant nuclear capture, even weak
depletions are governed by relatively distribution-insensitive
processes.

\section{Diagnosing DM self-interactions}
\label{sec:selfints}

\begin{figure}[h!]
\begin{subfigure}{0.5\textwidth}
  \hspace{-0.5cm}
  \includegraphics[width=1.0\linewidth]{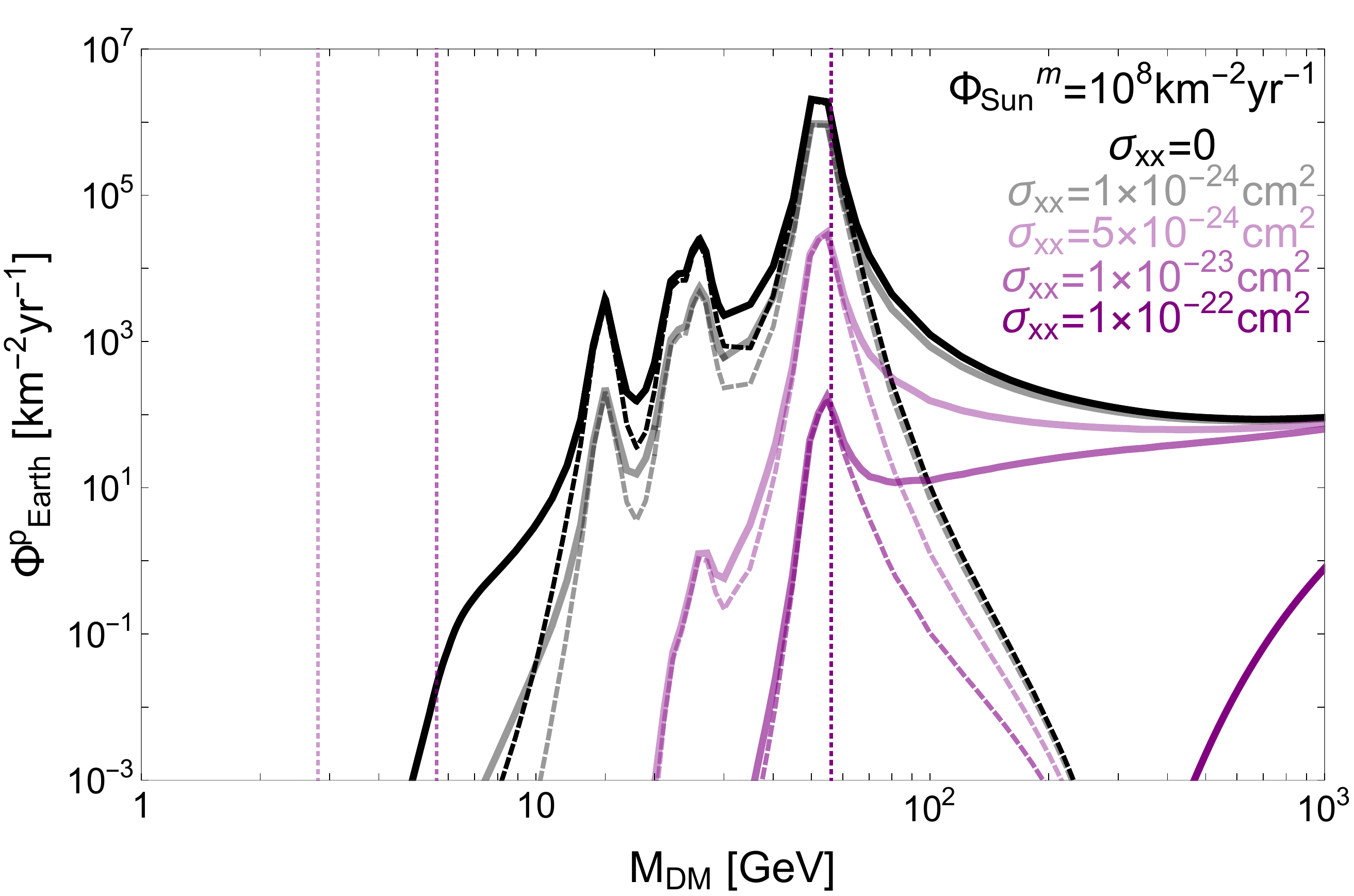}
\end{subfigure}%
\begin{subfigure}{0.5\textwidth}
  \hspace{-0.5cm}
  \includegraphics[width=1.0\linewidth]{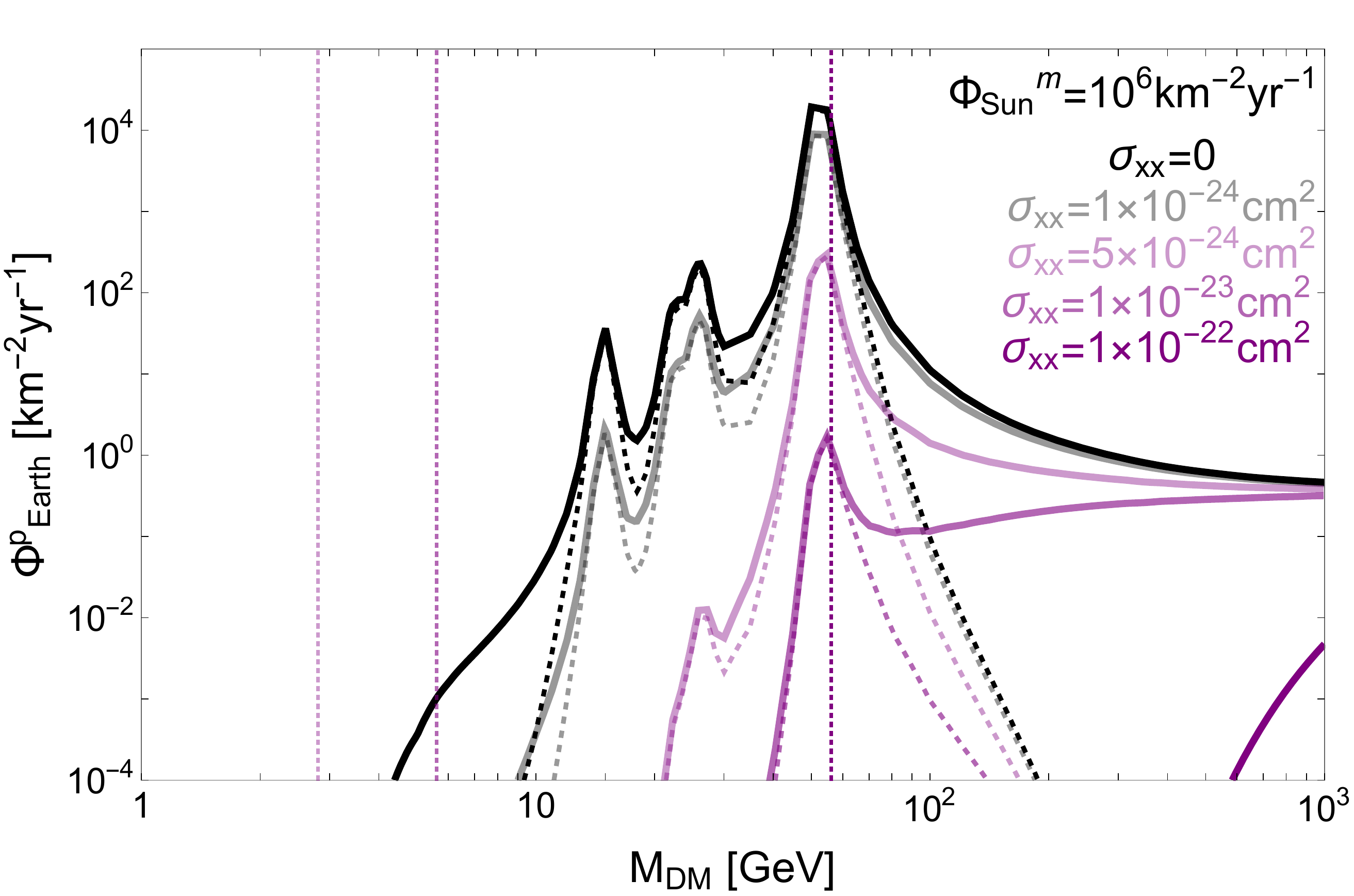}
\end{subfigure}
\caption{Predicted annihilation flux in the Earth for constant
  self-interaction cross-section and $s$-wave annihilations, in
  $\text{km}^{-2}\text{yr}^{-1}$, as a function of $M_{\text{DM}}$.
  Colored lines show different values of $\sigma_{\text{xx}}$, as
  indicated by opacity. In the left plot we fix
  $\Phi^{\text{m}}_\odot= 10^8\text{ km}^{-2}\text{yr}^{-1}$ and in
  the right plot $\Phi^{\text{m}}_\odot= 10^6\text{
    km}^{-2}\text{yr}^{-1}$. Solid and dashed contours correspond to
  Free Space and Direct Capture for the Earth, respectively. The
  regions to the left of the vertical dotted lines are excluded by the
  Bullet Cluster observations for the self-interactions of the
  corresponding color.}
\label{plot:SunAndEarth_Annihilation_Flux_1}
\end{figure}

We are now ready to demonstrate how comparing Earth and Sun
annihilation fluxes can reveal the presence of dark matter
self-interactions.  In most scenarios, the numerically larger Sun
annihilation flux is likely to be observed before the Earth
annihilation flux.  Suppose the Sun flux is measured to be
$\Phi^{\text{m}}_\odot$. If we neglect self-interactions, then we can
solve
\beq
\Phi_\odot^{\text{m}} = \Phi_\odot(M_{\text{DM}}, \sigma_{\text{p}}, \sigma_{\text{xx}}=0)
\eeq
to obtain the value of the nuclear cross-section
$\sigma_{\text{p}}^{(0)}=\sigma_{\text{p}} (M_{\text{DM}};
\Phi^{\text{m}}_\odot)$ needed to explain the solar measurement in the
absence of self-interactions, as a function of $M_{\text{DM}}$.  The
deduced $\sigma_{\text{p}}^{(0)}$ in turn predicts a specific Earth
annihilation flux as a function of $M_{\text{DM}}$. We call this the
``null" prediction,
\begin{align}
\Phi_\Earth^{(0)} =  \Phi_\Earth\left(M_{\text{DM}}, \sigma^{(0)}_{\text{p}}(M_{\text{DM}}; \Phi^{\text{m}}_\odot)\right).
\end{align}
This prediction is shown by the black lines in
Fig.~\ref{plot:SunAndEarth_Annihilation_Flux_1}.

Self-interactions, if present, enhance the Sun population while
suppressing the Earth population, leading to a smaller Earth flux than
the null prediction, sometimes by many orders of magnitude.  In the
presence of self-interactions, the predicted Earth flux can be
obtained from a measured Sun flux as a function of $M_{\text{DM}}$ and
$\sigma_{\text{xx}}$, again by finding the value of the nuclear
cross-section that yields the desired flux from the Sun,
\begin{align}
\sigma^{\text{(p)}}_{\text{p}} = \sigma_{\text{p}}\left(M_{\text{DM}}, \sigma_{\text{xx}};\Phi^{\text{m}}_\odot \right).
\end{align}
The corresponding Earth flux is then simply
\begin{align}
\Phi^{\text{(p)}}_\Earth = \Phi_\Earth\left(M_{\text{DM}},\sigma^{\text{(p)}}_{\text{p}}\left(M_{\text{DM}}, \sigma_{\text{xx}};\Phi^{\text{m}}_\odot \right), \sigma_{\text{xx}}\right).
\end{align}
As shown in Fig.~\ref{plot:SunAndEarth_Annihilation_Flux_1},
self-interactions of astrophysical interest can suppress the predicted
Earth flux by many orders of magnitude, far more than can be
accommodated by uncertainties in the Earth's composition.  This
suppression is sizeable for cross-sections orders of magnitude below
the Bullet Cluster bounds, making this solar system comparison a
leading test of dark matter self-interactions.

Fig.~\ref{plot:SunAndEarth_Annihilation_Flux_1} shows results for both
Free Space and Direct Capture regimes in the Earth, and demonstrates
that for heavy DM, $M_{\text{DM}}\gtrsim 100 \,\text{GeV}$, the choice
of DM velocity distribution at the Earth's orbit is numerically more
consequential than the suppression from self-interactions.  Thus to
make strong statements about self-interactions in this regime, some
additional information about the DM velocity distribution seen by the
Earth would be required, perhaps from detailed numerical simulations.
In the resonant capture regime, however, the impact of different
velocity distributions can easily be a much smaller effect than the
overall numerical suppression from self-interactions.

\begin{figure}[t]
\begin{subfigure}{0.56\textwidth}
  \hspace{-1.2cm}
  \includegraphics[width=1.0\linewidth]{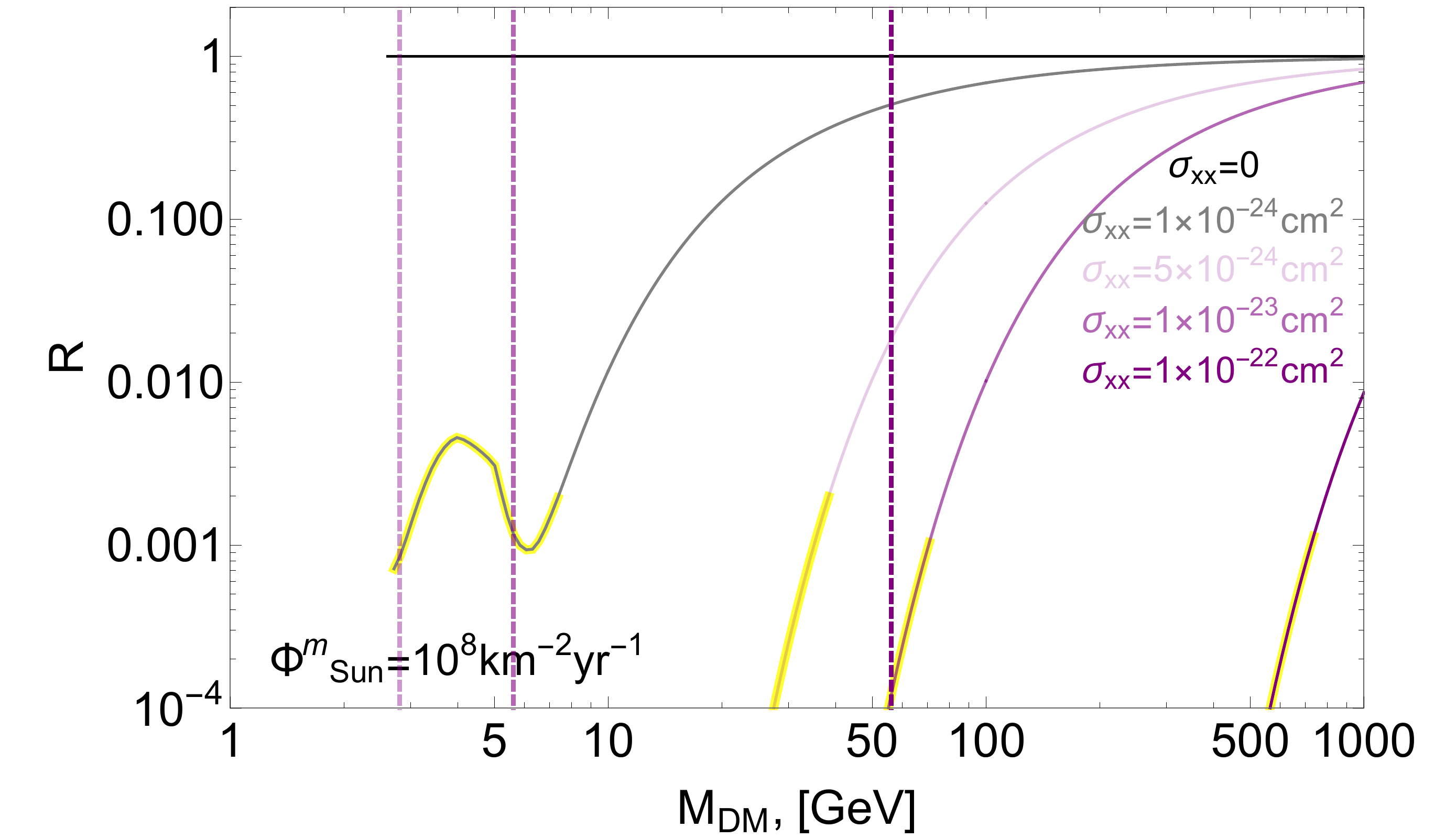}
\end{subfigure}%
\begin{subfigure}{0.56\textwidth}
  \hspace{-1.4cm}
  \includegraphics[width=1.0\linewidth]{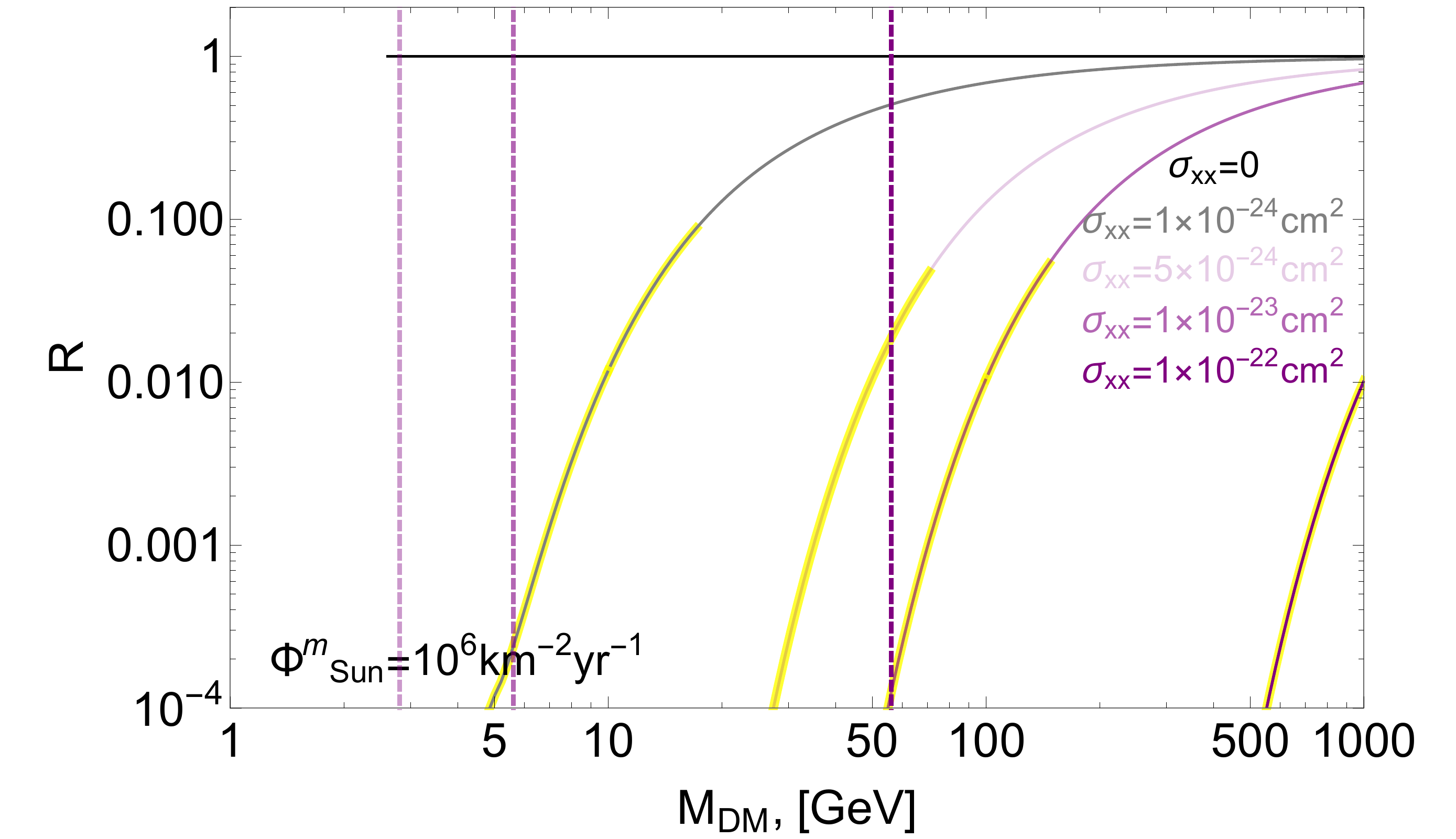}
\end{subfigure}
\caption{Plots of the ratio $R$ of self-interacting to
  non-self-interacting Earth annihilation fluxes as a function of DM
  mass, with fixed $s$-wave annihilation rate. Curves of different
  colors correspond to different choices of $\sigma_{\text{xx}}$. In
  the left plot we fix $\Phi^{\text{m}}_\odot=10^8
  \text{km}^2\text{yr}^{-1}$, while in the right plot we fix
  $\Phi^{\text{m}}_\odot=10^6 \text{km}^2\text{yr}^{-1}$. Points to
  the left of the vertical dashed lines are in tension with Bullet
  Cluster constraints for the self-interactions of the corresponding
  color. The thick yellow lines indicate parameter space with nuclear
  cross-sections below the values that guarantee Sun or Earth
  thermalization. All parameter space shown is consistent with current
  direct detection results.}
\label{plot:Earth_Flux_Ratio_Plots}
\end{figure}

Comparing the left and right panels of
Fig.~\ref{plot:SunAndEarth_Annihilation_Flux_1} shows that the
relative suppression of the Earth flux is not strongly dependent on
the value of the Sun flux.  We can understand this straightforwardly,
and derive an explicit analytical approximation for the suppression as
follows.  We begin by defining the ratio between predicted Earth
fluxes with and without self interactions,
\begin{align}
\label{eq:r}
R \equiv \frac{\Phi^{\text{(p)}}_\Earth(M_{\text{DM}}, \sigma_{\text{xx}};\Phi^{\text{m}}_\odot)}{\Phi^{(0)}_\Earth(M_{\text{DM}};\Phi^{\text{m}}_\odot)},
\end{align}
which quantifies the suppression of the expected Earth flux; this
ratio is shown in Fig.~\ref{plot:Earth_Flux_Ratio_Plots}.  At fixed
$C_{\text{ann}}$, we must always have $R\leq 1$, simply as a
consequence of solar system kinematics.
%
%
To find an analytical approximation to $R$, we begin with the simpler
regime where neither Earth nor Sun populations have attained steady
state.  This is generic in the Earth, and for the Sun it holds for
self-interaction cross-sections $\sigma_{\text{xx}}\lesssim
10^{-24}\,\text{cm}^2$.  The DM population in the Earth can thus be
approximated as
\beq
N^{(0)}_\Earth = C_{\Earth\text{c}}(\sigma^{(0)}_{\text{p}})\tau_\Earth
\eeq
in the absence of self-interactions, and
\beq
\quad N_\Earth^{\text{(p)}} = C_{\Earth\text{c}}(\sigma^{(p)}_{\text{p}})\tau_\Earth\frac{1}{1+\frac{C_{\Earth\text{se}}}{2}\tau_\Earth}
\eeq
with self-interactions included.   Since $C_{\text{c}}\propto \sigma_{\text{p}}$, we then have
\begin{align}
R = \left(\frac{ \sigma^{(p)}_{\text{p}}   }{\sigma^{(0)}_{\text{p}}}   \frac{1}{  1+\frac{C_{\Earth\text{se}}}{2}   \tau_\Earth    }\right)^2.
\end{align}
Meanwhile, for a Sun-captured DM population that is far away from
equilibrium, $\sigma^{(p)}_{\text{p}}$ and $\sigma^{(0)}_{\text{p}}$
are determined by solving
\begin{align}
\Phi^{\text{m}}_\odot&=\frac{1}{4\pi D^2}C_{\odot\text{ann}}\left(C_{\odot\text{c}}(\sigma^{(0)}_{\text{p}})\tau_\odot\right)^2 \\
\Phi^{\text{m}}_\odot&= \frac{1}{4\pi D^2}C_{\odot\text{ann}}\left(C_{\odot\text{c}}(\sigma^{(p)}_{\text{p}})\tau_\odot\frac{1}{1-\frac{C_{\odot\text{sc}}}{2}\tau_\odot}\right)^2, 
\end{align}  
yielding 
\begin{align}
\left(\frac{ \sigma^{(p)}_{\text{p}}   }{\sigma^{(0)}_{\text{p}}}\right)^2 = \left( 1-\frac{C_{\odot\text{sc}}}{2}   \tau_\odot \right)^2, 
\end{align} 
so that the expected suppression from self-interactions is 
\begin{align}
R \approx \left(\frac{ 1-\frac{C_{\odot\text{sc}}}{2}\tau_\odot}{ 1+\frac{C_{\Earth\text{se}}}{2}   \tau_\Earth}\right)^2.
\end{align}
At larger values of $\sigma_{\text{xx}}$, self-interactions become
sufficiently strong that the Sun-captured population reaches
steady-state. This will re-introduce a dependence on
$\Phi^{\text{m}}_\odot$ in $R$. However, since most of the Sun DM
population self-captures in this case, very small $\sigma_{\text{p}}$
values can be accommodated, pushing much of this region below the
thermalization floor at a given $\Phi^{\text{m}}_\odot$.

Finally, it is worth remarking that this technique to extract both
the self-interaction cross-section and the nuclear cross-section from
two solar system flux measurements works only for $M_{\text{DM}}$ above the
mass where evaporations become important in both the Sun and, more
restrictively, the Earth.  For DM masses below this evaporation mass,
the annihilation flux of non-self-interacting DM no longer depends on
$\sigma_\text{p}$ \cite{Busoni:2013kaa}.  Thus for DM lighter than the
evaporation mass, the degeneracy in predictions as well as the much
reduced annihilation fluxes make it very challenging to extract useful
information about DM self-interactions through this comparison of
annihilation fluxes from the Earth and Sun.  Additionally, as Fig.~\ref{plot:Earth_Flux_Ratio_Plots}
 demonstrates, for any given value of $\sigma_{\text{xx}}$, there is a minimum
DM mass for which the assumption of thermalization is valid, and the procedure here is thus self-consistent.

\section{Conclusions}
\label{sec:conclusions}

We have demonstrated that the annihilation of DM captured by the Sun
and the Earth can provide a leading test of DM self-interactions.
DM's self-scattering cross-section enhances the DM capture rate in the
Sun while suppressing it in the Earth, thanks simply to the differing
kinematics of scattering in the two potential wells.  Thus comparing
the DM populations captured by both bodies provides a powerful
diagnostic of the presence of DM self-interactions.  We examined DM
capture and annihilation in detail in the case of a constant
spin-independent nuclear cross-section and an isotropic
self-interaction cross-section, and showed that self-interactions can
suppress the Earth flux by many orders of magnitude, far more than
could be accounted for by uncertainties in the Earth's interior
composition.  While the absolute size of the annihilation flux from
the Earth depends on the velocity distribution of DM at the Earth's
orbit, the fractional suppression of the flux from self-interactions
is largely insensitive to the details of this distribution.  In
particular, self-interaction cross-sections that are large enough to
affect the properties of dwarf galaxies have a sizeable quantitative impact on the
predicted Earth annihilation flux.  Moreover, depending on the DM
mass, significant suppressions are possible for DM self-scattering
cross-sections orders of magnitude below current astrophysical
bounds. We further derived a semi-analytical approximation to the
fractional suppression of the Earth flux that holds over most of the
parameter space still testable by direct detection
experiments for spin-independent nuclear cross-sections.

Our calculations assume that captured DM thermalizes with the nuclei
in the massive body, and thus can be described by a time-independent
spatial distribution.  As direct detection experiments place ever more
stringent constraints on DM-nuclear scattering, the surviving
parameter space where this assumption is valid is rapidly shrinking.
As we discuss in detail in the Appendix, concerns about thermalization
are especially acute for dark matter with strong self-interactions.
For the constant self-interaction cross-sections investigated here, we find that energy
transfer from halo DM to bound DM is a subdominant barrier to thermalization, but that self-interactions can increase the thermalization timescale in the Sun simply through enhancing the total DM population.  Without thermalization, one would need a full numerical
treatment of the trajectories of captured DM particles to obtain a
prediction for the annihilation flux.  However, the essential result
of this paper---that the differing kinematics of DM capture within the
Sun and the Earth lead to differing and potentially observable impacts
of DM self-interactions on the respective bound populations---remains
true even in this case.  Similarly, while we have assumed for
simplicity a constant spin-independent DM-nuclear scattering
cross-section, our essential conclusion holds for any DM-nuclear interaction that allows significant
capture rates in the Earth as well as the Sun.

The DM annihilation fluxes from the Sun and the Earth depend on the kinematics assumed for scattering as well as on the matrix elements governing DM interactions, and thus may also be
affected by departures from the standard halo distribution, e.g. a
(thick) dark disk \cite{Read:2008fh,Bruch:2009rp}.\footnote{Thin dark
  disks are already highly constrained by Gaia data
  \cite{Schutz:2017tfp,Buch:2018qdr}.} 
In particular the low relative velocity of a
disk component of DM would tend to enhance, rather than suppress, the
Earth signal relative to the Sun signal, while suppressing direct
detection event rates \cite{Bruch:2009rp,Fan:2013bea}.  High velocity DM streams,
on the other hand, would dramatically suppress Earth annihilation fluxes while enhancing 
direct detection rates.  Critically, for nuclear cross-sections that allow for thermalization,
direct detection experiments would typically be able to provide a direct and independent probe of the
 velocity distribution of DM at Earth's orbit.  Thus one may anticipate that, combining information from
DM direct detection experiments, studies of DM diffusion through the solar system, and observed annihilation fluxes from Sun and Earth, DM particle physics could readily be disentangled from astrophysics.
 Sizeable
changes to DM annihilation cross-sections, relative to the reference
thermal value used here, can also impact the relative size of the
Earth and Sun annihilation fluxes \cite{Delaunay:2008pc}.
Significantly enhanced and environment-dependent annihilation rates
are often realized in models of self-interacting dark matter, the
consequences of which we will return to in future work.

\paragraph{Acknowledgements.} We gratefully acknowledge useful
conversations with A.~Peter and P.~Tanedo.   The work of CG and
JS was supported in part by DOE grants DE-SC0015655 and Early Career
DE-SC0017840.

\appendix

\section{Sun and Earth models}
\label{sec:models}

In this Appendix we describe the density profile and elemental
composition of our adopted Sun and Earth models. Additionally, we
collect the values of relevant DM halo parameters and Solar System
velocities in Table~\ref{table:Parameters}.

\begin{table}
\centering
\small
\begin{tabular}{| c | p{7.8cm} | p{2.3cm}|}
\hline
$\rho_\odot$ & \centering{DM density in the halo}&\centering{0.4 $\text{GeV}/\text{cm}^3$} \tabularnewline \hline
${\bf{v}}_{\text{LSR}}$ & \centering{Local Standard of Rest at Sun's position } &\centering{(0,235,0)\text{ km/s}} \tabularnewline \hline
${\bf{v}}_{\text{pec}}$ & \centering{Sun's peculiar velocity } &\centering{(11,12,7)\text{km/s}} \tabularnewline \hline
$\bar{v}$ & \centering{DM rms velocity in the halo} & \centering{288 \text{km/s}} \tabularnewline \hline
$v_\text{c}$ & \centering{Sun's escape velocity at Earth's position}  & \centering{42 \text{km/s}} \tabularnewline \hline 
$\tilde{v}_{\text{E}}$ & \centering{Average of Earth's speed around the Sun} & \centering{30 \text{km/s}} \tabularnewline \hline
$\tau_{\odot}$ & \centering{Age of the Sun} & \centering{$5\times 10^9$ \text{years}} \tabularnewline \hline
$\tau_{\Earth}$ & \centering{Age of the Earth} & \centering{$4.5\times 10^9$ \text{years}} \tabularnewline \hline
\end{tabular}
\caption{DM halo and solar system parameters relevant for DM capture in the Sun and Earth.}
\label{table:Parameters}
\end{table}

\subsection{The Sun}
\label{sec:appsun}

For the Sun's density profile we use the BS05(AGS, OP) model of
Ref.~\cite{Bahcall:2004pz}.  We divide the profile region into two
regions: region 1 for $ 0\leq r \leq 0.85\times R_{\odot}$ and region 2
for $0.85\times R_{\odot} \leq r \leq R_{\odot}$ and fit a quadratic
polynomial in $r$ to the density profile in region 2 and the exponent
of a quadratic polynomial in region 1. The density is defined as a
piecewise function in the two regions. For the elemental abundance we
use the elements included in the BS05(AGS, OP) model, complemented
with the abundance of heavy elements from
\cite{asplund2009chemical}. Note that we use the same isotope
abundance for $N_i$ in the Sun and the Earth
\cite{brault1981isotopes}. The detailed heavy element content used in
the Solar model is shown in Table \ref{table:Sun_Elem_Abund}.
We take the age of the Sun to be $\tau_\odot = 5.0\times 10^9$ yrs.

\begin{figure}[th!]
\begin{center}
\includegraphics[width=0.85\textwidth]{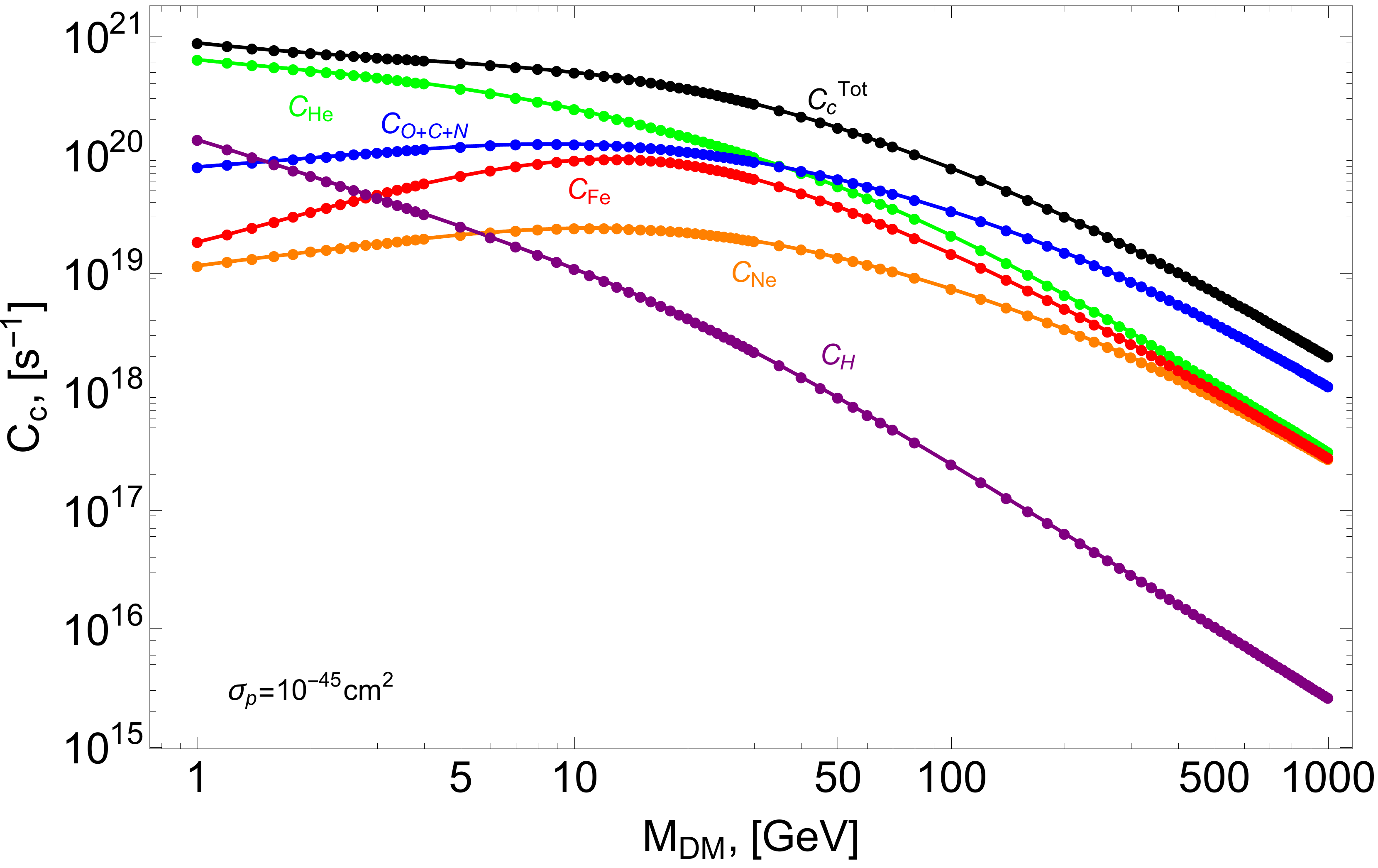}
\caption{ Nuclear capture rate $C_c$ in the Sun as a function of DM
  mass. The green line corresponds to capture due to
  $\text{He}_4$. The blue line corresponds to combined capture due to
  oxygen, carbon and nitrogen. The red, purple and orange lines
  correspond to contributions due to iron, hydrogen and neon,
  respectively. Finally, the black line shows the total capture rate
  from all elements. We show values for a reference $\sigma_{\text{p}}
  = 10^{-45} \text{cm}^2$.}
\label{plot:Sun_CCapture}
\end{center}
\end{figure}

\begin{table}[th]
\centering
\small
\begin{tabular}{| c | p{3.3cm} | p{2.3cm}|}
\hline
Element & \centering{Isotope Abundance [by mass \%]}& \centering{$\log (\epsilon)$} \tabularnewline \hline
$^{20}\text{Ne}$ & \centering{100} & \centering{7.93} \tabularnewline \hline
$^{56}\text{Fe}$ & \centering{100} &\centering{7.50} \tabularnewline \hline
$^{58}\text{Ni}$ & \centering{68}  & \multirow{2}{*}{\centering{\hspace{8mm}6.22}} \tabularnewline \cline{1-2}
$^{60}\text{Ni}$ & \centering{26} &  \tabularnewline \hline
\end{tabular}
\caption{List of heavy elements used in the Solar model, where $\log (\epsilon_i)=\log \left(\frac{n_i}{n_\text{H}}\right)+12$ is the log-10 number density of the $i^{th}$ element as a fraction of the hydrogen number density $n_\text{H}$.}
\label{table:Sun_Elem_Abund}
\end{table}

\subsection{The Earth}
\label{sec:appearth}

We use the Preliminary Reference Earth Model (PREM)
\cite{dziewonski1981preliminary} for the Earth's density profile
$\rho(r)$, where $r$ is the distance from the center of the Earth. The
density profile in PREM consists of 10 regions. We retain the two
inner regions: region 1 (inner core) from $0\leq r \leq1221.5 \text{
  km}$ and region 2 (outer core) from $1221.5\text{ km} \leq r \leq
3480\text{ km}$, but, for simplicity, collect the remaining ones into
two larger domains: region 3 (mantle) from $3480\text{ km} \leq r \leq
5701 \text{ km}$ and region 4 (crust) from $5701\text{ km} \leq r \leq
R_{\Earth}$. The density profiles in regions 1 and 2 are fit to
quadratic polynomials, while the profiles in regions 3 and 4 are fit
to linear polynomials. The fit functions are then used in constructing
a piecewise density function $\rho(r)$.  For the elemental composition
we include the core and mantle weight concentrations of the most
abundant elements reported in Ref.~\cite{mcdonough1995composition},
together with the isotope abundance of Ref.~\cite{rosman1998isotopic}.
The elemental content used in the Earth model is presented in Table
\ref{table:Earth_Elem_Abund} below. It is worth observing that nuclei with non-zero spin are relatively rare in the Earth, and thus the nuclear capture rate of DM in the Earth for spin-dependent cross-sections is highly suppressed.  We take the age of the Earth to
be $\tau_\Earth = 4.5\times 10^9$ yrs.

\begin{figure}[th]
\begin{center}
\includegraphics[width=0.45\textwidth]{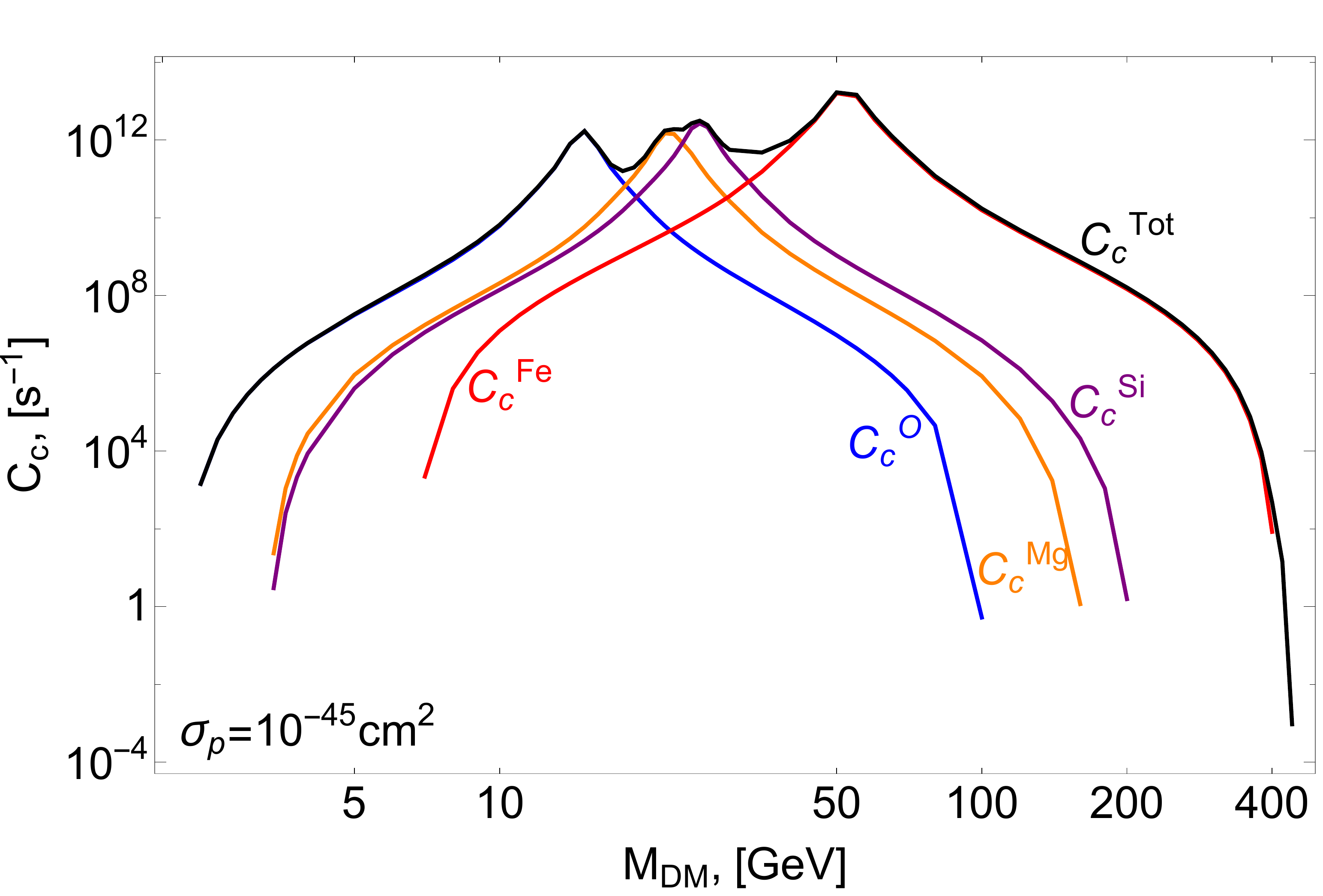}
\includegraphics[width=0.45\textwidth]{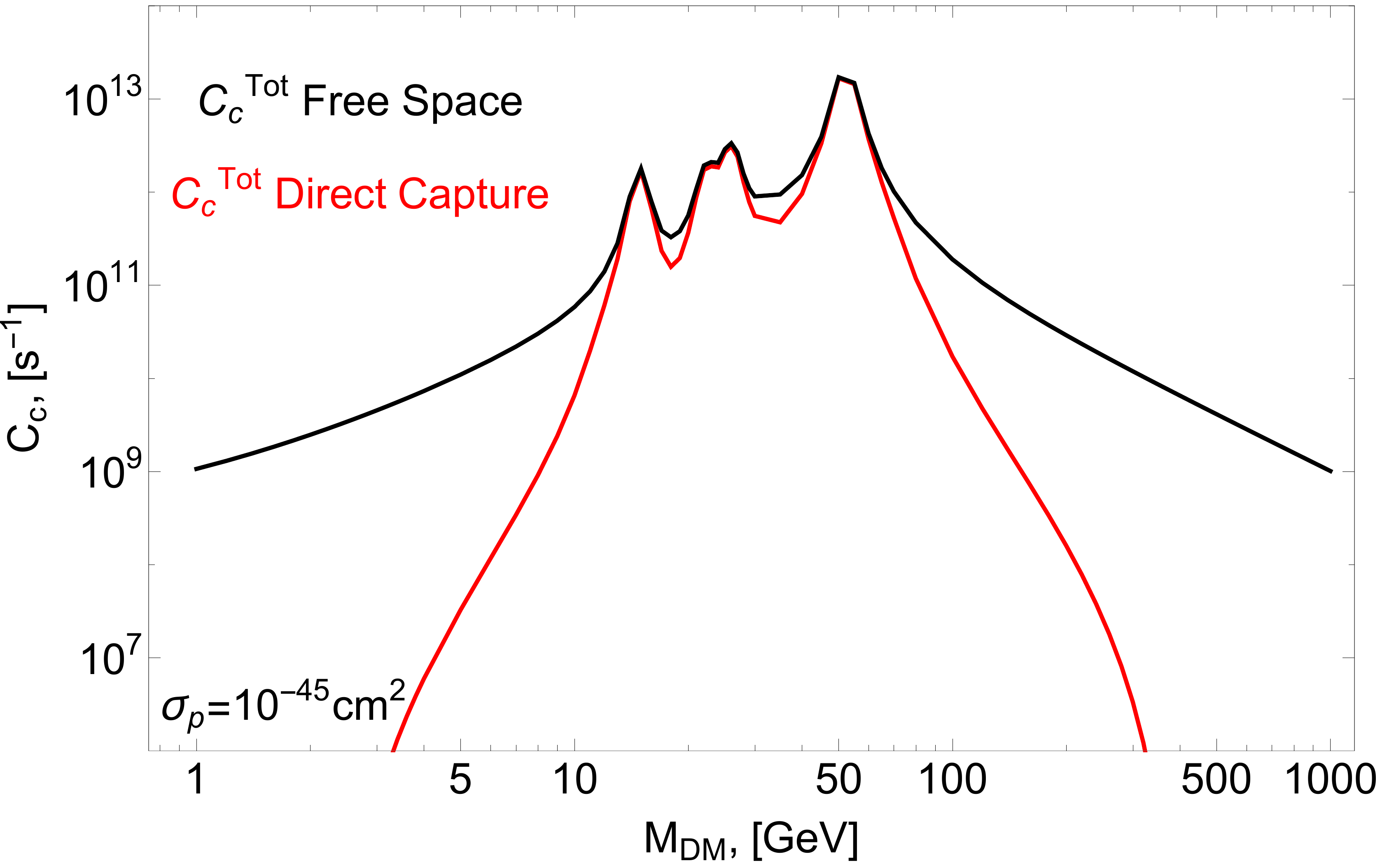}
\caption{Nuclear capture rate $C_c$ in the Earth as a function of DM
  mass for a reference $\sigma_{\text{p}} = 10^{-45} \text{cm}^2$. {\bf Left}:
  Contributions of oxygen (blue), magnesium (orange), silicon
  (purple), and iron (red) to the total capture rate (black), for
  direct capture from the unbound DM population. {\bf Right:}
  Comparison of the total capture rate from direct capture (red) and
  from the free space distribution (black).}
\label{plot:Earth_CCapture_FD_vs_Dir_Capt}
\end{center}
\end{figure}

\begin{table}[th!]
\centering
\small
\begin{tabular}{| c | p{3.3cm} | p{2.3cm}| p{3.2cm}|}
\hline
Element & \centering{ Isotope Abundance [by mass \%]}  & \centering{Core abundance [by mass \%]} & \centering{Mantle abundance [by mass \%]} \tabularnewline \hline
$^{16}\text{Ox}$  & \centering{100} & \centering{0} & \centering{44}  \tabularnewline \hline
$^{23}\text{Na}$  & \centering{100} & \centering{0} & \centering{0.27}\tabularnewline \hline
$^{24}\text{Mg}$  & \centering{79} & &  \\ \cline{1-2}
$^{25}\text{Mg}$ & \centering{1} & \centering{0} &\centering{22.8} \tabularnewline \cline{1-2}
$^{26}\text{Mg}$  &\centering{ 11} & &  \\ \hline
$^{27}\text{Al} $  & \centering{100} & \centering{0} & \centering{2.35}\tabularnewline \hline
$^{28}\text{Si} $  & \centering{100} & \centering{6} & \centering{21} \tabularnewline \hline
$^{32}\text{S} $  & \centering{100} & \centering{1.9} & \centering{0} \tabularnewline \hline
$^{40}\text{Ca} $  & \centering{100} &\centering{0} & \centering{2.53}\tabularnewline \hline                     
$^{52}\text{Cr} $  & \centering{83}.8 &\multirow{2}{*}{{\hspace{8mm} 0.9}} & \multirow{2}{*}{\hspace{9mm} 0.2625} \\ \cline{1-2}
$^{53}\text{Cr} $  & \centering{9.5} &  &  \\ \hline
$^{55}\text{Mn} $  &\centering{100} &\centering{0.3} & \centering{0}\tabularnewline \hline
$^{56}\text{Fe} $  & \centering{100} & \centering{85}  & \centering{6.26}\tabularnewline \hline
$^{59}\text{Co} $  & \centering{100} & \centering{0.25} &\centering{0}\tabularnewline \hline
$^{58}\text{Ni} $  & \centering{68} & \multirow{2}{*}{\centering{\hspace{9mm}5.2}}  &\multirow{2}{*}{\hspace{11mm}0.196} \\ \cline{1-2}
$^{60}\text{Ni} $  & \centering{26} &  &  \\ \hline
\end{tabular}
\caption{Elemental abundances used in the Earth model. All abundances are given in percentages with respect to the total mass. For the elements with multiple isotopes the listed core and mantle abundances refer to the total abundance of the element and not its individual isotopes. Isotope abundances are assumed spatially constant within the two indicated regions.}
\label{table:Earth_Elem_Abund}
\end{table}

\begin{figure}[th]
\begin{center}
\includegraphics[scale=0.3]{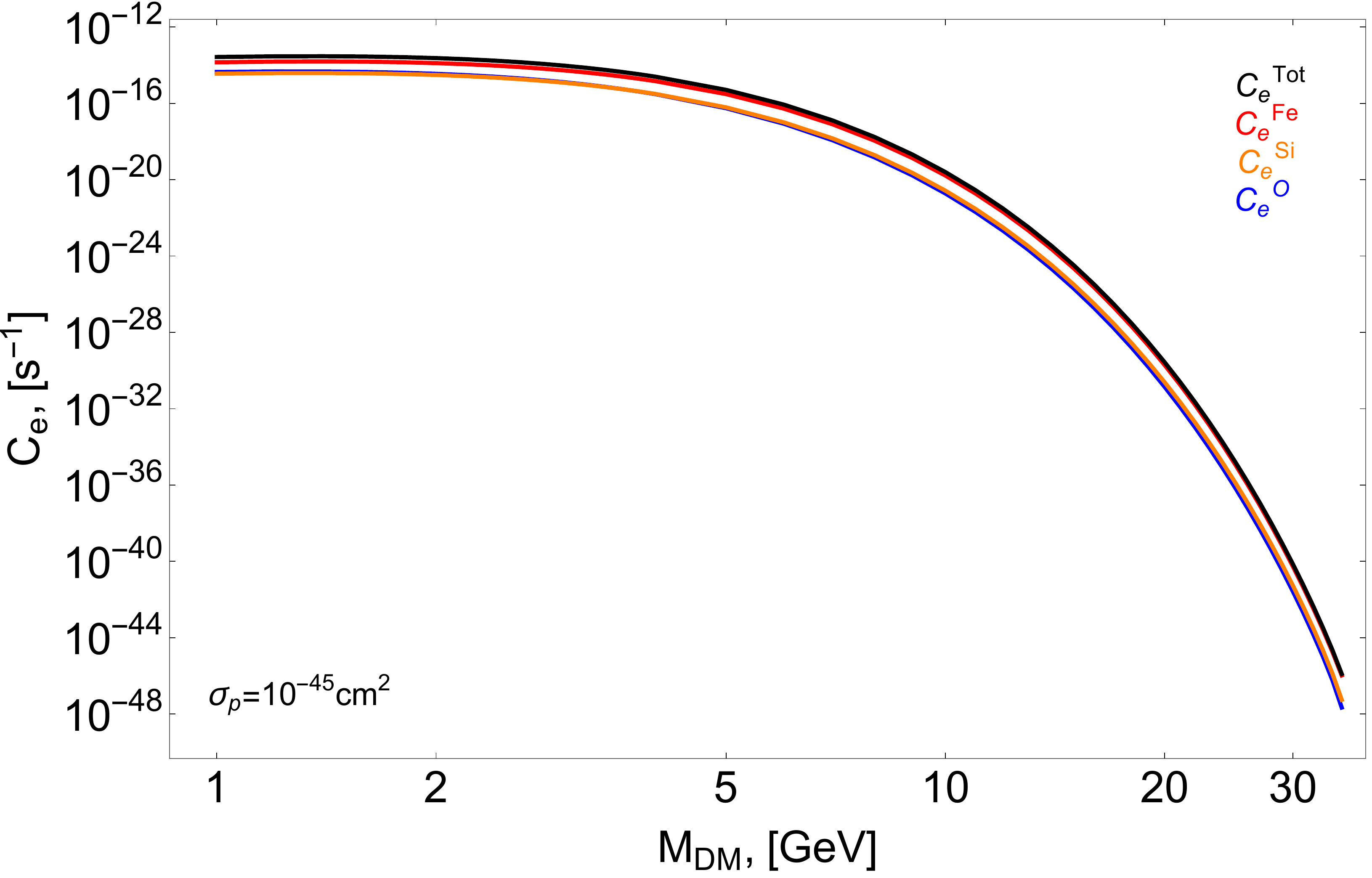}
\caption{Evaporation coefficient in the Earth, as a function of DM
  mass $M_{\text{DM}}$. The black curves show the total evaporation
  coefficient. The blue, orange, and red curves correspond to
  evaporation due to oxygen, silicon, and iron, respectively. We show
  results for a reference $\sigma_{\text{p}} = 10^{-45} \text{cm}^2$.}
\label{plot:Earth_CEvaporation}
\end{center}
\end{figure}

\section{Thermalization of captured DM}
\label{sec:therm}

The simplicity of the analysis in this work relies critically on the
assumption that captured DM quickly thermalizes with the nuclei of the
capturing body.  As direct detection experiments push allowed values
of the spin-independent DM-nucleon cross section $\sigma_{\text{p}}$
ever closer to the regime where thermalization in solar system objects
is no longer assured, a careful treatment of the thermalization of
captured DM becomes ever more important for understanding the validity
of this approach to gravitational capture.  Here we provide an updated
estimate of the thermalization of DM with solar nuclei through
spin-independent interactions, and examine how this estimate is
affected by the presence of DM self-interactions.  Our treatment here
is not exact; rather, we point out several generic effects that should
be taken into account in any discussion of thermalization and estimate
their effects. In particular, here we observe that: (1) DM-nuclei
scatterings allow the DM to shed ever smaller fractions of its
momentum as it loses energy; (2) requiring the thermalization
timescale $\tau_{\text{th}}$ to be shorter than the age of the solar
system is a {\it necessary but not sufficient} condition for the
captured DM population to be well described by a thermal distribution;
and (3) in the presence of DM self-interactions, energy exchange with
the halo DM population is potentially important, although, for constant
self-interaction cross-sections, not the dominant process limiting
thermalization.  Our results are summarized in
Figs.~\ref{fig:SunAndEarthThermalization}--\ref{plot:SunAndEarth_Thermalization_4}
below.

\subsection{Thermalization with nuclei}

The average fractional energy loss of a DM particle in a collision
with a nucleus is determined by the kinematics of the scattering,
$\widebar{\Delta E}/E = \mu_{i}/(2 \mu_{i,+}^2)$. Thus,
barring resonances where $M_{\text{DM}}\approx m_{i}$, DM does
not efficiently shed momentum in a typical scattering process, and
numerous scatterings are required in order to cool captured DM down to
the temperature of the massive body. A necessary condition for
thermalization is $\tau_\text{a} \geq \tau_{\text{th}}$, where
$\tau_{\text{a}}$ is the age of the capturing body and
$\tau_{\text{th}}$ is the amount of time needed for a typical captured
DM particle to down-scatter to the thermal energy in the core of the
capturing body. The interaction rate for elastic DM-nucleon collisions
is
\begin{align}
 \Gamma_i = v_i \sum_{j} n_j \sigma_j= v_i\sum_{j}\frac{1}{\lambda_j},
\end{align}
where $v_i$ is the DM particle velocity between collisions $i$
and $i+1$, $\sigma_j$ is the DM cross-section with nucleus $j$,
and the summation is performed over all nuclear species\footnote{Here
  we have assumed that the time that captured DM particles spend
  outside the volume of the capturing body is negligible in comparison
  to $\tau_{\text{th}}$ and $\tau_\text{a}$.  On the other hand, we
  will also use the volume-averaged density of nuclei in our estimate,
  which underestimates the scattering rate as DM settles into the
  dense core.  A full treatment of these issues would require orbital
  simulations and is beyond the scope of this paper; see also
  \cite{Widmark:2017yvd}.}. We have also introduced the mean free path
$\lambda_j$ for collisions with species $j$. Let $t_i$
denote the average time the DM particle spends between the two
collisions. Then our thermalization statement reads
\begin{equation}
\label{eq:taucond}
 \tau_\text{a} \geq \tau_{\text{th}} = \sum_{i=0}^{N_{\text{th}}-1} t_i =  \sum_{i=0}^{N_{\text{th}}-1}\frac{1}{\Gamma_i},
\end{equation}
where $N_{\text{th}}$ is the number of collisions required for the DM
particle to reach the thermal energy of nuclei in the core. We write
the nuclear cross-section as (neglecting form factors)
\be
\sigma_j = \sigma_{\text{p}}A^2_j\frac{M_{\text{DM}}^2 m^2_j}{(M_{\text{DM}}+m_j)^2}\frac{(m_{\text{p}}+M_{\text{DM}})^2}{m_{\text{p}}^2M_{\text{DM}}^2} \equiv \sigma_\text{p}K_j.
\ee
We can then obtain the average mean free path,
\be
\label{eq:meanlambda}
 \left\langle\frac{1}{\lambda}\right\rangle = \sum_j\left \langle\frac{1}{\lambda_j}\right\rangle = \sigma_{\text{p}} \sum_j \langle n_jK_j\rangle,
\ee
where the brackets denote averaging over the volume of the capturing
body, so that $\langle n_j K_j\rangle =
\frac{1}{V_\text{a}}\int d^3r\, n_j K_{j}$.  It is
useful to define a reduced $\bar \lambda$ that depends only on the
properties of the capturing body by pulling out a factor of
$\sigma_\text{p}$:
\beq
\frac{1}{ \langle\bar{\lambda}\rangle} \equiv\frac{1}{ \sigma_\text{p}} \langle\frac{1}{\lambda}\rangle
\eeq
Using Eq.~\ref{eq:meanlambda}, we can then rewrite
Eq.~\ref{eq:taucond} as
\begin{equation}
\label{eq:sigmabound0}
\sigma_\text{p} \geq \frac{\langle\bar{\lambda}\rangle}{\tau_\text{a}}\sqrt{\frac{M_{\text{DM}}}{2}}\sum_{i=0}^{N_{\text{th}}-1}\sqrt{\frac{1}{E_i}},
\end{equation}
where $ E_i = \frac{1}{2}M_{\text{DM}}v^2_i$ is the DM
kinetic energy.  To obtain a lower bound on $\sigma_{\text{p}}$ we now
compute the average energy lost by the DM in the $i$th collision.  In
a collision with nucleus $j$, the mean (i.e., angle-averaged) energy
loss is
\beq
\label{eq:qij}
Q_{i,j} = \frac{\mu_j}{2\mu^2_{+,j}}E_{i-1}
\eeq
(we neglect the thermal motion of the nucleus).  Since we have
multiple nuclear species in the bulk, we obtain the representative
energy loss by a weighted average of $Q_{i,j}$ over the
number fraction of the nuclear species and the elastic cross section:
\begin{equation}
\langle Q_i \rangle = \frac{\sum_j\langle f_j\sigma_jQ_{i,j}\rangle}{\sum_j \langle f_j\sigma_j\rangle}.
\end{equation}
This quantity $\langle Q_i \rangle$ represents the energy lost
in a typical collision. Using Eq.~\ref{eq:qij},
\begin{equation}
\langle Q_i \rangle = \langle \bar{\mu}\rangle E_{i-1}
\end{equation}
where we define
\begin{equation}
 \quad \langle \bar{\mu}\rangle = \frac{\sum_j \langle f_j K_j \frac{\mu_j}{2\mu^2_{+,j}}\rangle}{\sum_j\langle f_jK_j\rangle}.
\end{equation}
Given a model for the composition of an astrophysical body, $\langle
\bar{\mu} \rangle$ is completely specified as a function of DM
mass. This allows us to express the average DM energy after the $i$th
collision as a fraction of the DM energy immediately after capture,
$E_\text{0}$,
\begin{equation}
\label{eq:ei}
E_i = E_{i-1} - \langle Q_i \rangle =  (1 - \langle \bar{\mu} \rangle)^iE_\text{0}.
\end{equation}
After $N_{\text{th}}$ collisions, the DM particle's energy is equal to the
core's thermal energy. Hence
\begin{equation}
\label{eq:nth}
\frac{3}{2}T_{\text{core}} = E_{N_{\text{th}}} =  (1 - \langle \bar{\mu} \rangle)^{N_{\text{th}}}E_\text{0} .
\end{equation}
Using Eqs.~\ref{eq:ei} and~\ref{eq:nth} in Eq.~\ref{eq:sigmabound0} lets us sum the series, yielding
\begin{equation}
\sigma_{\text{p}} \geq  \sqrt{\frac{M_{\text{DM}}}{2}}\frac{\langle \bar{\lambda}\rangle}{\tau_\text{a}} E_\text{0}^{-\frac{1}{2}}\sum_{i=0}^{N_{\text{th}}-1}\big((1-\langle\bar{\mu}\rangle)^{-\frac{1}{2}}\big)^i = \sqrt{\frac{M_{\text{DM}}}{2E_\text{0}}}\frac{\langle \bar{\lambda}\rangle}{\tau_\text{a}}\frac{\sqrt{\frac{E_\text{0}}{E_{\text{th}}}}-1}{(\, 1-\langle \bar{\mu}\rangle )\,^{-\frac{1}{2}} -1 }.
\end{equation}
Finally, we need to estimate the initial energy after capture, $E_\text{0}$.
In an individual capture event, the DM particle will have an energy
after capture $E_\text{0} = \frac{1}{2}M(u^2 + v^2_{\text{esc}}(r)) -\Delta
E$, where $\Delta E$ is the energy lost in scattering against a
nucleus.  We can find the average $E_0$ by dividing the energy
injection rate by the capture rate $C_{\text{c}}$,
\begin{align}
E_\text{0} = \frac{\dot{E}_{\text{c}}}{C_{\text{c}}},
\end{align}
where $\dot{E}_{\text{c}}$ is constructed by the same logic as the
capture coefficient. Specifically,
\begin{align}
\dot{E}_{\text{c}} = \int_0^{R} 4\pi r^2dr \sum_j\frac{d\dot{E}_{\text{c},j}}{dV},
\end{align}
where the summation is performed over all relevant nuclear species and
the differential energy injection rate is
\begin{align}
\frac{d\dot{E}_{\text{c},j}}{dV} = n_{j}(r)n_{\text{DM}}\int_0^{\infty}\frac{d^3uf_\eta(u)}{u}w(r)\Omega^{\text{c}}_{u,j}(r),
\end{align}
with
\begin{align}
\Omega^{\text{c}}_{u,j}(r) =
w(r)\Theta\left(\frac{\mu_{j}}{\mu^2_{j,+}}
  -\frac{u^2}{w^2(r)}\right)\int^{\Delta E_{\text{max}}}_{\Delta
  E_{\text{min}}}\left(\frac{d\sigma}{d\Delta
    E}\right)_{j}\left(\frac{1}{2} M_{\text{DM}} w^2(r)-\Delta E\right)d(\Delta E).
\end{align}
Here $\Delta E_{\text{max}}$ and $\Delta E_{\text{min}}$ are
determined by the kinematics of the scattering and the requirement of
gravitational capture (Eqns~\ref{eq:ncapmin} and \ref{eq:ncapmax}),
and an explicit expression for $d\sigma/d\Delta E$ can be found in
Eq.~\ref{eq:dsigmadE}.

\begin{figure}[t]
\begin{subfigure}{0.5\textwidth}
  \hspace{-0.5cm}
  \includegraphics[width=1.0\linewidth]{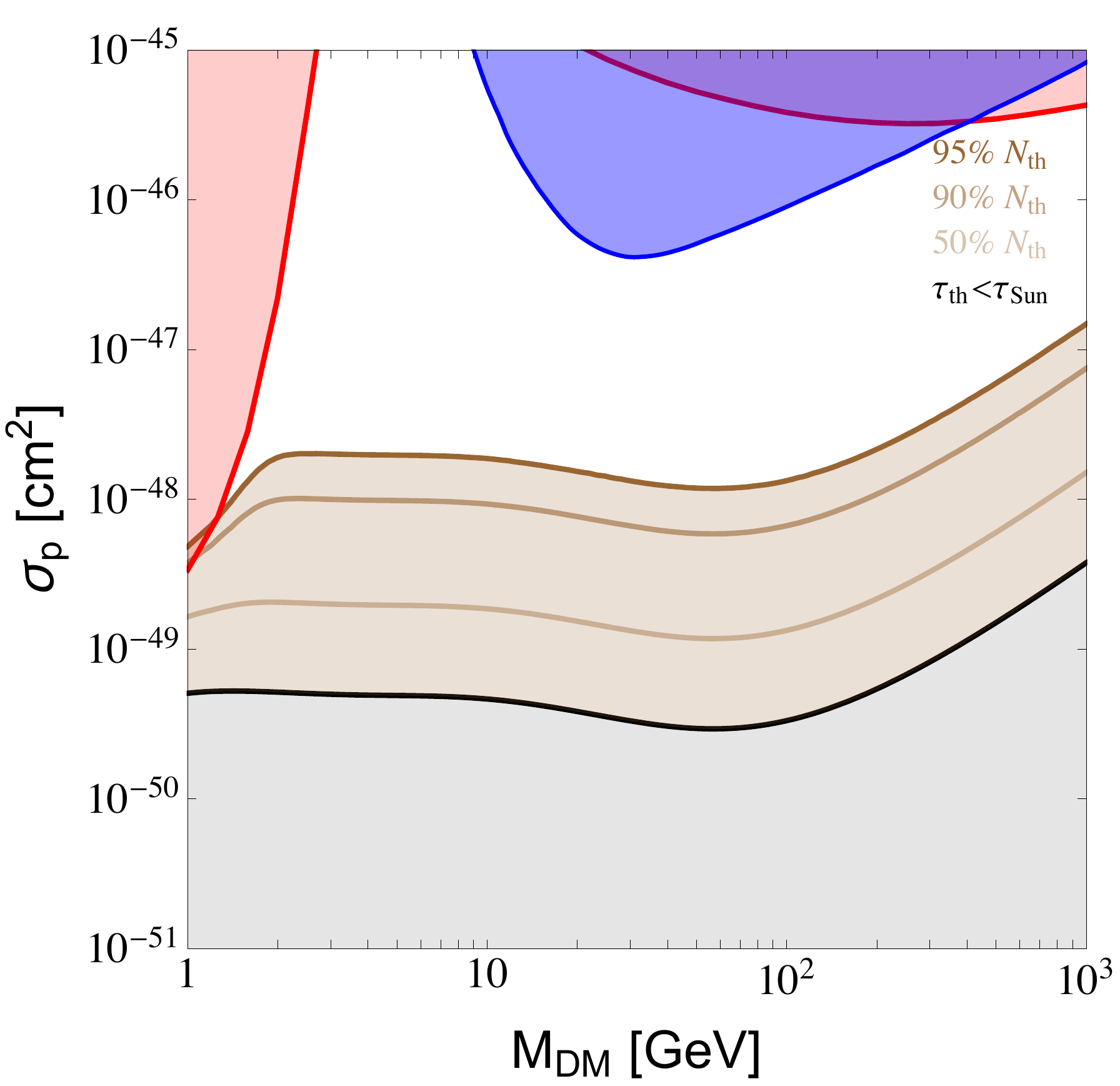}
\end{subfigure}%
\begin{subfigure}{0.5\textwidth}
  \hspace{-0.5cm}
  \includegraphics[width=1.0\linewidth]{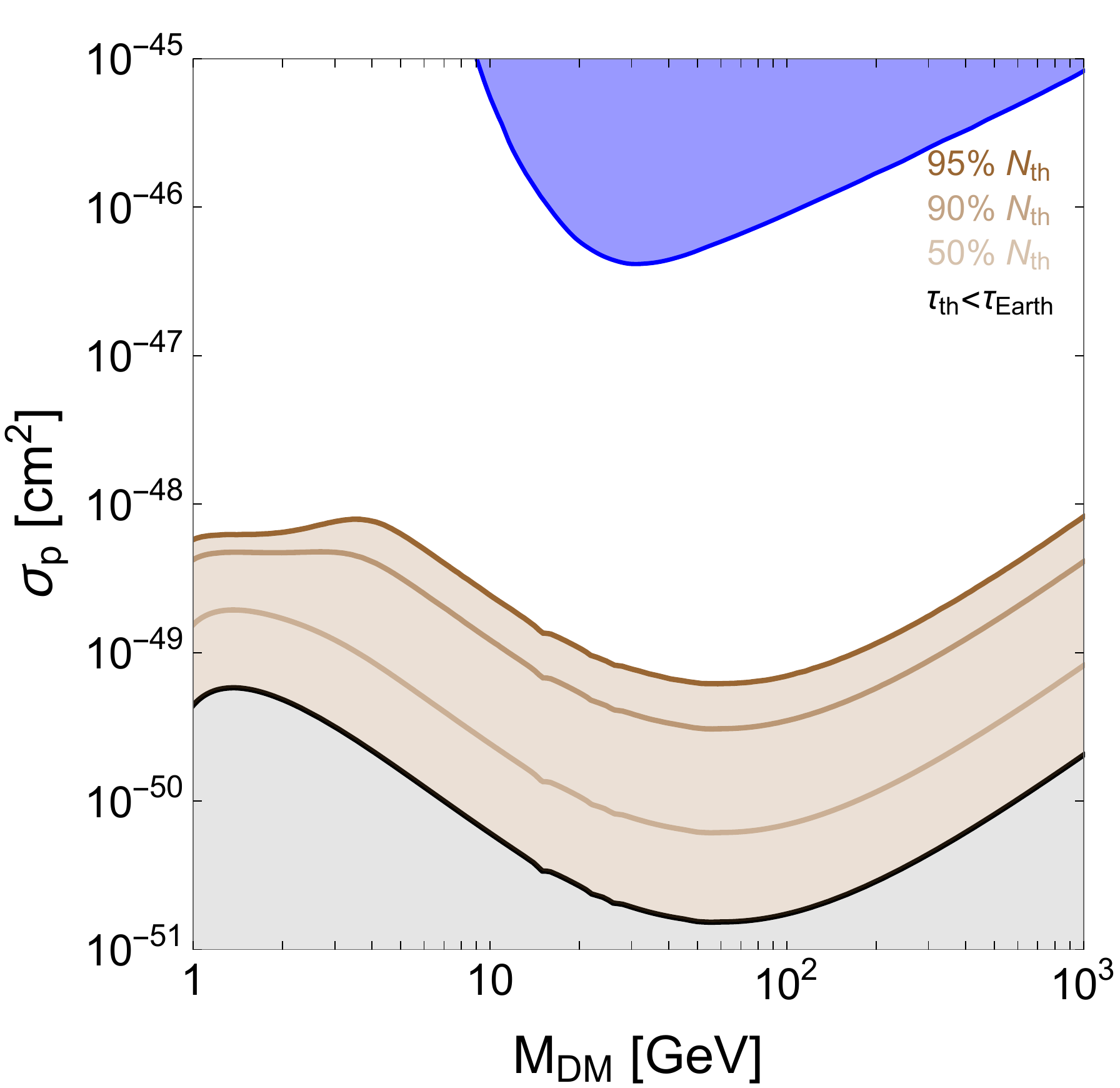}
\end{subfigure}
\caption{Estimated thermalization bound on $\sigma_{\text{p}}$ as a
  function of DM mass $M_{\text{DM}}$ in the Sun (left) and Earth
  (right). Above the solid brown lines the corresponding percentage of
  the captured population is thermalized. Below the black lines
  $\tau_{\text{th}} > \tau_\odot \text{ or }\tau_\Earth$. The blue shaded region is excluded by
  the PandaX, LUX and XENON1T direct detection experiments. In the left
  plot we also show in red regions that attain a steady-state
  population in the Sun, $\xi < \tau_\odot$; the corresponding regions
  for the Earth are off the plot.}
\label{fig:SunAndEarthThermalization}
\end{figure}

The resulting lower bound on $\sigma_{\text{p}}$ is shown in black in
Fig.~\ref{fig:SunAndEarthThermalization} for both the Sun and the
Earth.  Overall, thermalization in the Sun is more restrictive than in
the Earth, despite its higher overall core temperature, thanks to the
Sun's more energetic population of captured DM particles and its
composition of lighter nuclei.  Note that our neglect of nuclear form
factors in this estimate is reasonable for the Sun, where energy loss
on heavy species like iron is subdominant, but can lead to 
underestimates in the lower bound for the Earth, where iron is a
central contributor to DM down-scattering.  As the requirement of
thermalization in the Sun is more stringent at the higher DM masses
for which energy loss on iron is important, this underestimate will
not be important for our purposes.

However, the lower bound on $\sigma_{\text{p}}$ constructed above is a
necessary but not sufficient condition for the captured DM population
to be well-described by a thermal distribution.  If the thermalization
time is not much smaller than the age of the solar system, then only
the earliest captured DM particles will be thermalized. These
particles will only be a small fraction of the total population today,
rendering a thermal description of the population
inaccurate. Therefore, we introduce the following condition for the
thermal distribution to be a reasonably self-consistent approximation:
\begin{align}
\label{eq:ftherm}
N_{\text{th}} \geq f_{\text{th}} N(\tau_\odot),
\end{align}
which demands that at least a fraction $f_{\text{th}} \gtrsim 90\%$ of
the Sun-captured DM population today is thermalized. We impose an
analogous condition for the Earth-captured DM population today
$N(\tau_{\Earth})$. In order for the population $N_{\text{th}}$ to be
thermalized, each of these particles must spend a time
$\tau_{\text{th}}$ inside the Sun (or Earth) after capture. Hence,
\begin{align}
N_{\text{th}} = N(\tau_\odot - \tau_{\text{th}}),
\end{align}
where $\tau_{\text{th}}$ is the thermalization time computed
above. Imposing this condition on the captured DM population results
in a stronger constraint on $\sigma_{\text{p}}$ as a function of
$M_{\text{DM}}$. We show these lower bounds for the Sun and Earth in
blue in Figure \ref{fig:SunAndEarthThermalization}.  For reasonable
choices of $f_{\text{th}}$ this results in a lower bound on
$\sigma_{\text{p}}$ that is larger by about an order of magnitude
compared to the bound obtained simply by requiring $\tau_{\text{th}} >
\tau_\odot$.

\subsection{Thermalization in the presence of DM self-interactions}

Adding dark matter self-interactions changes the above discussion in
two ways.  First and most importantly, self-interactions provide
additional mechanisms for energy transfer into, among, and out of the
captured DM population \cite{Chen:2015poa}.  As our aim here is to
establish the region where our ansatz describing the captured DM
population as thermalized at the core temperature of the capturing
body is self-consistent, the most important of these processes for our
purposes is the net rate of energy injected into the captured dark
matter population from scattering off of DM particles in the halo. For
captured DM to remain at the nuclear temperature, the rate of this
energy injection should not exceed the rate at which bound DM can
transfer energy to the nuclei.  Secondarily, the presence of
self-interactions changes the evolution of the total captured DM
population with time, and hence can alter how long it takes for a
given fraction $f_{\text{th}}$ of the total DM population to
thermalize even in the absence of changes to the thermalization
timescale $\tau_{\text{th}}$.

Given a typical energy injection $\langle E_{\text{Halo, el}} \rangle$
from elastic scattering with halo DM, we can estimate
the rate at which this energy is transferred to the nuclei:
\begin{equation}
\label{eq:dotEnuc}
\dot{E}_{\text{el, ave}} = \frac{\langle E_{\text{Halo, el}} \rangle - E_{\text{th}} }{\tau_{\text{th}}},
\end{equation}
where, as in the previous discussion, $\tau_{\text{th}}$ is the
timescale for a DM particle to downscatter from its initial energy to
$E_{\text{th}}$.  Once we determine the characteristic initial energy
$\langle E_{\text{Halo, el}} \rangle$ from elastic scatterings between
bound and halo DM, the technology of the previous subsection lets us
directly evaluate this rate.

We begin by estimating the per-particle rate of energy transfer from
halo DM particles to captured DM particles, $\dot{E}_{\text{Halo,
    el}}$.  Per unit volume, this rate can be expressed as
\begin{align}
\frac{d\dot{E}_{\text{Halo, el}}}{dV} = n_{\text{DM}}\int_{0}^{\infty}\frac{d^3uf_{\eta}(u)}{u}w(r)\Omega^{\text{el}}(w),
\end{align}
where $\Omega^{\mathrm{el}}(w)$ is the energy transfer rate per halo
DM particle in a single scattering, given in terms of the local
(thermal) density of captured DM particles $n_{\text{c}}(r)$ as
\beq
\Omega^{\text{el}}(w) = n_{\text{c}}(r) w \int d\cos\theta \,\Delta E(\cos\theta) \frac{d\sigma_{\text{xx}}}{d\cos\theta} \times \Theta( f(\cos\theta)).
\eeq
Here the theta-function $\Theta (f(\cos\theta))$ enforces the
conditions on the scattering angle required for the collision to
result in one bound and one unbound DM particle.  Two regions of phase
space contribute:
\begin{align}
\label{eq:theta1}
\cos{\theta}\leq \text{Min}\left( \frac{v^2_{\text{esc}}(r)-u^2}{w^2(r)}, \frac{u^2 -v^2_{\text{esc}}(r)u^2}{w^2(r)}\right),
\end{align}
 where the two particles switch places, and
\begin{align}
\label{eq:theta2}
\cos{\theta}\geq \text{Max}\left( \frac{v^2_{\text{esc}}(r)-u^2}{w^2(r)}, \frac{u^2 -v^2_{\text{esc}}(r)u^2}{w^2(r)}\right),
\end{align}
where the initially bound particle remains bound.  The energy
transferred to the captured DM population is $\Delta E = M_{\text{DM}}
w^2 (1\pm\cos\theta)/4$, with the $+$ sign pertaining to the first
case and the $-$ sign to the second.

It is useful to split $\Omega^{\text{el}}(w)$ into two pieces:
\begin{align}
\Omega^{\text{el}}(w) = w(r)\left( \Theta(u - v_{\text{esc}}(r))\Omega_1 +  \Theta(v_{\text{esc}}(r)- u)\Omega_2\right).
\end{align}
When $u>v_{\text{esc}}$, the maximum scattering angle allowed by
Eq.~\ref{eq:theta1} is $\cos\theta_{\text{max}} = (v^2-u^2)/w^2$, and
the minimum allowed scattering angle of Eq.~\ref{eq:theta2} is
$\cos\theta_{\text{min}}=-\cos\theta_{\text{max}}$.  This yields for
$\Omega_1$
\begin{eqnarray}
\Omega_1 &=& \frac{1}{4} m w^2\left[ \int_{-1}^{\cos\theta_{\text{max}}} d\cos\theta
\frac{d\sigma_{\text{xx}}}{d\cos\theta}(1-\cos\theta)+\int^{1}_{-\cos\theta_{\text{max}}} d\cos\theta
\frac{d\sigma_{\text{xx}}}{d\cos\theta}(1+\cos\theta)\right] \nonumber \\
        &= & \frac{1}{2} m w^2 \int_{-1}^{\cos\theta_{\text{max}}} d\cos\theta
\frac{d\sigma_{\text{xx}}}{d\cos\theta}(1-\cos\theta),
\end{eqnarray}
where the second equality holds for any cross-section that is even in
$\cos\theta$, and similarly for $\Omega_2$.  Specializing to a
constant cross-section, the rate of energy exchange with the halo is
\begin{align}
\dot{E}_{\text{Halo,el}} = \frac{1}{8}\rho_{\odot}\sigma_{\text{xx}}\int^{R}_{0} 4\pi dr r^2 n_{\text{c}}(r)\left[\int_{0}^{v_{\text{esc}}(r)}\frac{d^3uf_\eta(u)}{u}f_{+}(r,u) + \int^{\infty}_{v_{\text{esc}}(r)}\frac{d^3uf_\eta(u)}{u}f_{-}(r,u) \right],
\end{align} 
where
\begin{align}
f_{\pm}(r,u) = w^4(r) \pm 2w^2(r)(u^2 - v^2_{\text{esc}}(r)) + (u^2 -v^2_{\text{esc}}(r))^2.
\end{align}
Similarly, the rate for these elastic scattering events to occur is given by
\begin{align}
\Gamma_{\text{Halo, el}}= 2n_{\text{DM}}\sigma_{\text{xx}}\int_{0}^{R}4\pi r^2dr\, n_\text{c}(r)
\left[\int_{0}^{v_{\text{esc}}(r)}\frac{d^3uf_\eta (u)}{u}u^2 + \int^{\infty}_{v_{\text{esc}}(r)}\frac{d^3uf_\eta(u)}{u}v^2_{\text{esc}}(r) \right].
\end{align}
This allows us to construct the average energy deposited in the captured
population following an elastic scattering with DM in the halo:
\begin{align}
\langle E_{\text{Halo, el}} \rangle = \frac{\dot{E}_{\text{Halo,el}}}{\Gamma_{\text{Halo, el}}}.
\end{align} 

We can now compare the rate of energy deposition from the halo
$\dot{E}_{\text{Halo,el}}$ to the rate at which this energy is
transferred to the nuclei, Eq.~\ref{eq:dotEnuc}.  Our self-consistency
criterion is
\begin{align}
\label{eq:selfintconsistent}
\dot{E}_{\text{el, ave}} = \dot{E}_{\text{Halo,el}}.
\end{align} 
%
\begin{figure}[h!]
\includegraphics[width=0.5\linewidth]{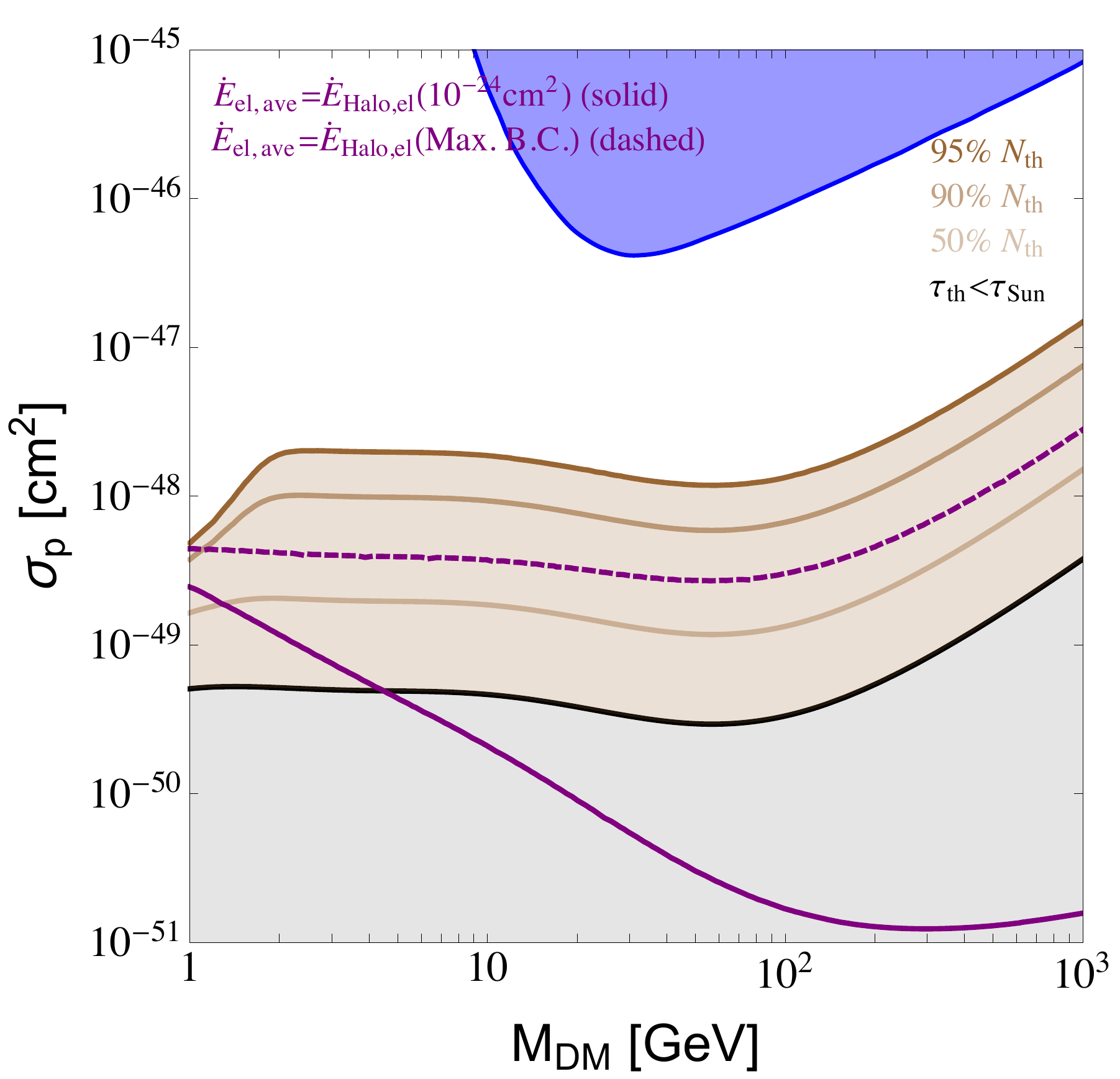}
\caption{Thermalization bound on $\sigma_{\text{p}}$  as a
  function of DM mass $M_{\text{DM}}$ in the Sun, for constant
  self-interaction cross-section. Above the solid (dashed) purple line
  $\dot{E}_{\text{el, ave}}^{1/2} >\dot{E}_{\text{Halo,el}}$, for
  $\sigma_{\text{xx}} = 10^{-24}$ $\text{cm}^2$ (maximum
  $\sigma_{\text{xx}}$ allowed by the Bullet Cluster bound). The
  corresponding bounds for the Earth are everywhere below
  $\sigma_{\text{p}}=10^{-51} \text{cm}^2$, and thus the Earth plot is not
  shown. The blue shaded region is excluded by
  the PandaX, LUX and XENON1T direct detection experiments.}
\label{plot:SunAndEarth_Thermalization_1}
\end{figure}
%
The elastic nuclear rate depends on $M_{\text{DM}}$ and
$\sigma_{\text{p}}$, while the elastic halo rate depends on
$M_{\text{DM}}$ and $\sigma_{\text{xx}}$. At fixed
$\sigma_{\text{xx}}$ the condition of Eq.~\ref{eq:selfintconsistent}
becomes a contour in the $(M_{\text{DM}}, \sigma_{\text{p}})$
plane. Above the contour the thermal assumption is self-consistent:
captured particles, while continuously being heated via
$\dot{E}_{\text{Halo,el}}$, shed their energy sufficiently
quickly. These contours are shown in
Fig.~\ref{plot:SunAndEarth_Thermalization_1}.

\begin{figure}[t]
\begin{subfigure}{0.5\textwidth}
  \hspace{-0.5cm}
  \includegraphics[width=1.0\linewidth]{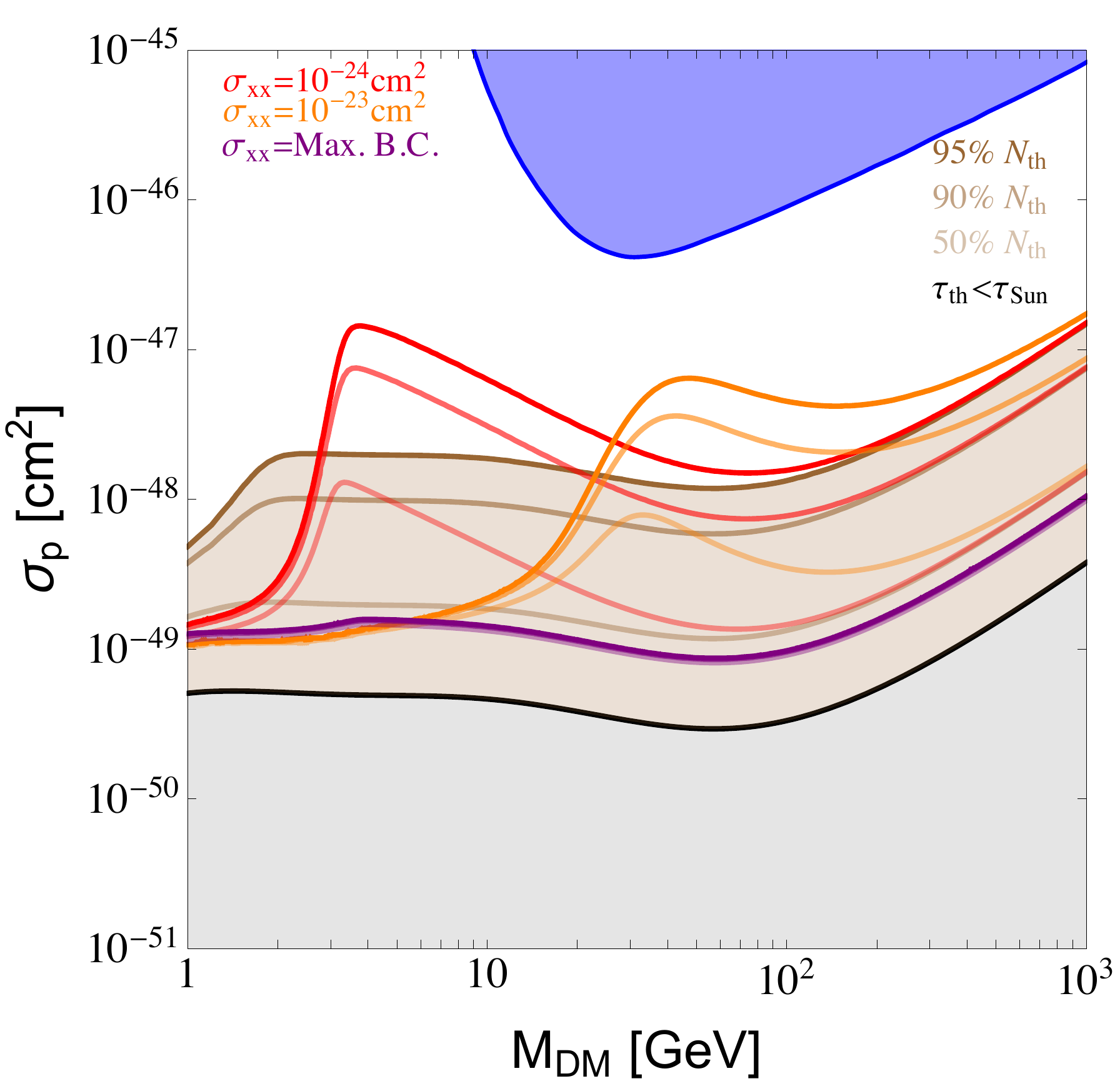}
\end{subfigure}%
\begin{subfigure}{0.5\textwidth}
  \hspace{-0.5cm}
  \includegraphics[width=1.0\linewidth]{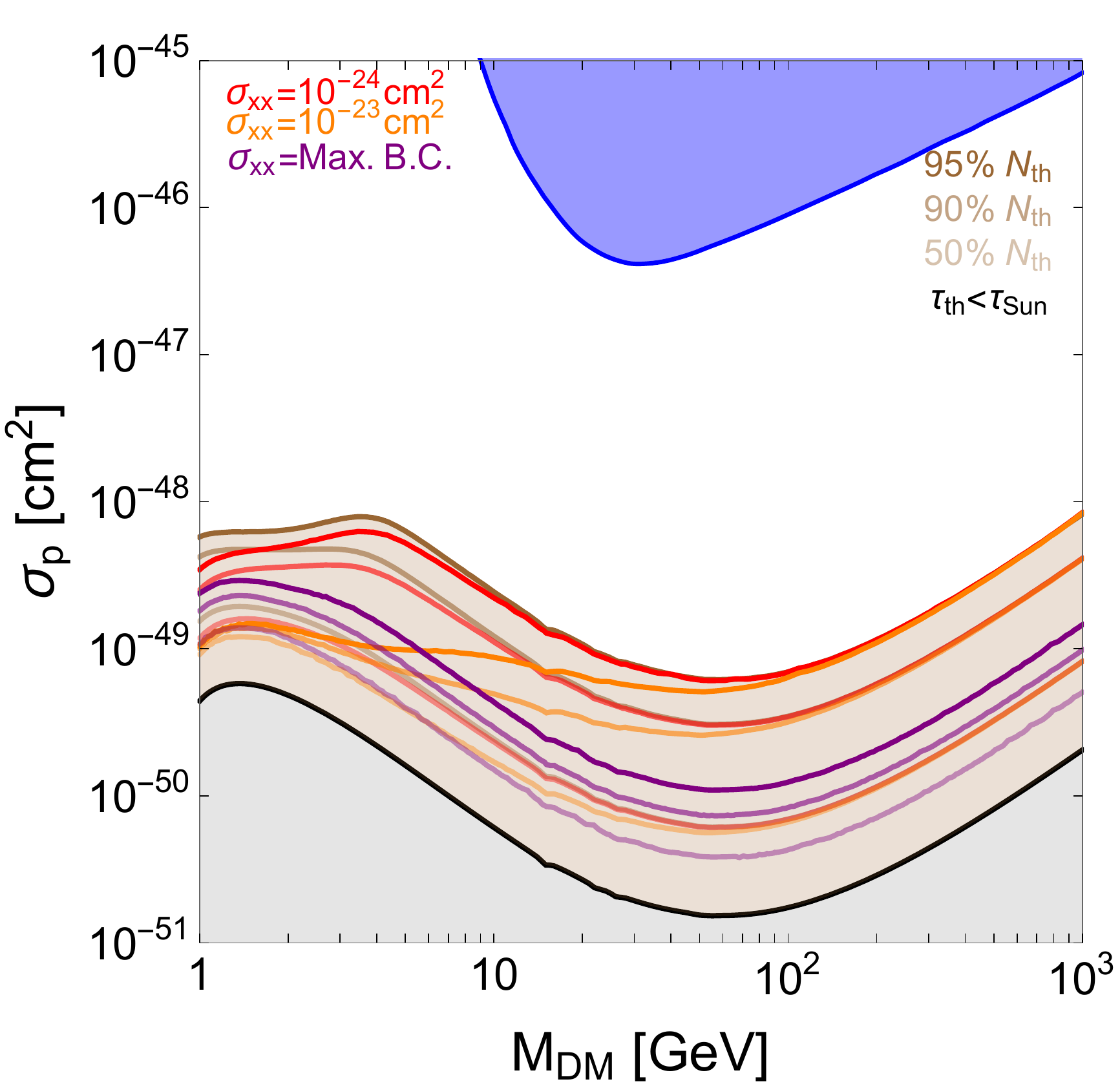}
\end{subfigure}
\caption{Thermalization bound on $\sigma_{\text{p}}$ as a function of
  DM mass $M_{\text{DM}}$ in the Sun (left) and Earth (right), for
  constant self-interaction cross-sections. The colors of the contours
  correspond to different choices of $\sigma_{\text{xx}}$ as
  indicated. The brown contours are constructed assuming no
  self-interactions. For each choice of $\sigma_{\text{xx}}$ we show
  three curves of differing opacities which correspond to different
  thermalization percentages: above the most opaque contour 95\% of
  the population is thermalized, above the least opaque contour 50 \%
  of the population is thermalized. The blue shaded region is excluded by
  the PandaX, LUX and XENON1T direct detection experiments.}
\label{plot:SunAndEarth_Thermalization_4}
\end{figure}

We must additionally consider how the size of the thermal population
is impacted by the introduction of self-interactions. We recall the
thermalization criterion introduced in the previous subsection,
Eq.~\ref{eq:ftherm}, which demands that at least a fraction
$f_{\text{th}} \gtrsim 90 \%$ of the total captured DM population
today $N(\tau_\odot)$ had enough time to thermalize. In the presence
of self-interactions the evolution of $N(t)$ will change because of
self-capture or self-ejection.  This effect by itself will change the
timescale for thermalization through nuclear interactions.  These
effects are shown in Fig.~\ref{plot:SunAndEarth_Thermalization_4}.

Here we discuss the qualitative features shown in
Fig.~\ref{plot:SunAndEarth_Thermalization_4}. We begin with the
Sun. For small and moderate $\sigma_{\text{xx}}$ the equilibration
time is still small compared to the age of the Sun. Because of
self-interactions, the number of captured particles in the Sun is
always larger than in the absence of self-interactions. Hence $N(\tau
- \tau_{\text{th}})$ is larger in the presence of self-interactions
than in their absence. The thermalization constraint we impose is
essentially that $N(\tau - \tau_{\text{th}})$ be sufficiently close to
$N(\tau)$. Now, in the time interval $[ \tau -\tau_{\text{th}}, \tau
]$ more particles would be captured in the presence of
self-interactions, driving $N(\tau - \tau_{\text{th}})$ to lower
values compared to $N(\tau )$. So, in order to satisfy this constraint
$\sigma_{\text{p}}$ must increase. This is shown by the red and orange
curves of the left plot of
Fig.~\ref{plot:SunAndEarth_Thermalization_4}. At large $M_{\text{DM}}$
the self-capture coefficient is negligible so that the two sets of
contours merge into the contours which assume no self-interactions.

One would then naively expect $\sigma_{\text{p}}$ to be larger for
larger $\sigma_{\text{xx}}$ in order for the thermalization criterion
to be satisfied. However, at sufficiently large $\sigma_{\text{xx}}$
the equilibration time becomes small enough (because $C_{\text{sc}}$
increases with $\sigma_{\text{xx}}$) so that the population today is
at equilibrium, while the population at $\tau -\tau_{\text{th}}$ is
very near equilibrium. In this case $\sigma_{\text{p}}$ can be smaller
because the captured population at the two times $\tau
-\tau_{\text{th}}$ and $\tau $ are forced to be proximal by the
equilibration mechanism. This behavior is shown by the purple
line. For intermediate $\sigma_{\text{xx}}$, the population
equilibrates at small $M_{\text{DM}}$ and the behavior is the same as
previously described (purple line). At large $M_{\text{DM}}$, the
population is not equilibrated, but the self-capture coefficient
remains important. In this region we have the same behavior as for
moderate $\sigma_{\text{xx}}$ at moderate masses.

For the thermalization of the Earth-captured population
self-interactions act in the opposite direction: the number of
captured particles in the Earth is always smaller that in the absence
of self-interactions. So, in the time interval $[ \tau
-\tau_{\text{th}}, \tau ]$ particles would be ejected in the presence
of self-interactions, driving $N(\tau)$ closer to $N(\tau -
\tau_{\text{th}})$. In turn, $\sigma_{\text{p}}$ can be smaller for
the thermalization criterion to be satisfied. Hence, for larger
$\sigma_{\text{xx}}$, the bound on $\sigma_{\text{p}}$ is lower (red
and orange curves of right plot of Figure
\ref{plot:SunAndEarth_Thermalization_4}). For sufficiently large
self-interactions, the equilibration time is so small that the
population is at equilibrium today and near equilibrium at $\tau
-\tau_{\text{th}}$. This is similar to the Sun-captured population,
where the thermalization criterion is satisfied because the captured
populations at the two times $\tau -\tau_{\text{th}}$ and $\tau$ are
forced to be proximal by the equilibration mechanism.  We have checked
that for the Earth the bounds do not depend on the choice of incident
velocity distribution.

Finally, we comment that another consequence of introducing
self-interactions is that captured DM acquires an additional energy
loss mechanism via scattering against other captured DM particles.
Thus in general we would expect the introduction of self-interactions
to reduce the thermalization timescale.  We have conservatively
neglected this reduction of the thermalization timescale.
Additionally, DM particles which become gravitationally bound via
self-capture have a different characteristic initial energy because of
the differing kinematic constraints, but this effect is a small
correction.


\bibliographystyle{jhep}
\bibliography{references_list}

\end{document}